\documentclass[preprint2]{aastex}    
\usepackage{psfig}
\usepackage{apjfonts}

\slugcomment{Draft of \today}

\shortauthors{Strickland \etal}
\shorttitle{Hot gas in the halos of star-forming galaxies}

\newcommand{\eg}{{\rm e.g.\ }}

\newcommand{\ie}{{\it i.e.\ }}
\newcommand{\cf}{{\rm cf.\ }}
\newcommand{\etal}{{\rm et al.\thinspace}}

\newcommand{\cm}{{\rm\thinspace cm}}
\newcommand{\km}{{\rm\thinspace km}}
\newcommand{\pcc}{\hbox{$\cm^{-3}\,$}}
\newcommand{\s}{{\rm\thinspace s}}
\newcommand{\yr}{{\rm\thinspace yr}}
\newcommand{\erg}{{\rm\thinspace erg}}
\newcommand{\ps}{\hbox{\s$^{-1}\,$}}
\newcommand{\pyr}{\hbox{\yr$^{-1}$}}
\newcommand{\ergps}{\hbox{$\erg\s^{-1}\,$}}

\newcommand{\kmps}{\hbox{$\km\s^{-1}\,$}}
\newcommand{\pcmsq}{\hbox{$\cm^{-2}\,$} }

\newcommand{\halpha}{H$\alpha$}

\newcommand{\hi}{H{\sc i}}
\newcommand{\hii}{H{\sc ii}}
\newcommand{\nii}{[N{\sc ii}]}

\newcommand{\nH}{\hbox{$N_{\rm H}$}}
\newcommand{\Mdot}{\hbox{$\dot M$}}

\newcommand{\pc}{{\rm\thinspace pc}}
\newcommand{\kpc}{{\rm\thinspace kpc}}
\newcommand{\Mpc}{{\rm\thinspace Mpc}}
\newcommand{\keV}{{\rm\thinspace keV}}

\newcommand{\Lsol}{\hbox{$\thinspace L_{\sun}$}}
\newcommand{\Msol}{\hbox{$\thinspace M_{\sun}$}}
\newcommand{\Zsol}{\hbox{$\thinspace Z_{\sun}$}}

\newcommand{\photons}{{\rm\thinspace photons}}
\newcommand{\photsurfb}{{\photons\ps\pcmsq{\rm\thinspace arcsec}$^{-2}\thinspace$}}
\newcommand{\ergsurfb}{{\ergps\pcmsq{\rm arcsec}$^{-2}\thinspace$}}

\begin{document}

\title{A high spatial resolution X-ray and H$\alpha$
study of hot gas in the halos of star-forming disk galaxies. 
I. Spatial and spectral properties of the diffuse X-ray emission}

\author{David K. Strickland,\altaffilmark{1,2,3}
	Timothy M. Heckman,\altaffilmark{3}
	Edward J.M. Colbert,\altaffilmark{3}
	Charles G. Hoopes,\altaffilmark{3} and
	Kimberly A. Weaver.\altaffilmark{4}}

\altaffiltext{1}{{\it Chandra} Fellow.}
\altaffiltext{2}{Guest investigator of the UK Astronomy Data Centre.}
\altaffiltext{3}{Department of Physics and Astronomy, 
	The Johns Hopkins University,
	3400 North Charles Street, Baltimore, MD 21218}

\altaffiltext{4}{NASA/Goddard Space Flight Center, 
	Code 662, Greenbelt, Maryland 20771}

%

\begin{abstract}
We present arcsecond resolution {\it Chandra} X-ray and ground-based optical
\halpha~imaging of a sample of ten edge-on star-forming disk
galaxies (seven starburst and three ``normal'' spiral galaxies),
a sample which covers the full range of star-formation intensity found
in disk galaxies.
The X-ray observations make use of the unprecedented
spatial resolution of the {\it Chandra} X-ray observatory to
more robustly than before remove X-ray emission from point sources, and hence
obtain the X-ray properties of the diffuse thermal emission alone.
We have combined the X-ray observations
with existing, comparable-resolution, ground-based \halpha~and R-band
imaging, and present a mini-atlas of images on a common spatial
and surface brightness scale to aid cross-comparison. 
In general, the morphology of the extra-planar diffuse X-ray
emission is very similar to the extra-planar
\halpha~filaments and arcs, on both small and 
large scales (scales of 10's of pc 
and kiloparsecs respectively). The most spectacular cases of this
are found in NGC 1482 (for which we provide the first published 
X-ray observation) and NGC 3079. 
We provide a variety of quantitative measures of how the spectral hardness and
surface brightness of the diffuse X-ray emission varies with increasing
height $z$ above the plane of each galaxy. Of the eight
galaxies in which diffuse X-ray emitting halos are found (the starbursts
and the normal spiral NGC 891), 
significant spatial variation in
the spectral properties of the extra-planar emission ($|z|\ge2$ kpc) is
only found in two cases: NGC 3628 and NGC 4631. 
In general, the vertical distribution of the halo-region 
X-ray surface brightness is best described as an exponential, with
the observed scale heights of the sample galaxies lying in the range
$H_{\rm eff} \sim 2$ -- 4 kpc.
The presence of extra-planar X-ray emission is alway associated with the
presence of extra-planar optical line emission of similar vertical extent.
No X-ray emission was detected from the halos
of the two low mass normal spiral galaxies NGC 6503 and NGC 4244.
AGN, where present, appear to play no role in powering or shaping the
outflows from the the starburst galaxies in this sample.
The {\it Chandra} 
ACIS X-ray spectra of extra-planar emission from all these galaxies
can be fit with a common two-temperature spectral model with
an enhanced $\alpha$-to-iron element ratio. This is consistent
with the origin of the X-ray emitting gas being \emph{either}
metal-enriched merged SN ejecta \emph{or} shock-heated
ambient halo or disk material with moderate levels of metal depletion
onto dust.  Our favored model is that SN feedback in the disks
of star-forming galaxies create, via blow out and venting of hot
gas from the disk, tenuous exponential atmospheres
of density scale height $H_{\rm g} \sim 4$ -- 8 kpc.
The soft thermal X-ray emission observed in the halos of the starburst
galaxies is either this pre-existing halo medium, which has been 
swept-up and shock heated by the starburst-driven wind, or
wind material compressed near the walls of the outflow by reverse shocks
within the wind. In either case the X-ray emission provides us with
a powerful probe of the properties of gaseous halos around star-forming
disk galaxies.
\end{abstract}


\section{Introduction}
\label{sec:introduction}

Massive stars exercise a profound influence over the baryonic component
of the Universe, through the return of ionizing radiation, and
via supernovae (SNe), kinetic energy and metal-enriched gas, back into the
interstellar medium (ISM) from which they form --- 
usually called ``feedback''. 
Feedback influences gas-phase conditions in the immediate
environment of the clusters within which the massive stars 
form \citep{mckee95,hollenbach97,wiseman98,pudritz00}, 
on galactic scales the phase structure and energetics of the ISM 
\citep{mckee77,cox81,norman89,norman96},
and on multi-Mpc scales the thermodynamics and enrichment of
the inter-galactic medium (IGM --- \eg see \citealt*{chiang88,ham90,shapiro94,voit96,heckman99,aguirre01}).

The vast range of spatial scales involved is only one of the difficulties 
encountered in attempting to study feedback. Restricting the discussion
to purely mechanical feedback from SNe and stellar winds (commonly 
termed SN feedback), a further 
difficulty is the broad range of
complicated gas-phase physics -- (magneto)hydrodynamic
effects such as shocks and turbulence, thermal conduction, and non-ionization
equilibrium emission processes. 
A final complication is that much of the energy and
metal-enriched material involved is in the hard-to-observe coronal 
($T \ga 10^{5}$ K) and hot ($T \ga 10^{6}$ K) gas phases, necessitating the
use of space-based FUV and X-ray telescopes.

X-ray observations of the halo-regions of nearby edge-on disk galaxies
provide the best single probe of the action of mechanical feedback 
on the galactic scale. Optical spectroscopy and imaging, 
sensitive to warm neutral and ionized gas with $T \sim 10^{4}$ K, 
are a less direct probe of mechanical
feedback as much of the excitation energy is supplied by ionizing radiation
from massive stars (radiative feedback), even when the kinematics
and spatial distribution
of this gas are controlled by SN feedback.
If mechanical feedback
processes transport mass, newly-synthesized metals, and energy from galaxies
into the IGM, then this material must pass through the halo of the host galaxy
on its way into the IGM, where its properties can be ascertained
before it expands and fades into observational invisibility.

Diffuse thermal X-ray emission is seen to extend to large
distances ($5$ to 20 kpc) above the planes of edge-on {\em starburst}
galaxies (e.g. \citealt*{fhk90,armus95,rps97,dwh98}),
which also show unambiguous optical emission line evidence for energetic
galactic scale outflows --- ``superwinds''
\citep{ham90}. These outflows are potentially very significant
in enriching and heating the IGM \citep{heckman99}. 

That superwinds can be driven by the collective power of large numbers of
supernovae is beyond doubt --- the observational evidence for the link between
starburst regions and kinematic and/or morphological 
evidence of outflow is overwhelming
(\eg see \citealt{bland88,ham90,mckeith95,lehnert95,lehnert96,dahlem97},
among many others). 
Nevertheless, the question of whether Seyfert nuclei or low-luminosity
AGN (LLAGN) can also contribute significantly to these
outflows, or can directly drive poorly-collimated superwind-like outflows
in the absence of any starburst, has been raised many times 
\citep*{schiano85,veilleux94,colbert96a,colbert96b,williams99,pietsch98,veilleux02b}.
This is a potentially serious complication for assessing SN feedback, as
many of the ``classical'' starburst superwind galaxies also host weak Seyferts
or LLAGN.

Both fully-numerical and semi-analytical
theoretical modeling
\citep{suchkov94,silich98,babul} suggest that the properties
of the gaseous halos these winds expand into play a major role in
determining the fate of the material in superwinds
(i.e. escape into the IGM, or confinement and eventual return to the disk).
Although simulations of starburst-driven winds should be treated
with caution, given the lack of calibration against observational
data, one robust result is the sensitivity of the wind properties
(in particular mass and metal loss to the IGM) on the
disk and halo ISM properties 
\citep[\eg][]{deyoung94,suchkov94,silich98,ss2000,silich01}. 
In our view, the emphasis given to
galactic mass as a primary parameter influencing gas and metal ejection
efficiencies \citep[\eg][]{dekel86,maclow99,ferrara_tolstoy00,fps00}
blinds non-expert readers to the more fundamental role played by the
poorly-known disk and halo ISM distributions.
The gravitational potential only enters at a secondary level, by
influencing the gas distribution and finally via the escape
velocity.
Redressing the lack of observational knowledge about the 
gaseous halos that superwinds must pass through 
is one motivation for studying
the halos of \emph{normal} star-forming
galaxies.

Modern theories of the ISM in normal
spiral galaxies also predict hot gas in the halo, due to the
interchange of material between the disk and halo through 
``galactic fountains'' \citep{shapiro76,bregman80,avillez00},
or in a more-localized form through ``chimneys'' \citep{norman89}.
Energy feed back into the ISM from supernovae is believed to 
create bubbles of hot, metal-enriched,  gas, which
blow out of the disk and vent their contents into the halo of the
galaxy. After $\ga 10^{8}$ years or so, this
material cools and falls back to the disk,
possibly in a form analogous to the Galactic high velocity clouds.

Unfortunately, many of the fundamental aspects of these fountain/chimney
disk/halo interaction
theories are observationally unconstrained, fundamentally due to
our {\em current} lack of knowledge about the X-ray properties
of spiral galaxy halos. 
Studies, with {\it Einstein} and the {\it ROSAT PSPC}, 
of nearby edge-on star-forming galaxies
revealed {\em only one} genuinely 
normal spiral, NGC 891 \citep{bregman_and_pildis},
with X-ray-emitting gas in its halo (although it is common to find starburst
galaxies with hot halos mistakenly clasified as normal spirals, \eg NGC 4631).
Although edge-on 
normal spiral galaxies did not receive as much attention with {\it Einstein}
\& {\it ROSAT} as other classes of galaxy, 
those observations that were performed
failed to produce believable detections of hot halo gas
in other edge-on  normal 
star-forming galaxies 
\citep*{bregman82,vogler95,bregman97,fabbiano97,benson00}.

It is also not clear to what extent fountains are a distinct phenomenon
from superwinds. It is not unreasonable to think that physical
conditions and kinematics of  
the gas making up a fountain may be quite different from
that in a superwinds, despite the general similarity in driving mechanism.
For example, the filling factor of the dominant 
X-ray-emitting plasma might be high in a fountain \citep{bregman80}, 
as compared to superwinds where the filling factor is low 
\citep{chevclegg,suchkov94,ss2000,strickland00}. 
Intriguingly, in a {\it ROSAT} PSPC-based sample of both normal
spiral galaxies and starburst galaxies of all inclinations, 
\citet{read01} find that the X-ray-emitting gas ``in the coronal systems
[\ie galaxies assumed to have a hydrostatic hot gas corona/halo]
appears to be cooler than that seen in the wind systems [\ie starburst
galaxies with superwinds]''.

\subsection{A description of the present study}

In this paper we report the results of a small, but representative,
survey of the diffuse X-ray emission from 10, approximately edge-on, 
star-forming disk galaxies using the ACIS imaging CCD spectrometer
on the {\it Chandra} X-ray Observatory. 

Compared to previous ROSAT and/or {\it ASCA}-based studies,
or even current work using {\it XMM-Newton}, the use of {\it Chandra}
offers significant advantages.  Firstly,
{\it Chandra}'s unequaled spatial resolution\footnote{The on-axis 
angular resolution of ACIS is $\sim0\farcs8$ (FWHM), which
corresponds to a physical scale of 
$\sim 7.8 \, (D/2 {\rm Mpc})\pc$ for a source at distance $D$.
By way of comparison, the resolution of the {\it ROSAT} HRI was
$\sim 5\arcsec$, {\it XMM-Newton} $\sim 6\arcsec$ 
and {\it ROSAT} PSPC $\sim30\arcsec$.}
allows the robust detection and removal of X-ray emission from 
the extremely numerous X-ray point sources found in star-forming galaxies,
and the foreground and background point sources found over the entire sky,
without the loss of large areas of sky.
Secondly, accurately quantifying the spatial morphology and surface
brightness of the diffuse emission
requires a resolution well matched to the physical sizes of interest,
\ie equivalent to the ground-based optical imaging.

 In \S~\ref{sec:sample} we discuss simple
methods of assessing the intensity of star formation in galaxies,
and the sample of galaxies in this paper.
The data analysis techniques we have used are presented in 
\S~\ref{sec:data_analysis}.
The properties of the diffuse X-ray emission in the halos of 7 starburst
and 3 normal spiral galaxies we observed are presented in 
\S~\ref{sec:results}. 
We provide a detailed discussion of the
location of the X-ray emission with respect to other properties of the
galaxies in \S~\ref{sec:results:gal_notes}, a quantitative study
of the spatial variation of the spectral hardness and surface brightness
along the minor axis (\S\ref{sec:results:extent}), 
and the spectral properties of the halo X-ray
emission (\S\ref{sec:spectra}). In \S~\ref{sec:discussion} we
discuss and interpret some of these results, and summarize
the results and conclusions of this paper
in \S~\ref{sec:conclusions}.

In \citet[henceforth referred to as Paper II]{strickland03}
we investigate how
the empirical properties of the hot gas, in particular
the extra-planar emission, correlate with the size, mass, star formation rate
and star formation intensity in the host galaxies.
We find that SN-feedback models best explain the observed diffuse
X-ray emission. We use 
this data to empirically test basic, but fundamental,
aspects of wind and fountain theories, in particular the critical energy
required for disk ``blow-out'' \citep[\eg][]{maclow88}, and
then discuss the issue of halo blow-out. 

We shall present a quantitative
study of the
relationship between the \halpha~surface brightness, 
optically-derived gas kinematics,
and the physical origin and properties of the 
halo thermal X-ray emission in a separate future paper.

\begin{deluxetable}{llccrrrrrrrrrrrr}
 \tabletypesize{\scriptsize}%
\rotate
\tablecolumns{16} 
\tablewidth{0pc} 
\tablecaption{Basic physical properties of the sample galaxies
	\label{tab:galaxies}} 
\tablehead{ 
\colhead{Galaxy} & \colhead{$\alpha, \delta$ (J2000)} 
	& \colhead{$i$}
	& \colhead{PA}
	& \colhead{$v_{\rm rot}$}
	& \colhead{D}
	& \colhead{scale}
	& \colhead{$f_{60}$}
        & \colhead{$f_{60}/f_{100}$}
	& \multicolumn{3}{c}{$L/(10^{10} \Lsol)$}
	& \colhead{$L_{H\alpha}$} 
	& \colhead{$M_{\rm TF}$} 
	& \colhead{$SFR_{\rm IR}$}
	& \colhead{$\log \tau_{\rm gas}$}  \\
\colhead{} & \colhead{(h m s, $\degr$ $\arcmin$ $\arcsec$)}
	& \colhead{($\degr$)}
	& \colhead{($\degr$)}
        & \colhead{(km/s)}
        & \colhead{(Mpc)}
        & \colhead{(pc)}
        & \colhead{(Jy)}
	& \colhead{}
	& \colhead{($L_{\rm IR}$)}
	& \colhead{($L_{\rm K}$)}
	& \colhead{($L_{\rm B}$)} 
	& \colhead{($10^{40}\ergps$)}
	& \colhead{($10^{10} \Msol$)} 
	& \colhead{($\Msol$ yr$^{-1}$)} & \colhead{} \\
\colhead{(1)} & \colhead{(2)}
	& \colhead{(3)}
	& \colhead{(4)}
	& \colhead{(5)} & \colhead{(6)}
	& \colhead{(7)} & \colhead{(8)} 
	& \colhead{(9)} & \colhead{(10)}
	& \colhead{(11)} & \colhead{(12)} & \colhead{(13)}
	& \colhead{(14)} & \colhead{(15)} & \colhead{(16)}
	}
\startdata

M82 & 09 55 51.9 \phm{-}+69 40 47.1$^{a}$
	& 73$^{b}$, 82$^{c}$ & 65$^{d}$
	& 137$^{a}$ & 3.6$^{e}$ & 17.5 & 1313.5$^{f}$ & 0.97 
	& 5.36 & 0.75 & 0.33 & 7.0$^{g}$ & 1.9 
	& 9.2 & 2.04 \\
NGC 1482 & 03 54 39.3 \phm{+}-20 30 08.9$^{h}$
	& 58$^{i}$ & 103$^{j}$ 
	& 165$^{k}$ & 22.1 & 107.1 & 35.3$^{f}$ & 0.77
	& 5.00 & 0.84 & 0.38 &  N 9.4$^{l}$ & 3.6 
	& 8.6 & 1.74 \\
NGC 253 & 00 47 33.2  \phm{+}-25 17 16.2$^{m}$ 
	& 79$^{n}$, 72$^{o}$ & 52$^{d}$, 49$^{o}$
	& 225$^{m}$ & 2.6$^{p}$ &  12.6 & 936.7$^{f}$ & 0.50 
	& 2.10 & 0.89 & 0.58 & 3.6$^{q}$ & 10.6 
	& 3.6 & 2.11 \\
NGC 3628 & 11 20 16.95 \phm{-}+13 35 20.1$^{r}$
	& 87$^{s}$, 80$^{r}$ & 104$^{d}$
	& 229$^{t}$ & 10.0$^{u}$ & 48.5 & 51.6$^{f}$ & 0.50
	& 1.74 & 1.58 & 1.06 & N 2.3$^{v}$ & 11.3 
	& 3.0 & 2.33 \\
NGC 3079 & 10 01 57.8 \phm{-}+55 40 47.2$^{w}$
	& 85$^{w}$ & 165$^{d}$
	& 244$^{t}$ & 17.1 & 82.9 &  50.2$^{f}$ & 0.49 
	& 4.76 & 1.55 & 0.98 & N 9.1$^{d}$ & 14.1 
	& 8.1 & 2.62 \\
NGC 4945 & 13 05 25.3 \phm{+}-49 29 09.0$^{x}$
	& 78$^{y}$ & 43$^{y}$ 
	& 184$^{t}$ & 3.7 & 17.9 & 588.1$^{z}$ & 0.42
	& 2.70 & 0.94 & 0.73 & N 2.4$^{d}$ & 5.2 
	& 4.6 & 2.88 \\
NGC 4631 & 12 42 07.2 \phm{-}+32 32 31.9$^{aa}$
	& 85$^{ab}$, 81$^{ac}$ & 86$^{ac}$, 83$^{ac}$
	& 150$^{ab}$ & 7.5$^{ab}$ & 36.4 & 82.9$^{z}$ & 0.40 
	& 1.74 & 0.62 & 0.98 & N 14.3$^{ad}$ & 2.6 
	& 3.0 & 2.48 \\
\tableline
NGC 6503 & 17 49 26.4 \phm{-}+70 08 39.7$^{ae}$
	& 75$^{af}$ & 123$^{af}$ 
	& 120$^{ag}$ & 5.2$^{ah}$ & 25.2 & 10.2$^{f}$ & 0.35 
	& 0.12 & 0.14 & 0.18 & 0.76$^{ai}$ & 1.2 
	& 0.20 & 2.97 \\
NGC 891 & 02 22 33.2 \phm{-}+42 20 56.2$^{aj}$
	& 89$^{aj}$ & 23$^{aj}$ 
	& 225$^{t}$ & 9.6 & 46.5 & 61.1$^{z}$ & 0.31
	& 2.47 & 1.66 & 0.78 & 3.3$^{ae}$ & 10.6 
	& 4.2 & 3.19 \\
NGC 4244 & 12 17 29.7 \phm{-}+37 48 20.4$^{ak}$
	& 85$^{ak}$ & 48$^{ak}$ 
	& 100$^{ak}$ & 3.6 & 17.5 & 4.2$^{z}$ & 0.26 
	& $\le 0.02$ & 0.04 & 0.08 & 0.42$^{ad}$ & 0.6 
	& 0.04 & 4.53 \\
\enddata 
\tablecomments{Column 2: Coordinates of the dynamical center of the
        galaxy, where available.
	Column 3: Inclination.
	Column 4: Position angle of the galactic disk. Multiple values
        for the inclination and position angle are given when
        several values are often reported in the literature.
	Column 5: Inclination-corrected circular velocity within the disk.
        Column 6: Assumed distance. Unless a specific reference is
	given, distances were calculated using the galaxy recessional
	velocity with respect to the microwave background \citep{rc3},
	and $H_{0} = 75 \kmps \Mpc^{-1}$.
	Column 7: Physical distance 
	corresponding to an angular size of 1 arcsecond.
	Column 8: IRAS $60\micron$ flux in Janskys.
	Column 9: IRAS $60$ to $100\micron$ flux ratio.
	Column 10: Total IR luminosity based on the observed IRAS fluxes, 
	$L_{IR} = 5.67\times10^{5} 
	D^{2}_{\rm Mpc} \times  (13.48 f_{12} + 5.16 f_{25} + 2.58 f_{60} + f_        {100}) \Lsol$, where $D_{\rm Mpc}$ is the distance to the galaxy
        in Mpc \citep{sander96}. Given the large angular size of these local
	galaxies we use IRAS fluxes from extended source analyses, 
	the exact source of the IRAS data used
	is given in a case by case basis above. Where multiple measurements
	were available for any galaxy, we used the following sources in 
	order of preference: \citet{rice88}, and \citet{soifer89}.
	Column 11: K-band luminosity, calculated from the K-band total
	magnitudes given in \citet{2mass_largegals} and converted to
	the Cousins-Glass-Johnson system using the transformations
	presented in \citet{carpenter01}.
	Zero points for conversion from magnitude to	
	flux were taken from \citet*{bessel98}. Note that the units
	are $\Lsol$, not $L_{\rm K, \odot}$.
	Column 12: B-band luminosity from RC3 catalog $B_{T}$
	magnitude \citep{rc3}, and corrected for Galactic extinction
	using reddening values from \citet{burstein82}. 
	Note that no correction for extinction internal
	to the specific galaxy has been made.
	Column 13:  Observed \halpha-luminosity in units of 
	$10^{40} \ergps$. Note these values have {\em not} been
	corrected for extinction. Measurements
	that have not been corrected for \nii~emission are
	marked with a preceding $N$.
	Column 14: Baryonic mass (stellar plus \hi) 
	estimated from the K-band Tully Fisher relationship of \citet{bell01},
	where $M_{\rm TF} = 10^{9.79} \times (v_{\rm rot}/100 
	\kmps)^{3.51} \Msol$.
	Column 15: Total galactic star formation rate in
	units of $\Msol \pyr$ estimated from the IR luminosity 
	\citep{kennicutt98b}, 
	where $SFR_{\rm IR} = 
	4.5\times 10^{-44} L_{\rm IR}$ [$\ergps$].
	Column 16: Logarithm of the gas consumption timescale in Myr,
        where $\log \tau_{gas} = \log \Sigma_{gas} - \log \Sigma_{\rm SFR}$,
        and $\Sigma_{gas}$ and $\Sigma_{\rm SFR}$ are the surface density
        of gas ($\Msol$ pc$^{-2}$) and star-formation rate ($\Msol$
        yr$^{-1}$ kpc$^{-2}$) respectively. Data are
        from \citet{kennicutt98a}, except for NGC 1482,
        NGC 3628, NGC 4244, NGC 4631
        and NGC 4945. For these galaxies the star formation rate was
        estimated from the IRAS fluxes. Gas masses were obtained from
        \citet{elfhag96} for NGC 1482 (using the CO brightness to $H_{2}$
        column density scaling in \citet{kennicutt98a}, and a
        radius of 2 kpc, as ISO observations by \citet{dale2000}
        demonstrate that 80\% of the
        star-formation occurs within this radius),
        \citet{irwin96} for NGC 3628 (central, $R \le 340$ pc, $H_{2}$
        gas mass), from \citet{olling96} for NGC 4244 (total H{\sc I}
        gas mass), from \citet{golla94b} for NGC 4631 ($H_{2}$ gas mass
        for $R \le 2.5$ kpc) and from \citet{dahlem93} for NGC 4945
        (molecular gas mass within central 1.1 kpc).
	}
\tablerefs{(a) \citet{weliachew84}, (b) \citet{ichikawa95}, 
	(c) \citet{lynds63}, (d) \citet{lehnert95}, (e) \citet{freedman94},
	(f) \citet{soifer89}, (g) \citet{mccarthy87},
	(h) 2MASS Second Incremental Release,
	(i) \citet{hameed99}, 
	(k) Calculated assuming $v_{\rm rot} = W_{20}/2\sin^{2} i$, 
	using the $W_{20}$ value from \citet*{roth91},
	(l) \citet{hameed99},	
	(m) \citet{sorai2000}, (n) \citet{pence81},
	(o) \citet{puche91}, 
	(p) \citet{puche88}, 
	(q) From \halpha~images published in \citet{strickland02},
 	(r) \citet{douglas96},
	(s) \citet{irwin96}, (t) \citet{sofue97},  (u) \citet{soifer87}
	(v) \citet{fhk90}, 
	(w) \citet{irwin91}, 
	(x) \citet{ott2001},
	(y) \citet{koorneef93}, 
	(z) \citet{rice88},
	(aa) Dynamical center, \citet{golla94b}, (ab) \citet{rand94},
	(ac) \citet{golla99},  	(ad) \citet{hoopes99},
	(ae) Central optical \& X-ray 
	source, \citet*{lira02},	
	(af) \citet{rc3},  (ag) \citet{bottema97},
	(ah) \citet{karachentsev97}, 
	(ai) \citet{kennicutt98a}. 
	(aj) Nuclear 1.4 GHz continuum point source, \citet{rupen91},
	(ak) \hi~dynamical center, \citet{olling96}.
	}
\end{deluxetable} 

\begin{figure}[!t]
\plotone{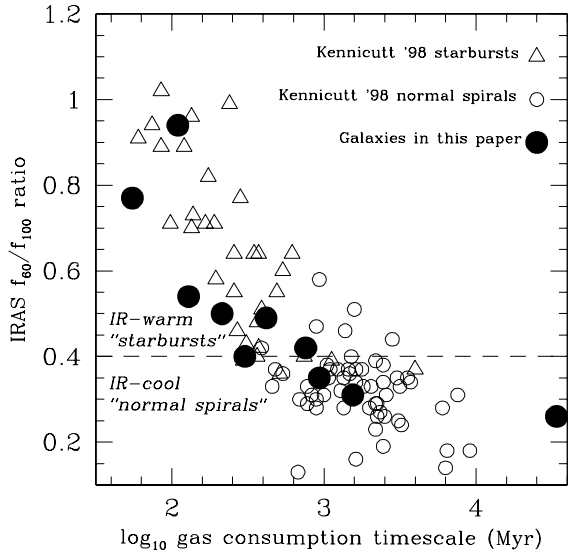}
\caption{Infrared warmth and gas consumption timescales for our edge-on 
star-forming galaxy sample (filled circles), compared to other 
starburst (open triangles) and normal spiral (open circles) galaxies
from the \protect\citet{kennicutt98a} sample. See discussion in 
\S~\ref{sec:sample}.}
\label{fig:sample_taugas}
\end{figure}

\section{Star formation in disk galaxies, and our edge-on disk galaxy
sample}
\label{sec:sample}

Our sample of edge-on ($i \ga 60\degr$) star-forming galaxies is formed
from seven starburst galaxies and three  ``normal'' 
star-forming galaxies. Basic physical properties of the
sample galaxies are shown in Table~\ref{tab:galaxies}.
We have chosen to study only
moderate luminosity ($L_{\rm BOL} \sim 10^{10} \Lsol$), moderate-mass 
($M \sim 10^{10}$ -- $10^{11} \Msol$) disk galaxies within
a distance of $D \la 20 \Mpc$, and not include any {\it Chandra}
observations of star-forming dwarf galaxies or 
ultra-luminous infrared galaxies (ULIRGs).
Given that the physical conditions in these other systems are somewhat 
distinct from star-forming disks,  and that
they rarely present the clean ``edge-on'' geometry necessary to
separate disk X-ray emission from halo X-ray emission, 
it makes sense to study them separately, at least initially. 
Furthermore, the X-ray fluxes
from both the dwarf starbursts and the ULIRGs are considerably 
lower than those of the typical galaxies within this sample.

This work can be considered as an 
updated version of earlier {\it ROSAT} PSPC and {\it ASCA}
based surveys of diffuse X-ray emission in 
star-forming galaxies (\citealt*{rps97}; 
\citealt*{dwh98}; \citealt{read01}), 
with which our sample shares many galaxies in common.

\subsection{Star formation rates and star-formation intensities}

Our sample spans the full range of star formation activity
found in disk galaxies
(see Fig.~\ref{fig:sample_taugas}), although we emphasize that it
is not complete in any flux or volume-limited sense. 
In Fig.~\ref{fig:sample_taugas} we plot the
estimated gas consumption time-scales
for the galaxies in our sample, along with the normal and
starburst galaxies from \citealt{kennicutt98a}, against the
IRAS 60 to $100\micron$ flux ratio.
The gas consumption time-scale is simply
the ratio of the total gas mass (atomic plus molecular gas) divided
by the star formation rate, $\tau_{\rm gas} = M_{\rm gas}/\Mdot_{SFR}$.
Starbursts are traditionally defined as those galaxies 
that would totally consume the gas reserves within the star forming region
in a time significantly less than a Hubble time (typically within 
a few hundred Myr), 
were star formation to continue at the currently observed rate.
More normal spiral galaxies, such as NGC 891 and the Milky Way
have gas consumption time-scales of order 
several Gyr \citep*{kennicutt94,demello02}.

The gas consumption time-scales
were calculated directly from the gas surface densities and star-formation
rate surface densities given in \citet{kennicutt98a}, except for 
NGC 1482, NGC 3628, NGC 4244, NGC 4631 \& NGC 4945, which were not 
part of Kennicutt's sample. 
For these galaxies used the same
methodology of \citet{kennicutt98a} to calculate the nuclear-region gas
consumption time-scale. Gas masses were obtained from the CO and \ion{H}{1}
data of \citet{elfhag96}, \citet{irwin96}, \citet{olling96}, \citet{rand94},
\citet{golla94b} and \citet{dahlem93}, and 
the star-formation rates were calculated from the IR luminosity
using the relationship given in \citep{kennicutt98b}. 
Values for the calculated star formation rates and 
gas consumption time scales are given in Table~\ref{tab:galaxies}.

Fig~\ref{fig:sample_taugas} also validates the definition
of galaxies with a IRAS flux ratio $f_{60}/f_{100} > 0.4$
as being starburst galaxies. 
Physically, the dust temperature in a star-forming galaxy 
(and hence the IRAS $f_{60}/f_{100}$ ratio)
is related to the energy density of the inter-stellar radiation field
(in particular the energy density in the UV, from 900 to 3000 \AA, see \citealt*{dust_temp}), and hence is related to 
the star-formation rate per unit area of the disk\footnote{Based on the
data presented in \citet{dust_temp}, then to first order 
$f_{60}/f_{100} \sim 0.32 \times U_{\rm ISRF}^{1/3}$,
where $U_{\rm ISRF}$ is the energy density in the interstellar radiation field
in units of eV cm$^{-3}$. Note that this only applies
{\em when massive stars
dominate the dust heating}.}. 

Theoretically, the fraction of the mechanical
energy returned to the ISM by massive stars
that is available to drive large scale flows (\ie that fraction that is
not radiated away) is a function of the SN rate per unit volume,
as this directly alters the porosity of the ISM, and hence the 
radiative losses SNe experience
\citep{larson74,mckee77,slavin93}. Feedback in a galaxy should then
depend on the star formation intensity
(SF rate per unit disk area or per unit volume), in addition to
the total SF rate. Therefore
we display the sample galaxies in all figures, and list them in all
further tables, in order of decreasing $f_{60}/f_{100}$ ratio.


\subsection{The starburst galaxies}

The starburst galaxies in our sample
are all good examples of the ``typical'' starburst galaxy
in the local universe, \ie infrared warm (IRAS 60 to 100 \micron~flux ratio
$f_{60}/f_{100} \ge 0.4$) galaxies with  a far-infrared 
luminosity $L_{\rm FIR} \approx$ 
a few $\times 10^{10} \Lsol$ \citep{soifer87}.
Minor-axis velocity-split optical emission lines
provide unambiguous kinematic evidence for 
$v \sim 300$ to $600 \kmps$ outflows --- galactic superwinds 
--- in M82, NGC 253, NGC 4945, NGC 3079 \& NGC 1482 
(\citealt{bland88,ham90,cecil01,veilleux02}).
Spectroscopic evidence for outflow is weaker, but still compelling,
for NGC 3628 (\citealt*{fhk90,irwin96}). NGC 4631 (where the star-formation
activity is more widely spatially distributed than in
the other starbursts of this sample)
currently lacks spectroscopic evidence for a outflow, apart for
one region to the south of the brightest \ion{H}{2} region \citep*{golla96}.
With the exception of NGC 4945 (which is at low Galactic latitude and
hence suffers high foreground obscuration), all of these galaxies have 
clearly-detected, $\sim 5$ to 20 
kpc-scale, extra-planar (``halo'') diffuse X-ray 
and \halpha~emission, again typical of galaxies with
$f_{60}/f_{100} \ge 0.4$ (\citealt{ham90,lehnert95,lehnert96,dahlem01}).

\subsection{The normal spiral galaxies}

At present there are relatively few {\it Chandra} observations aimed at 
normal (\ie non-starburst, non-Seyfert) spiral galaxies,
and the majority of these galaxies are not edge-on. Our sample includes
the only spiral galaxies with $i > 60\degr$ and IRAS $f_{60}/f_{100} < 0.4$
that have been observed
with {\it Chandra} for which the data was available as of 2002 September: 
NGC 891, NGC 4244, \& NGC 6503.

NGC 891 is a well-studied spiral galaxy, believed to be relatively similar
to our own galaxy, although somewhat more IR-luminous. It is the {\em only}
``normal'' spiral galaxy (apart from possibly our own) 
for which robust evidence of
extra-planar diffuse X-ray emission has been discovered
\citep{bregman_and_pildis}. Both NGC 6503 and NGC 4244 are relatively
low mass spiral galaxies. Individual galaxies are discussed in more
detail in \S~\ref{sec:results:gal_notes}.

\subsection{AGN contamination}

Only two of these galaxies host AGN luminous enough to possibly
bias the the star-formation rate and intensity diagnostics we employ
in this paper and in Paper II. These galaxies are
NGC 4945 (\eg \citealt{whiteoak86}; \citealt{marconi00}),
and to a lesser extent NGC 3079 (\eg \citealt{hawarden95}; 
\citealt{trotter98}).
The luminosity of these AGN,
their effect on the SF-rate indicators, 
and a quantitative investigation of
whether the AGN are partially responsible for the
outflows from these galaxies will be addressed in a future paper.
In \S~\ref{sec:discussion} we provide a preliminary, more qualitative, 
discussion of whether the AGN play a significant role driving the
outflows in these specific galaxies.  
Note that the luminosities and star formation rates given in 
Table~\ref{tab:galaxies} have not 
been corrected in any way for the (rather uncertain) AGN contribution.

\begin{deluxetable}{lclcrlcrll}
 \tabletypesize{\scriptsize}%
\tablecolumns{10} 
\tablewidth{0pc} 
\tablecaption{Chandra X-ray observation log
	\label{tab:xray_obs_log}} 
\tablehead{ 
\colhead{Galaxy} & \colhead{ObsID} & \colhead{Observation date,} 
	& \colhead{$\alpha$, $\delta$ (J2000)} 
	& \colhead{Roll}
	& \colhead{Mode} & \colhead{FP Temp.}
	& \colhead{Active}
	& \multicolumn{2}{c}{Good time} \\ 
\colhead{} & \colhead{} & \colhead{days since launch} 
	& \colhead{(h m s, $\degr$ $\arcmin$ $\arcsec$)}
	& \colhead{($\deg$)}
	& \colhead{} & \colhead{($\degr$C)} & \colhead{chips}
	& \multicolumn{2}{c}{(ks)} \\
\colhead{(1)} & \colhead{(2)}
	& \colhead{(3)}
	& \colhead{(4)}
	& \colhead{(5)} & \colhead{(6)}
	& \colhead{(7)} & \colhead{(8)} 
	& \colhead{(9)} & \colhead{(10)}
	}
\startdata
M82 	& 361+1302 & 1999 Sep 20, 55 & 09 55 55.1 \phm{-}+69 40 47.4
	& 30 & ACIS-I
	& -110 & {\bf 0123}67 &  48.51 I1 & 48.50 13 \\
M82 	& 2933  & 2002 Jun 18, 1061 & 09 55 53.6 \phm{-}+69 40 12.6
	& 288 & ACIS-S
	& -120 & 235{\bf 678} &  17.76 S2 & 17.99 S3 \\
NGC 1482 & 2932 & 2002 Feb 05, 928 & 03 54 39.9 \phm{+}-20 30 42.1
	& 289 & ACIS-S
	& -120 & 235{\bf 678} &  27.90 S2 & 23.46 S3 \\
NGC 253 & 969 & 1999 Dec 16, 146  & 00 47 34.0  \phm{+}-25 17 53.9 
	& 299 & ACIS-S
	& -110 & 5{\bf 678}9 & 13.50 S2 & 10.99 S3 \\
NGC 253 & 790 & 1999 Dec 27, 157  & 00 47 21.3  \phm{+}-25 13 00.6 
	& 300 & ACIS-S
	& -110 & 5{\bf 678}9 & 43.30 S2 & 38.30 S3 \\
NGC 3628 & 2039 & 2000 Dec 12, 508 & 11 20 18.4 \phm{-}+13 35 53.5
	& 71 & ACIS-S
	& -120  & 45{\bf 678}9 &  56.60 S2 & 54.70 S3 \\
NGC 3079 & 2038 & 2001 Mar 07, 593 & 10 01 53.3 \phm{-}+55 40 41.9
	& 200 & ACIS-S
	& -120 & 235{\bf 678} & 26.69 S2 & 26.52 S3 \\
NGC 4945 & 864 & 2000 Jan 27, 188 & 13 05 29.1 \phm{+}-49 27 51.5
	& 61 & ACIS-S
	& -120  & 23{\bf 678} & 40.66 S2 & 28.66 S3 \\
NGC 4631 & 797 & 2000 Apr 16, 268 & 12 41 56.9 \phm{-}+32 37 33.9
	& 201 & ACIS-S
	& -120  & 235{\bf 678} &   57.61 S2 & 55.67 S3 \\
\tableline
NGC 6503 & 872 & 2000 Mar 23, 244 & 17 49 28.4 \phm{-}+70 09 19.2
	& 95.6 & ACIS-S
	& -120 & 23{\bf 678} & 12.91 S2 & 10.64 S3 \\
NGC 891 & 794 & 2000 Nov 01, 467 & 02 22 22.2 \phm{-}+42 20 49.9
	& 187 & ACIS-S
	& -120  & 235{\bf 678} &   49.68 S2 & 30.72 S3 \\
NGC 4244 & 942 & 2000 Mar 20, 241 & 12 17 22.1 \phm{-}+37 46 50.6
	& 231 & ACIS-S 
	& -120  & 235{\bf 678} &   48.84 S2 & 47.51 S3 \\
\enddata 
\tablecomments{Column 2: {\it Chandra} observation identification number.
	Two separate observations of M82, taken on the same day and with the
	same telescope pointing, were merged.
	Column 4: Telescope axis pointing direction.
	Column 5: Telescope roll angle in degrees.
	Column 7: Focal plane CCD temperature. 
	Column 8: Identification numbers for the CCD chips active during
	the observation. Chips 5 and 7 (a.k.a. S1 and S3) are back-illuminated
	chips. The remaining chips are all front-illuminated chips.
        The images and spectra described in this paper are taken from
        only those chips shown in bold-face.
	Columns 9 and 10: The remaining exposure time left {\em after} removal
	of periods of high background count rate, for the two most
	important chips in each observation (the chosen chips are 
	specified alongside the exposure time). 
	}
\end{deluxetable}

\section{Data sources and data reduction}
\label{sec:data_analysis}

\subsection{X-ray data}

The {\it Chandra} ACIS data for the 10 sample galaxies comprises
11 ACIS-S observations and 2 ACIS-I observations 
(see Table~\ref{tab:xray_obs_log}). The 6 ACIS-S observations
of the starburst galaxies M82, NGC 253 (two separate observations),
 NGC 1482, NGC 3079 and NGC 3628 are all from guest observer programs
awarded to our group. The {\it Chandra}
data for the remaining galaxies was obtained from the {\it Chandra}
data archive.

All the {\it Chandra} data were reduced and analyzed in a uniform,
self-consistent, manner using a scripted analysis pipeline employing
the {\sc Ciao} (version 2.2.1) and {\sc Heasoft} (version 5.2, {\sc Xspec}
version 11.2.0h) software packages. For a more detailed description
of many of the data analysis techniques used we refer the reader to 
\citet{strickland02}. Unless otherwise stated, our
data processing follows the recommended guide lines released by the 
CXC and the ACIS instrument 
team\footnote{The {\sc Ciao} (v2.2) Science Threads
can be found at http://asc.harvard.edu/ciao/documents\_threads.html, and
the Penn State University ACIS instrument team's recipes 
at http://www.astro.psu.edu/xray/acis/recipes/.}.
Our analysis procedures take account of
 all currently known bugs and peculiarities
of the {\sc Ciao} software.
A few points of specific importance are discussed below.

\subsubsection{Calibration and spectral response files used}

The original {\it Chandra} data files were 
reprocessed to update the instrument gain and astrometric solution,
using version 2.12 of the {\it Chandra} Calibration Database (CALDB). 

The officially
recommended randomization of (spatially dependent) pulse height amplitude 
(PHA) channel values to (spatially-independent) PI channel value
was applied. The artificial $\sim0\farcs5$ spatial blurring applied
in the default {\it Chandra} pipeline was {\em not} applied in our 
reprocessing --- the resulting optical-axis
point source point spread function we measure in the data
is $0.8\pm{0.1}\arcsec$ FWHM.

The two ACIS-I observations of M82 (ObsIDs 361 \& 1302), taken on the
same day,
were combined into a single dataset using the {\sc Ciao} task 
{\sc reproject\_events}. We have not attempted to correct these ACIS-I
observations for the
radiation-damaged-induced charge transfer inefficiency problem (\eg see
\citealt{cti_correct}),
as the $\sim 20$\% in energy resolution and quantum efficiency this would
provide is small compared to other sources of uncertainty.

\subsubsection{Background flare removal}

Short periods of significantly-enhanced background count rate -- ``flares''
-- were removed using the iterative $3\sigma$
clipping technique described in \citet{strickland02}. The 
``good'' exposure times, \ie that remaining after clipping, 
are given in Table~\ref{tab:xray_obs_log}.

\begin{deluxetable}{lrrlrrrrr}
 \tabletypesize{\scriptsize}%
\tablecolumns{9} 
\tablewidth{0pc} 
\tablecaption{X-ray background rates and point source sensitivities
	\label{tab:xray_bgprops}} 
\tablehead{ 
\colhead{Galaxy} & \colhead{$\nH$} & \colhead{$\tau_{\rm Oviii}$}
	& \colhead{BG method}
	& \multicolumn{3}{c}{Background surface brightness} 
	& \multicolumn{2}{c}{$2\sigma$ PS sensitivity} \\ 
	 & & &
	& \colhead{0.3--1 keV}
	& \colhead{1--2 keV}
	& \colhead{2--8 keV}
	& \colhead{$f_{\rm X}$} & \colhead{$L_{\rm X}$} \\
\colhead{(1)} & \colhead{(2)}
	& \colhead{(3)}
	& \colhead{(4)}
	& \colhead{(5)} & \colhead{(6)}
	& \colhead{(7)} & \colhead{(8)} & \colhead{(9)}
	}
\startdata
M82$^{a}$  &  4.0 & 0.237 & local     &  2.45 & 1.63 &  6.28 & $2.4\times10^{-15}$ & $3.8\times10^{36}$ \\
M82$^{b}$  &  4.0 & 0.237 & blank sky  &  6.31 & 2.95 &  9.76 & $4.7\times10^{-15}$ & $7.3\times10^{36}$ \\
NGC 1482 &  3.7 & 0.219 & blank sky &  6.61 & 3.09 & 10.20 & $3.6\times10^{-15}$ & $2.1\times10^{38}$ \\
NGC 253$^{c}$  &  1.4 & 0.083 & blank sky &  6.36 & 2.97 &  9.80 & $7.2\times10^{-15}$ & $5.9\times10^{36}$ \\
NGC 253$^{d}$  &  1.4 & 0.083 & blank sky &  6.36 & 2.97 &  9.80 & $2.1\times10^{-15}$ & $1.7\times10^{36}$ \\
NGC 3628 &  2.2 & 0.130 & blank sky &  6.06 & 2.83 &  9.39 & $1.5\times10^{-15}$ & $1.8\times10^{37}$ \\
NGC 3079 &  0.8 & 0.047 & blank sky &  6.34 & 2.95 &  9.76 & $3.0\times10^{-15}$ & $1.0\times10^{38}$ \\
NGC 4945 & 15.7 & 0.930 & local     & 10.98 & 7.87 & 26.59 & $3.8\times10^{-15}$ & $6.1\times10^{36}$ \\
NGC 4631 &  1.3 & 0.077 & blank sky &  6.67 & 3.12 & 10.29 & $1.4\times10^{-15}$ & $9.6\times10^{36}$ \\
NGC 6503 &  4.1 & 0.242 & local     &  9.64 & 5.75 & 21.18 & $8.0\times10^{-15}$ & $2.6\times10^{37}$ \\
NGC 891  &  7.6 & 0.450 & local     &  9.78 & 6.53 & 20.33 & $3.0\times10^{-15}$ & $3.3\times10^{37}$ \\
NGC 4244 &  1.7 & 0.101 & local     &  5.06 & 2.80 &  9.62 & $1.7\times10^{-15}$ & $2.8\times10^{37}$ \\
\enddata
\tablecomments{Column (2): The foreground Galactic hydrogen column,
	in units of $10^{20} \pcmsq$ \citep{nhref}.
	(3): The optical depth at the energy of the O{\sc viii}
	Ly$\alpha$ line (E=0.65 keV) equivalent to the Galactic foreground
	column, assuming Galactic dust-to-gas
	ratios and Solar abundances. The optical depth at E=0.3 keV is 5.1
	times greater.
        (4): Method used to calculate the combined particle and
        X-ray background. See \S~\ref{sec:data:bgsub} for details.
	Columns 5,6 \& 7: Observed ACIS S3 chip background surface 
        brightness levels, given in
	units of $10^{-7}$ counts s$^{-1}$ arcsec$^{-2}$, for the
	0.3-1.0, 1.0-2.0 and 2.0-8.0 keV energy bands. For M82 the
	values correspond to the ACIS I3 chip instead of the S3 chip.
	(8): Intrinsic 0.3--8.0 keV flux of the faintest
	distinguishable individual point sources, assuming a 
	$\Gamma=1.7$ power law spectrum, absorbed by an intrinsic hydrogen
	column of $10^{21}\pcmsq$ in addition to the foreground Galactic column.
	See text for details.
	(9): The corresponding 0.3--8.0 keV X-ray luminosity.
	}
\tablenotetext{a}{ACIS-I observations of M82, ObsIDs 361 \& 1302.}
\tablenotetext{b}{ACIS-S observation of M82, ObsID 2933.}
\tablenotetext{c}{NGC 253 observation ObsID 969, nucleus and disk on chip S3, northern halo on S4.}
\tablenotetext{d}{NGC 253 observation ObsID 790, northern halo on chip S3, disk and nucleus on S2.}
\end{deluxetable}

\subsubsection{Astrometric accuracy}

We use the default astrometric solutions, based on the latest reprocessing
of all observations available as of June 2002. Typically, the 90\% 
uncertainty\footnote{From ``{\it Chandra} absolute astrometric 
accuracy'', \url{http://asc.harvard.edu/cal/ASPECT/celmon/index.html}} 
in the Chandra astrometric solution is $0\farcs6$.
We have only explicitly checked the astrometry for X-ray point
sources that have optical counterparts in the NGC 253 and NGC 3628
Chandra observations, which we find are accurate to within $0\farcs5$.

\subsubsection{Point source detection and removal}

We used the wavelet-based source detection algorithm {\sc wavdetect} 
to search for point-like objects in each X-ray observation. Each chip
was treated separately, searching for sources in images created in
the soft X-ray 0.3 -- 2.0 keV energy-band, hard X-ray 2.0 -- 8.0 keV
energy band and total 0.3 -- 8.0 keV energy band. Only sources with
signal-to-noise ratios $\ge 2$ were accepted as point sources,
in the initial iteration of the data analysis.

We screen out point sources from the data used for both 
imaging and spectral analysis to a radius equivalent 
to 4 Gaussian $\sigma$, based on a fit to the
radius of the {\it Chandra} PSF as a function of off-axis angle.
The holes left in any image by source removal are filled in using the
{\sc Ciao} task {\sc dmfilth}. The observed distribution of pixel
values in a background annulus around the point source, defines 
the probability  density function from which random values 
are chosen to fill in the 
source region (The {\sc dist} option in {\sc dmfilth}). 
Care was taken to ensure that the background annulus chosen
did not contain other point sources, 
which would have biased the interpolation, and this requirement largely 
controlled the width of the background annulus used. In very crowded fields
(\eg M82's nuclear region) a background annulus width 
of 2 ACIS pixels ($\sim 1\arcsec$) was used,
although more commonly an annular width of 8 ACIS pixels was used. 
In a few cases where it was impossible to create two source regions with 
non-overlapping background annuli,
we merged nearby point sources into a single source removal
region. A variety of tests were performed, using this
methodology on point-source-free diffuse emission regions, 
that
demonstrated that this procedure provided a Poisson-statistically accurate
interpolation of the diffuse emission.

For spectral analysis sources detected in any of the three bands (the soft, 
hard and total bands described above) are excluded when creating 
diffuse-emission-only spectra. When making images of diffuse emission,
the soft band source lists are used for any image created within the
energy range 0.3 -- 2.0 keV. The hard and total band source lists are used
for any image created within the 2.0 -- 8.0 keV energy band.

In practice, fully-automated point source detection is not perfect.
In a few cases bright features in the diffuse X-ray emission were identified
as point sources. Note that this is expected behavior for a wavelet-based
algorithm, which looks for distinct features on a variety of spatial-scales,
not just objects corresponding to the local PSF. In other cases,
low S/N features that most probably are genuine point sources
were missed by the source-detection algorithm. These sources became apparent
only once all brighter point sources had been removed. We assessed 
each such feature, and removed those that appeared to be point-like based on
our personal scientific judgment.

These observations are sensitive to point source fluxes of $f_{\rm X} \ga$ 
a few
times $10^{-15} \ergps \pcmsq$ in the 0.3 -- 8.0 keV energy band
(equivalent to a point source with a total of $\sim10$ photons detected
over the entire observation). If we assume that a typical point source
has a power-law X-ray spectrum with a photon index $\Gamma =1.7$,
and is seen through a total absorbing column equal to the sum of the
Galactic foreground (column 2 in Table~\ref{tab:xray_bgprops}) and
a $\nH = 10^{21} \pcmsq$ column intrinsic to the host galaxy, then
we are sensitive to point sources of intrinsic luminosity 
$\log L_{\rm X} \ge 36.5$ -- $37.5$, depending on the distance to
the galaxy and the length of the observation (Table~\ref{tab:xray_bgprops}). 
This is typically 10 -- 30 times fainter than the point sources 
in the {\it ROSAT}-based studies of \citet{rps97} and \citet{dwh98}.
For M82 and NGC 3079 the ACIS data probes sources two orders of magnitude
fainter than the PSPC observations.

\subsubsection{Background subtraction}
\label{sec:data:bgsub}

Accurate background subtraction is vital when studying faint diffuse 
X-ray emission. We experimented with various methods of background
level estimation and background subtraction for each observation,
before settling on the methods detailed below.

Whenever possible, we use the appropriate
summed blank-sky field provided as part of {\sc Ciao}. 
We rescale the blank-sky background data set so as to match the observed 
3.0 -- 6.0 keV energy band count rate in the data (having excluded all
point source X-ray emission when making the comparison). For M82 and NGC 253, 
the two brightest galaxies in the sample, we excluded
the area associated with the disk of the galaxy to avoid biasing the 
background rescaling factor with emission due to unresolved point 
sources in the disk, and possible diffuse hard X-ray emission from
the nuclear starburst regions (\eg \citealt{griffiths2000}). 
The assumption that there is negligible diffuse hard X-ray 
emission in the halos of these galaxies is justified on both an
empirical and theoretical basis (\eg \citealt{dwh98,ss2000,strickland02}).
The background correction is performed on a chip-by-chip
basis. Typically this requires a
correction of $\la 8$\%, in accordance with the observed
long-term variation in the background rate \citep{markevitch01}.

Use of the rescaled-blank sky datasets for background subtraction
is appropriate only if the spectral
shape of the background within each specific {\it Chandra} observation 
matches the ``typical''
background contained within the blank-sky fields. There are two factors that
might lead to this requirement not being fulfilled. Firstly, the intensity
of the soft X-ray background is known to vary by a factor $\sim 2$ over
the sky, based on the {\it ROSAT} All Sky Survey measurements
\citep{snowden95_rass}.
Secondly, the high radiation environment that {\it Chandra} 
experiences can lead to
observations being taken during periods of significantly higher than normal
particle backgrounds.

Using the online X-Ray Background Tool  
\footnote{The X-Ray Background Tool can be found at 
\url{http://heasarc.gsfc.nasa.gov/cgi-bin/Tools/xraybg/}} we find the
background in the summed {\it ROSAT} R1, R2 R4 and R5 bands 
(roughly the 0.12-1.21 keV energy band) 
for our targets deviates by at maximum 23\%
from the mean background in the sample. That the variation is smaller
than that found over the entire sky is to be expected, as we are observing
targets generally located well out of the plane of the galaxy.
In the R4+R5 band ($E = 0.47-1.21$ keV),
which is better choice for comparison with the Chandra 0.3-1.0 or 0.6-1.0 keV
energy bands (especially given the 
reduction in {\it Chandra}'s low energy sensitivity
due to molecular contamination of the optical blocking filter), the
extreme deviation is only $\pm{14}$\%. This number is similar to the
typical degree we have had to rescale the blank-sky background data.

All attempts to use rescaled blank-sky data for background subtraction
of the ACIS observations of NGC 4945, NGC 6503, NGC 891 and NGC 4244 
led to obvious over-or-under subtraction of the 
background (in particular in the
soft $E = 0.3$--1.0 keV energy band). 
For the first three targets, the particle background was significantly
higher than normal during a large fraction of each observation, even
after the strongest background ``flares'' had been removed. More aggressive
flare removal led to unacceptably-low remaining exposure times.
For NGC 4244, the
{\em low energy} 
count rates over the S3 chip (once point sources had been removed)
were significantly {\em lower} than expected based on comparison with
the blank-sky fields.
For these four datasets we calculated a local background surface brightness,
in any particular chip and energy band, based on the sum of 2 to 3 local
background regions. These regions were chosen to exclude any
diffuse X-ray emission from the host galaxy, while remaining
close enough to the galaxy to minimize the effect of any variation in the
background over the  chip. Such variations do exist, and have proved to
be particular troublesome at high energies in the ACIS observation of
NGC 891, and to a lesser extent in NGC 6503 and NGC 4244 observations.
Emission from all detected X-ray point 
sources was excluded when calculating the background count rate or
surface brightness. 

This local background estimate needs to be scaled appropriately when
calculating the background in any other region on the chip, given that
the background varies over any chip. For image-based analysis
we used the relative intensity from an exposure map (the product of the
effective exposure, and collecting area, at some X-ray photon energy for 
each point on each ACIS chip)
to map from an observed local background to a chip-wide model background
image. This was done on a chip-by-chip and energy band-by-energy band basis.
This method works well except when the background is high due to 
additional particle-based events, which are not vignetted by the
telescope structure in the same way as the X-ray photons are. 

For spectral analysis we do not use this exposure-map based correction,
instead rescaling the background so as to achieve a net zero
count rate in the point-source-subtracted halo spectra, in the
3.0 -- 8.0 keV energy band. This requires a very minor 
correction, in all cases $\la 8$\%. 

For the ACIS-I observation of M82 we used local background estimation 
for each of the four ACIS-I chips, even though the 
background appears well-behaved. This observation was taken much
earlier than the observation used to create the blank-sky fields, 
and we felt it better to use a direct, local, estimate of the background
rather than attempt to re-scale the blank-sky fields. The local
background regions were placed along the edges of the ACIS-I chips,
again avoiding any regions of known diffuse X-ray or optical emission.

Table~\ref{tab:xray_bgprops} specifies for which datasets blank-sky 
or local methods of background subtraction were used, along with the observed
background surface brightness in the S3 or I3 chips.
Simple estimates using the PIMMS count rate conversion software demonstrate
that the {\it ROSAT}-predicted background rates in 
the {\it Chandra} 0.3--1.0 and 1.0--2.0 keV bands
are lower than the background levels we find with {\it Chandra},
indicating that the particle contribution to the background is
significant, as expected.

The accuracy of the background subtraction for each data set was checked
by measuring the number of residual counts in a background-subtracted
image from regions free of point source
emission and any possible diffuse emission (based on visual inspection of
of the smoothed {\it Chandra} images, and comparison 
any existing X-ray images from {\it ROSAT} observations).
This was done in each of the four
following energy bands, 0.3--0.6, 0.6--1.0, 1.0--2.0 and 2.0-8.0 keV.
The size of these
regions was in all cases between 1 -- 2 square arcminutes, large enough that
there were between hundred and a  thousand counts due to the background
in each region and in each energy band.
In all data sets, and for all energies, the residual emission in 
these regions after background subtraction data lies within 1 standard 
deviation of zero. We conclude that our background subtraction methods 
are as good as any method that can be feasibly employed at this time.

\subsubsection{Diffuse emission X-ray spectra and spectral response files}

Background-subtracted spectra were manually created using the {\sc Ftool}
task {\sc Mathpha}, instead of letting the {\sc Xspec} 
spectral-fitting program
perform the background subtraction itself.
This allows us to both rebin the spectra to
ensure a minimum of 10 counts per bin after background subtraction 
(using {\sc Grppha}),
and associate more-realistic statistical errors with the data
(Poisson errors instead of $\surd N$ errors). This is necessary
as the surface brightness of the diffuse X-ray emission away from the
disk of these galaxies 
is typically at or below the background level. Thus the 
standard method of rebinning to a fixed number of counts per bin
before background subtraction does not ensure that there will be sufficient
counts per bin left after background subtraction to allow $\chi^{2}$ spectral
fitting.

Spectral response matrix files (RMFs) and auxiliary response files
(ARFs) for the diffuse emission spectra were constructed using the
{\sc Ciao} tools {\sc mkrmf} and {\sc mkwarf}, appropriately
weighted by the observed, point-source {\em and background-subtracted},
spatial distribution of the diffuse emission.

For those observations taken at an ACIS CCD operating temperature of
$-120\degr$ C (see Table~\ref{tab:xray_obs_log}) the standard
response files from CALDB v2.12 were used. The RMF files in
CALDB v2.12 for ACIS
observations taken with the CCDs at $-110\degr$ C are known to be
significantly less accurate, in particular with respect to
the energy resolution (see the discussion in \S~2.1.5 of 
\citealt{strickland02}). As updated responses for these observations
were not available at the time analysis was performed, 
we instead used the responses
for $-120\degr$ C for these observations, which is more
accurate than using the default $-110\degr$ C responses. 

In the case of the ACIS-I observation of M82 (the combined data from
ObsIDs 361 and 1302), appropriately-weighted RMF and ARF files were 
created separately for each ACIS-I chip. The {\sc Heasoft} tools
{\sc addrmf} and {\sc addrmf} were then used to combine these
separate response files into the single RMF and ARF file used in the 
spectral fitting.

\subsection{Optical narrow-band observations and data analysis}
\label{sec:optical_analysis}

For the majority of these galaxies we have used existing,
previously-published, narrow-band \halpha~imaging and narrow-or-broad-band
continuum imaging. For the purposes of comparing the spatial morphology
and extent of the optical emission with that of the X-ray emission, 
we found astrometric solutions for the optical images, typically using
between 10 -- 30 stars per image. Positions for these
stars were taken from either the
Automatic Plate Measuring (APM) survey \citep{apm}, or the
U.S.~Naval observatory A2.0 catalog.
The astrometric solutions were calculated using the {\sc Starlink}
software package {\sc Astrom}. The rms uncertainty in the absolute
positions for the various optical images lies in range $\sim 0\farcs3$
-- $0\farcs5$. We shall
present a quantitative X-ray-vs-\halpha~flux comparison in a future paper.

\subsubsection{Existing reduced optical data}

For M82 we used optical R-band and \halpha~images from the
``UV/Visible Sky Gallery on CDROM'' \citep{chen97}, originally
taken on the 0.9m telescope at KPNO. The images were
convolved with a Gaussian mask to achieve a common spatial
resolution of $1\farcs5$ FWHM, 
prior to using the R-band image to continuum-subtract
the \halpha~image

The continuum R-band and continuum-subtracted
\halpha~images of NGC 1482 were provided by
S. Hameed, and originally published in \citet{hameed99}.
For NGC 253 we use the
6649~\AA-continuum and continuum-subtracted \halpha~images 
originally published in 
\citet*{hoopes96}, in addition to the higher resolution
R-band and \halpha~images published in \citet{lehnert96} (these
images are the same as those used in \citealt{strickland02},
which gives more detail on their properties and origin).
The R-band and \halpha~images of NGC 3628 are those originally
obtained and presented by \citet{fhk90}. For NGC 3079 we use
the deep \halpha~image presented in \citet{ham90}, along with its
R-band counterpart, in addition to the new images discussed below.
The continuum-subtracted
\halpha~image of NGC 4945 is from \citet{lehnert96}. The continuum
image of NGC 4945 is from the Digitized Sky Survey. The continuum-subtracted
\halpha~and narrow-band continuum images of NGC 4631, NGC 891 and NGC 4244
are all from \citet{hoopes99}.

\subsubsection{Other optical data}

Narrow-band \halpha~and broad-band continuum images of NGC 3079
were obtained on the night of 2002 May 4, using the SPICAM instrument
on the 3.5m Apache Point Telescope. Four 300 s narrow-band images were
obtained through the 
6585 Hodge filter (center 6586~\AA, FWHM 22.3~\AA) filter,
in addition to two 60 s R-band images. The {\sc ccdproc} task in 
{\sc Iraf} was used to
perform standard overscan, bias subtraction, and flat-fielding corrections
and cosmic ray rejection. The images were convolved with a Gaussian
mask to achieve a uniform resolution of $1\farcs4$ FWHM prior to continuum
subtraction. The R-band image was subtracted off the 
median-combined narrow-band images, scaled so as to leave zero residual
emission associated with foreground Galactic stars.  
This continuum-subtracted \halpha~image has superior spatial resolution to
the more sensitive image from \citet{ham90}.

For NGC 6503 we obtained and reduced a dataset from the Isaac Newton
Group (ING) telescope archive. The observation was taken with the
1.0m Jacobus Kapteyn Telescope on La Palma on 1999 April 25. 
The same form of data reduction used on the NGC 3079 data was applied. 
The final effective resolution of the \halpha~and R-band
images is $1\farcs5$.


\section{Spatial analysis}
\label{sec:results}

\subsection{X-ray and optical images}
\label{sec:xray_and_optical_images}

Figures~\ref{fig:images:m82} to \ref{fig:images:n4244} show soft
and hard X-ray, optical R-band and continuum-subtracted \halpha~images
of all the galaxies in our sample. A common X-ray surface brightness
scale and physical region size (2 kpc, and 20 kpc square)
is used, as this provides a fairer comparison
of the relative brightness of the extra-planar diffuse emission
between the various galaxies.

In many cases the images are composites of data from several ACIS chips
--- generally the ACIS S2, S3 \& S4 chips (ACIS I0, I1, I2 \& I3 for the
ObsID 361 + 1302 observations of M82) --- chips which have different
sensitivities as a function of photon energy. To obtain a 
chip-and-instrument-independent measure of the X-ray surface brightness
we build an image in any broad energy band 
out of several images from narrower energy bands, 
appropriately weighted
by the effective area of the instrument for that sub-energy and
chip-type (created using the {\sc ciao} task {\sc mkexpmap}). 
This weighting is done after background (and any point-source) subtraction.
Finally, exposure times and pixel sizes are taken into account to
create an image in units of photons s$^{-1}$ cm$^{-2}$ arcsec$^{-2}$,
\ie an estimate of the X-ray photon flux in that energy band
immediately {\em prior} to passing through the telescope's optics. 
For example, each exposure-corrected
0.3--2.0 keV energy band image is built out of exposure-map-correct
images in the 
0.3--0.6, 0.6--1.0, 1.0--1.5 and 1.5--2.0 keV energy bands.
Inspection of the results of this
procedure in cases where we have separate front-illuminated and
back-illuminated chip exposures of the same object (\eg M82,
the disk and northern halo of NGC 253) shows that it works well,
even when relatively broad sub-energy bands are used (we use 0.3 and
0.5 keV-wide bands for E < 2 keV, and 2 keV-wide bands for E > 2 keV).

The X-ray images in Figs.~\ref{fig:images:m82} to \ref{fig:images:n4244}
have been adaptively smoothed to achieve a {\em local} S/N ratio of 3.
In some images, weak CCD-chip artifacts appear 
that proved impossible to totally
remove. These occur most noticeably along the edges of the chips.
The gaps between the chips are also visible, most notably in the ACIS-I
observations of M82.
For convenience we have labeled these features with ``CA'' (chip artifacts)
and outline the individual CCD chips with dotted lines.

We have not made any effort to show the
optical R and \halpha~images on a common surface brightness scale in this
paper. Instead,
we have chosen the minimum and maximum intensities to highlight 
specific features -- in the case of the R-band image to show the distribution
of stellar light and/or dust lanes, and in the case of the 
continuum-subtracted \halpha~images to show the extra-planar emission.
The location of the {\it Chandra} ACIS chips is also marked on the 
optical images.

\begin{figure*}[!t]
\epsscale{2.0}

\plotone{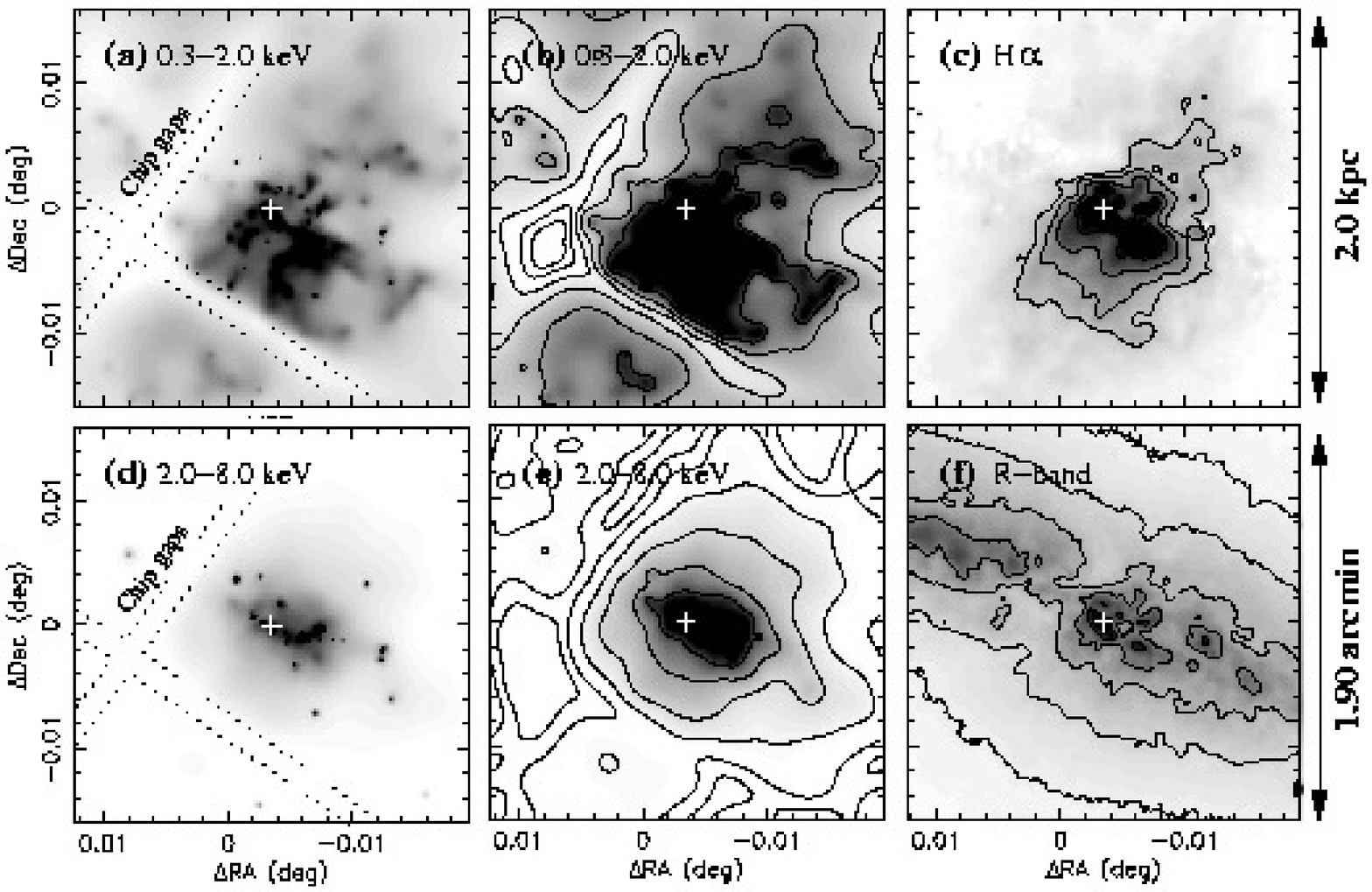}
\plotone{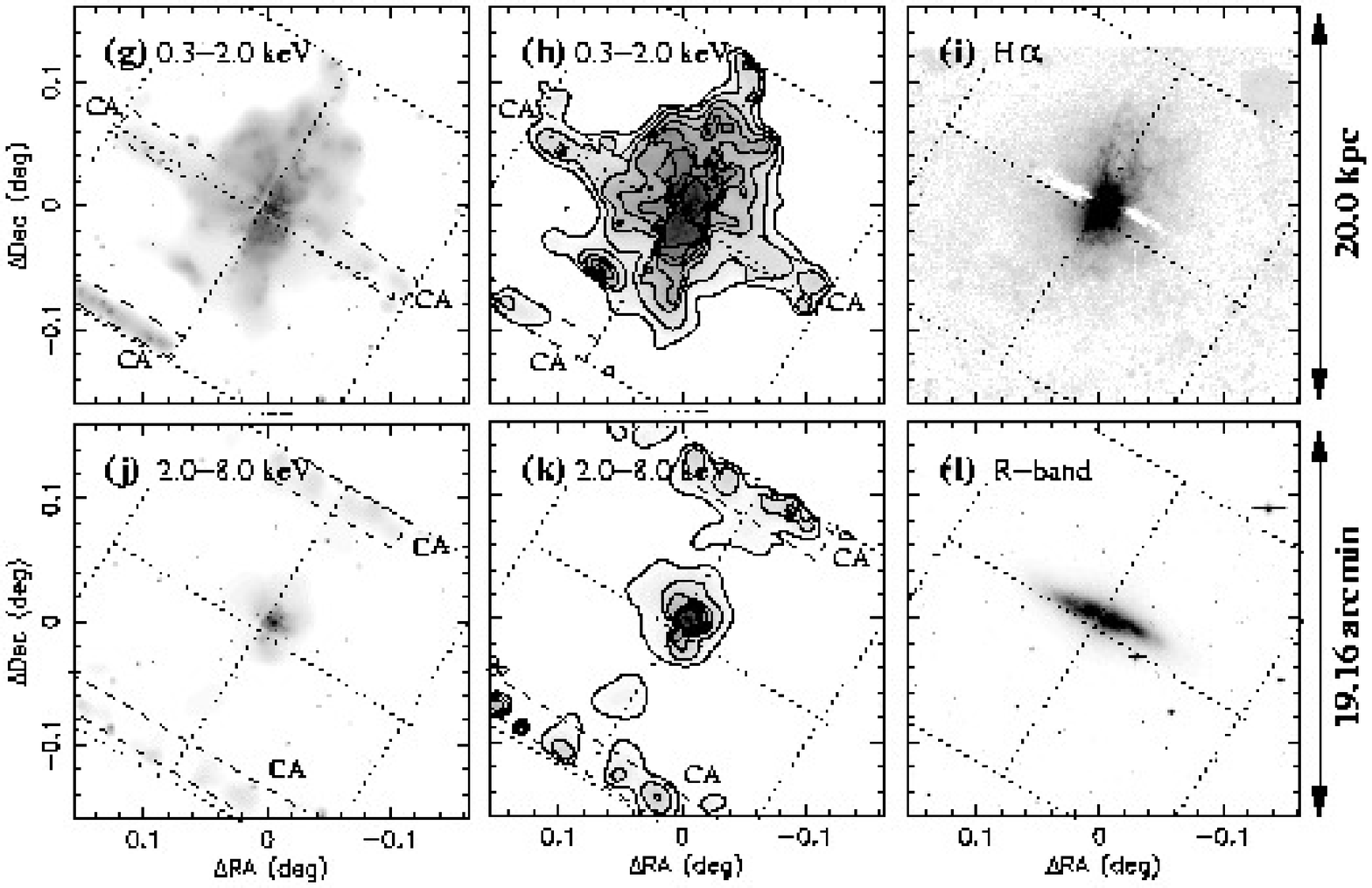}
\caption{{\it Chandra} X-ray and optical images of M82 
	(the ACIS-I observations).  Panels a through f 
	display the central 2 kpc of M82, with the
  dynamical center of the galaxy at the exact center of each image 
  (caption continued overleaf).
	}
\label{fig:images:m82}
\end{figure*}
\addtocounter{figure}{-1}
\begin{figure*}
\caption{ (continued).
  The nuclear region X-ray images, which include X-ray emission from
  point sources,
  (plots a \& d) are shown on
  a square-root intensity scale, from a surface brightness
  of $1\times10^{-8}$ counts s$^{-1}$ cm$^{-2}$ arcsec$^{-2}$
  (white) to $1\times10^{-5}$ counts s$^{-1}$ cm$^{-2}$ arcsec$^{-2}$
  (black). Contours are shown in intervals of 0.5 dex, between
  $2\times10^{-9}$ and $6\times10^{-7}$ counts s$^{-1}$ cm$^{-2}$ 
  arcsec$^{-2}$.
   Images of the diffuse X-ray emission (\ie all point sources removed)
  from the central 2 kpc 
  (plots b \& e) are also shown on
  a square-root intensity scale, from a surface brightness
  of $1\times10^{-8}$ counts s$^{-1}$ cm$^{-2}$ arcsec$^{-2}$
  (white) to $1\times10^{-6}$ counts s$^{-1}$ cm$^{-2}$ arcsec$^{-2}$
  (black). Contours are shown in intervals of 0.5 dex, between
  $2\times10^{-9}$ and $6\times10^{-7}$ counts s$^{-1}$ cm$^{-2}$ 
  arcsec$^{-2}$.
  Continuum-subtracted \halpha, and R-band images, are shown in 
  panels c \& f, respectively. The images are shown using
  a square-root intensity scale between arbitrary surface brightness
  levels. The position of the nucleus cited in
  Table~\ref{tab:galaxies} is marked by a black-or-white cross in
  panels a through f.
  Panels g through l are similar to panels 1 through f, but cover
  a $20 \times 20$ kpc region. Again, all plots are shown on a square
  root intensity scale. The contour levels for the X-ray images are the
  same as in panels a, b e \& f, although the gray-scale in shown
  between different surface brightness levels. In panels g \& j (X-ray 
  emission including the point source contribution),
  the gray-scale displays  surface brightnesses between
  $1\times10^{-9}$ (approximately half the background level)
  and $1\times10^{-5}$ counts s$^{-1}$ cm$^{-2}$ arcsec$^{-2}$.
  Panels h \& k (diffuse X-ray emission only),
  the gray-scale displays  surface brightnesses between
  $1\times10^{-9}$ (approximately half the background level)
  and $1\times10^{-6}$ counts s$^{-1}$ cm$^{-2}$ arcsec$^{-2}$.
  Note that all X-ray images have been background subtracted. 
  Artificial linear features in the X-ray images due to hot columns near
  the CCD node boundaries are marked CA (chip artifact), in 
  the cases where they
  could not be completely removed in the data processing. The location of the
  edges of the ACIS CCD chips are marked with thin dotted lines, and
  these locations are repeated on the optical images to ease cross-comparison
  between the X-ray and optical images.
	}
\end{figure*}

\subsection{Notes on individual galaxies}
\label{sec:results:gal_notes}

These are currently the highest resolution X-ray images of these
galaxies yet obtained, the first from which the point source contribution
can be largely removed, and they are likely to remain the only X-ray data
with spatial resolution comparable to ground based optical imaging
for the foreseeable future. We therefore believe it worthwhile to describe the
morphology of the diffuse X-ray emission we find, and its relationship to
the previously observed properties of these galaxies. 

In some cases the
{\it Chandra} data confirms the conclusions observers reached based on lower
spatial resolution {\it Einstein}, {\it ROSAT} HRI, PSPC and {\it ASCA} data,
while in other cases a revision of ideas is necessary.
In many cases structure is found within the diffuse emission
on all spatial scales down to the limits imposed by the combination
of the photon statistics and the ACIS PSF.

\subsubsection{M82}
\label{sec:results:gal_notes:m82}
\begin{figure*}[!ht]
\epsscale{2.0}

\plotone{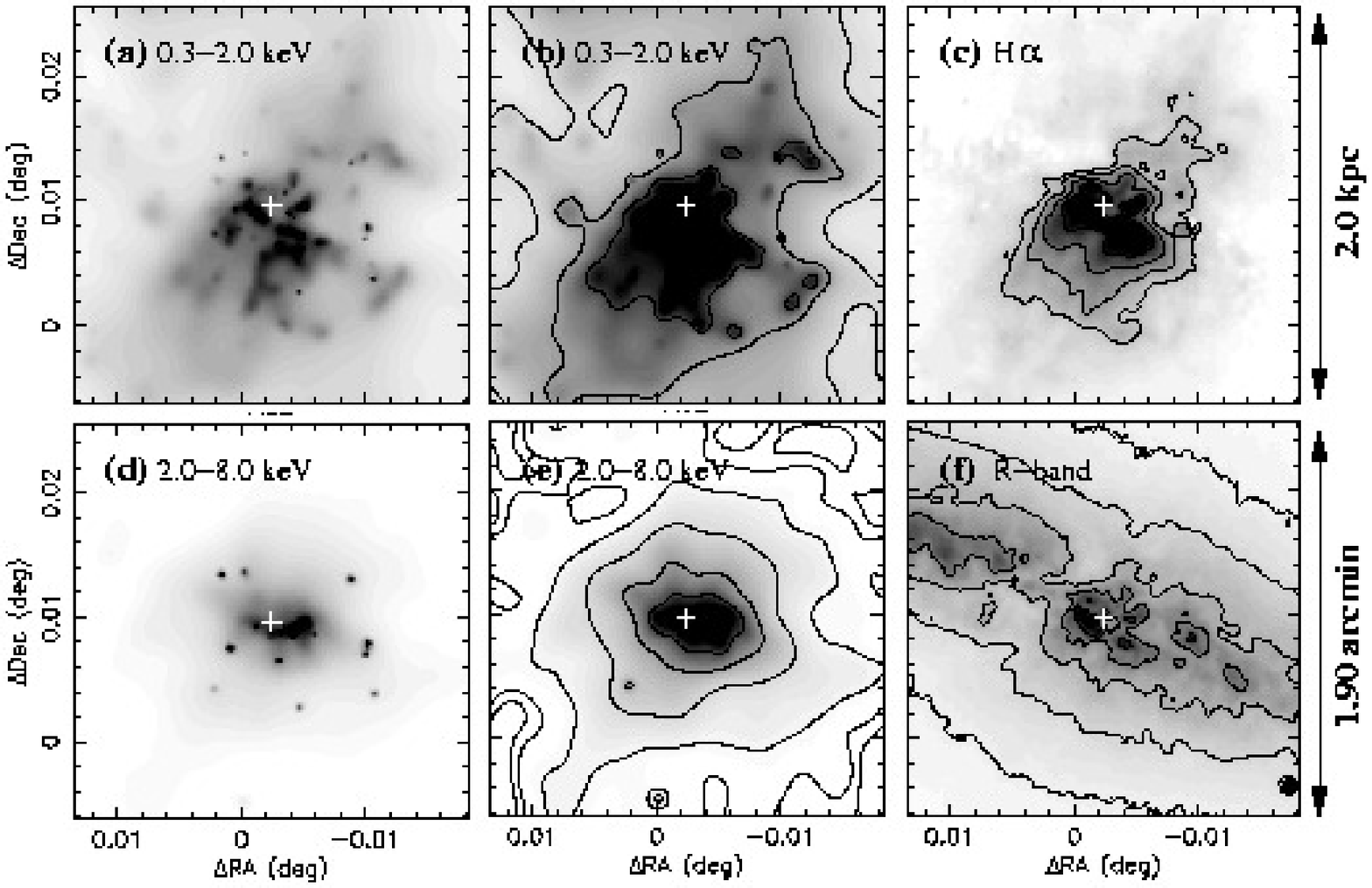}
\plotone{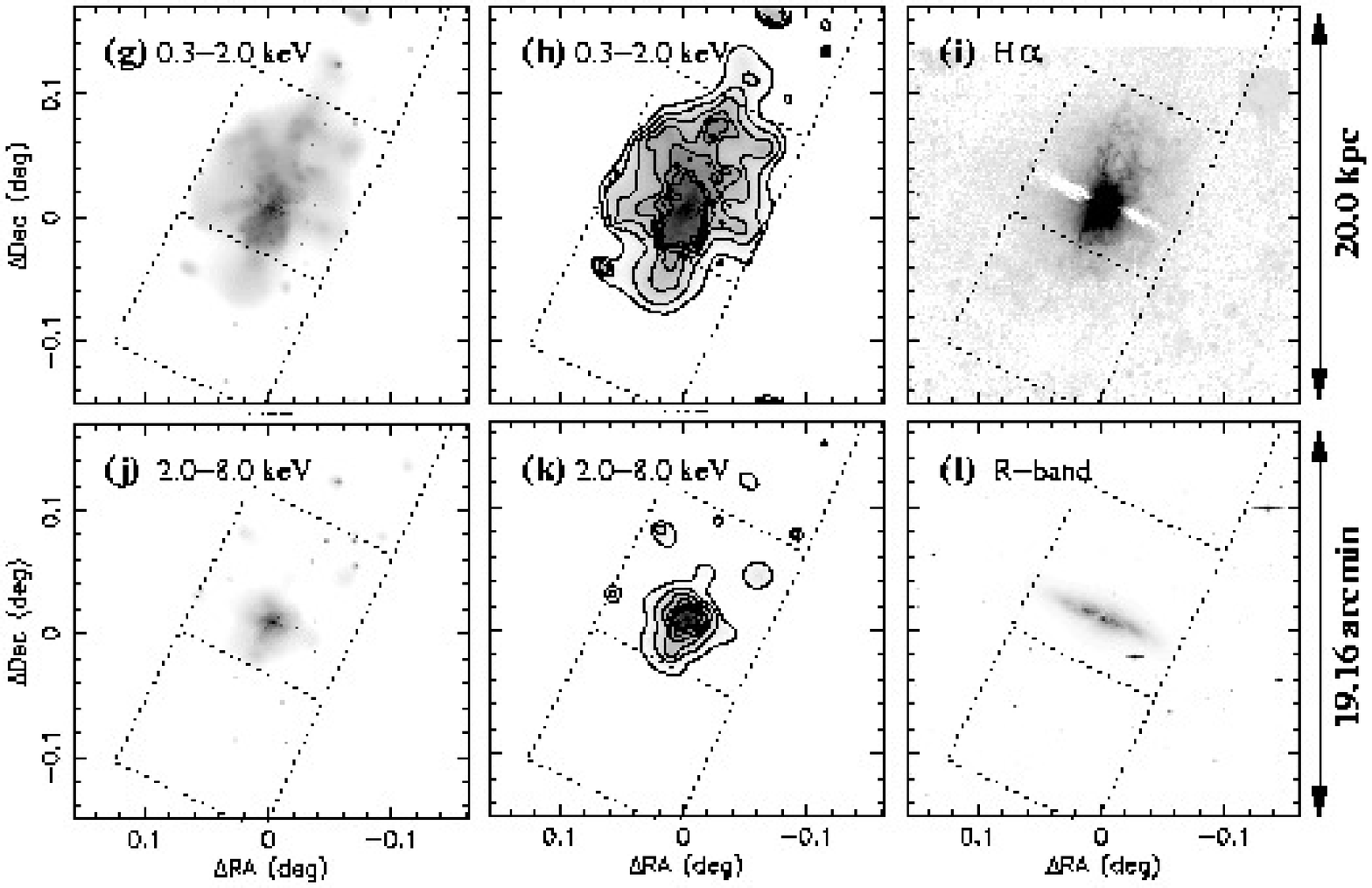}
\caption{{\it Chandra} X-ray and optical images of 
	M82 (the ACIS-S observation).
	The meaning of each panel, and the intensity scales and 
	contour levels used, are the same as those described 
	in Fig.~\ref{fig:images:m82}.
	}
\label{fig:images:m82dks}
\end{figure*}

Both the diffuse and point-source X-ray emission are sharply concentrated
within the central $\sim$ 900 pc, coincident with the central 
starburst region (\eg \citealt*{muxlow94,golla96b,forster01}). The brightest
pillars or filaments of diffuse soft X-ray emission emerge perpendicular
to the plane of the galaxy, toward the south east 
(Figs.~\ref{fig:images:m82}b, \ref{fig:images:m82dks}b). 
The diffuse emission to the north west 
is fainter, consistent with absorption by the intervening disk. 
In general, X-ray and \halpha~emission to the NW of the
major axis is fainter than that to the SE of the plane,
as expected based on the inclination of the galaxy.
Two relatively narrow X-ray-and-\halpha~bright
nuclear filaments to the north west of the nuclear disk are
notable as they also appear in the R-band image 
(Fig.~\ref{fig:images:m82}f). This to be most probably due to the
inclusion of \halpha~line emission in the R-band, as B-band images show
no similar structures.

On the larger scale, the ACIS images of the
diffuse emission show few differences from {\it ROSAT} HRI and 
PSPC images \citep*{bregman95,sps97}.
The filamentary nature of the brighter diffuse emission emission
is more apparent in the higher resolution, higher sensitivity ACIS
images than in the old ROSAT HRI images \citep{shopbell98}. 
As with NGC 253, the soft X-ray emission resembles the 
\halpha~emission on all scales.

Surrounding the filamentary X-ray emission is lower surface brightness
diffuse emission, which appears more uniformly distributed. The
larger spatial extent and greater uniformity might be purely
an artifact of lower photon statistics and heavier smoothing.
However, there is a smooth component of the \halpha~emission
\citep{bland88} with a similar spatial extent 
(\eg see Fig.~\ref{fig:images:m82}i). The smooth
\halpha~component is predominantly dust-scattered nuclear light, given
its polarization \citep* {schmidt76,bland88}, so the conditions
in the dusty medium and the X-ray emitting plasma must be somewhat
distinct from that in the  X-ray-and-\halpha~bright filaments.
 
The {\it Chandra} data also show diffuse X-ray emission associated with
the faint northern \halpha~filaments that stretch between 
the main X-ray/\halpha~nebula ($0 < z$ [kpc] $\la 5$), and the northern
``cloud'' or ``cap'' (at $z \sim 11$ kpc, see \citealt{devine99,lhw99}.
The northern cloud is not visible in  
Figs.~\ref{fig:images:m82} and \ref{fig:images:m82dks}, as it 
lies just beyond the top of these figures.).

To the south of the galaxy the diffuse X-ray emission can be traced out
to a distance of $\sim 7$ kpc below the plane of the galaxy, only slightly
further than seen with ROSAT. A notable feature is a bright
arc of diffuse X-ray emission, detached to the east of the main southern
outflow (centered at $\alpha = 09^{h} 56^{m} 38\fs3, 
\delta =  +69\degr 37\arcmin 47\farcs6$, J2000.0, 
see Fig.~\ref{fig:images:m82}h), possibly associated with
faint \halpha~emission at the same location (Fig.~\ref{fig:images:m82}i).
Given the apparent sharpness of the eastern edge of the southern outflow, the
presence of this diffuse feature further to the east is interesting.
This feature is not associated with any of the tidal \ion{H}{1} 
streamers around M82 \citep*{yun93}. The origin of this structure 
may be similar to that of the northern cloud discussed in \citet{lhw99}.

The diffuse hard X-ray emission, discovered by \citet{griffiths2000},
is preferentially extended along the major axis of the galaxy at high
X-ray surface brightness (Figs.~\ref{fig:images:m82}e, 
\ref{fig:images:m82dks}e). However, at lower surface brightness
it appears to also participate in the minor axis outflow
(see also Figs.~\ref{fig:images:m82}k and \ref{fig:images:m82dks}k).

There is no obvious enhancement in the diffuse X-ray emission, in
either the soft or hard energy bands, associated with the locations of the
4 expanding \ion{H}{1} shells described by \citet*{wills02}.
Several of the brighter X-ray point sources lie within the
confines of Wills \etal's shell 3 (previously detected in  CO emission, 
see \citealt{weiss99,matsushita00}).
It is worth noting that the off-nuclear intermediate luminosity X-ray object 
\citep{kaaret01,matsumoto01} lies {\em outside} shell 3,
weakening the argument by \citet{matsushita00} 
that the two are related.

\begin{figure*}
\epsscale{2.0}

\plotone{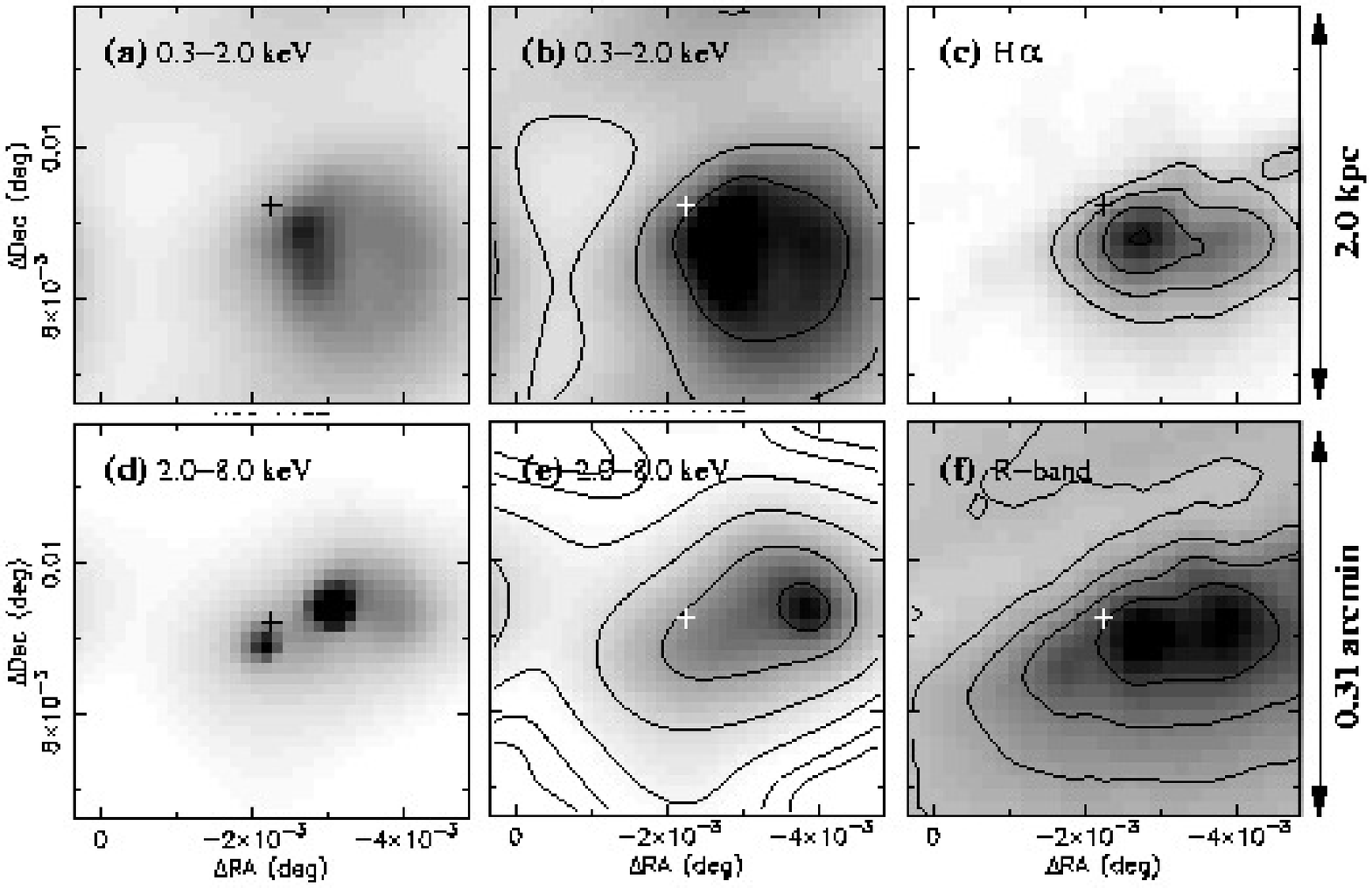}
\plotone{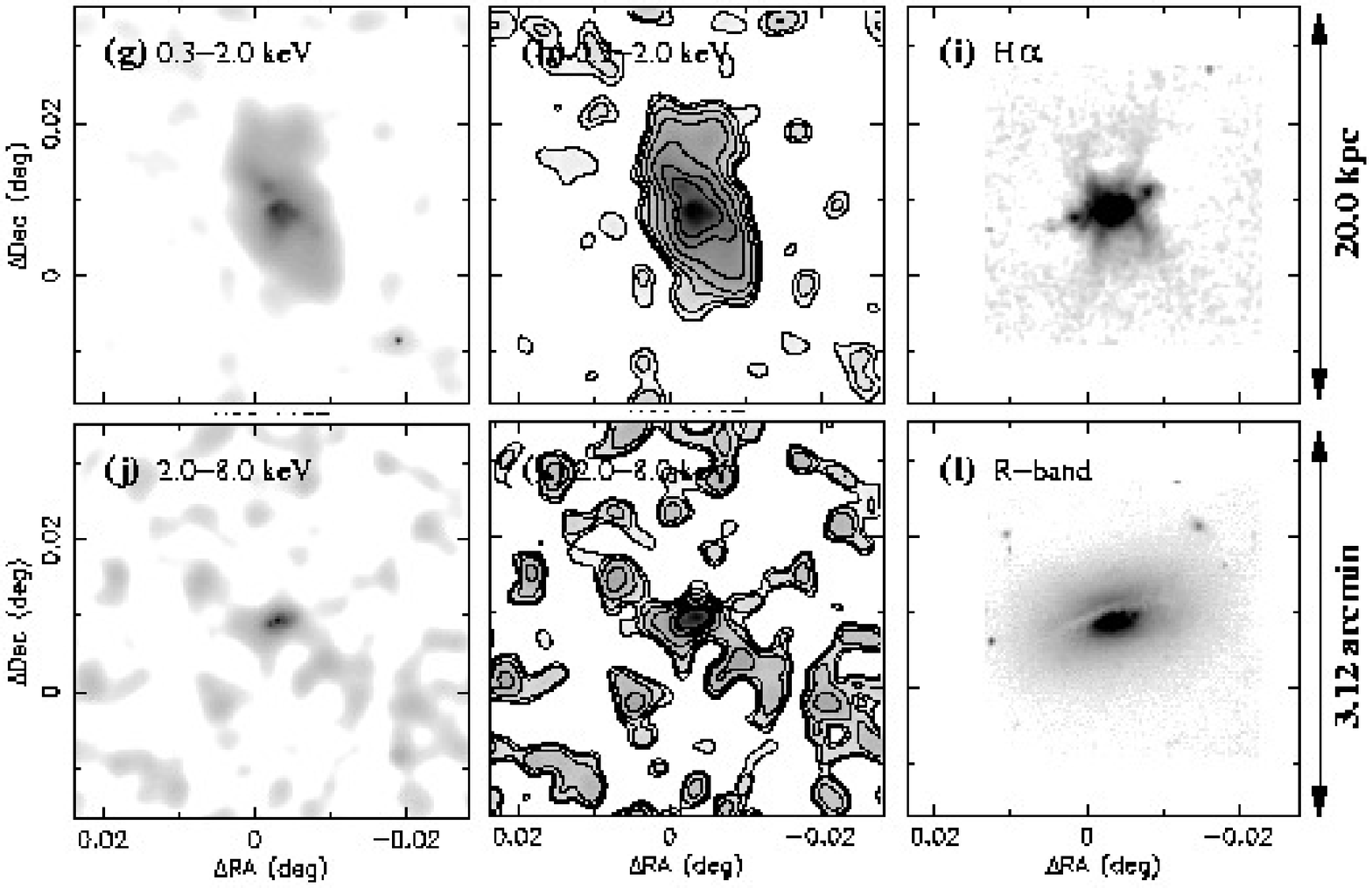}
\caption{{\it Chandra} X-ray and optical images of  NGC 1482.
	The meaning of each panel, and the intensity scales and 
	contour levels used, are the same as those described 
	in Fig.~\ref{fig:images:m82}.
	}
\label{fig:images:n1482}
\end{figure*}

\subsubsection{NGC 1482}
\label{sec:results:gal_notes:n1482}
NGC 1482 is a IR-warm starburst galaxy, of similar mass and size to 
that of M82, located at a distance of approximately 22 Mpc. Given its
southern declination, and only moderate inclination ($i \sim 58\degr$),
it has escaped serious observational attention until recently. 
\citet{hameed99} first pointed out the spectacular extra-planar
\halpha~filamentation (\eg Fig.~\ref{fig:images:n1482}i), prompting both our
{\it Chandra} observation (the first published X-ray study of this
galaxy), and an optical spectroscopic investigation
by \citet{veilleux02} that confirmed the presence of a $\sim 250$ km/s
outflow. There is no sign of any AGN activity, its optical 
spectroscopic classification being that of a standard
\hii~galaxy \citep{kewley2001}. 

The star formation appears to be concentrated in a $\sim 1$ kpc-diameter
nuclear disk, based on the location of both the \halpha~emission and 
the hard X-ray emission (Fig.~\ref{fig:images:n1482}c \& d). 
In the soft X-ray band (0.3--2.0 keV), a well collimated,
limb-brightened outflow emerges to the south of the nucleus 
(the beginnings of which can be seen in Fig.~\ref{fig:images:n1482}b), 
and extends out to $z \sim 5$ kpc below the plane of the galaxy 
(Fig.~\ref{fig:images:n1482}h), before disappearing at approximately the
same location at which the \halpha~emission disappears 
(Fig.~\ref{fig:images:n1482}i). The overall morphology is that
of the archetypal superwind -- a limb-brightened cone truncated at its 
base in the disk of the galaxy.

The northern outflow is very similar in morphology and size, although both
the X-ray and \halpha~emission from the innermost kiloparsec is 
heavily absorbed by the strong dust lane that is visible in the
R-band image (Fig.~\ref{fig:images:n1482}l) of the galaxy.

Close inspection of unsmoothed images shows that the
soft X-ray emission is very similar in morphology to the \halpha~emission,
and displays the same limb-brightened truncated conical geometry, despite
the filled-in appearance of the smoothed X-ray images presented in
Fig.~\ref{fig:images:n1482}. We will present a more detailed comparison
of the X-ray emission with the \halpha~emission in a future paper.

The surface brightness of the diffuse soft X-ray emission is
higher than in the majority of the galaxies in this sample, 
and is very similar
to that of the diffuse emission from M82.

\begin{figure*}
\epsscale{2.0}

\plotone{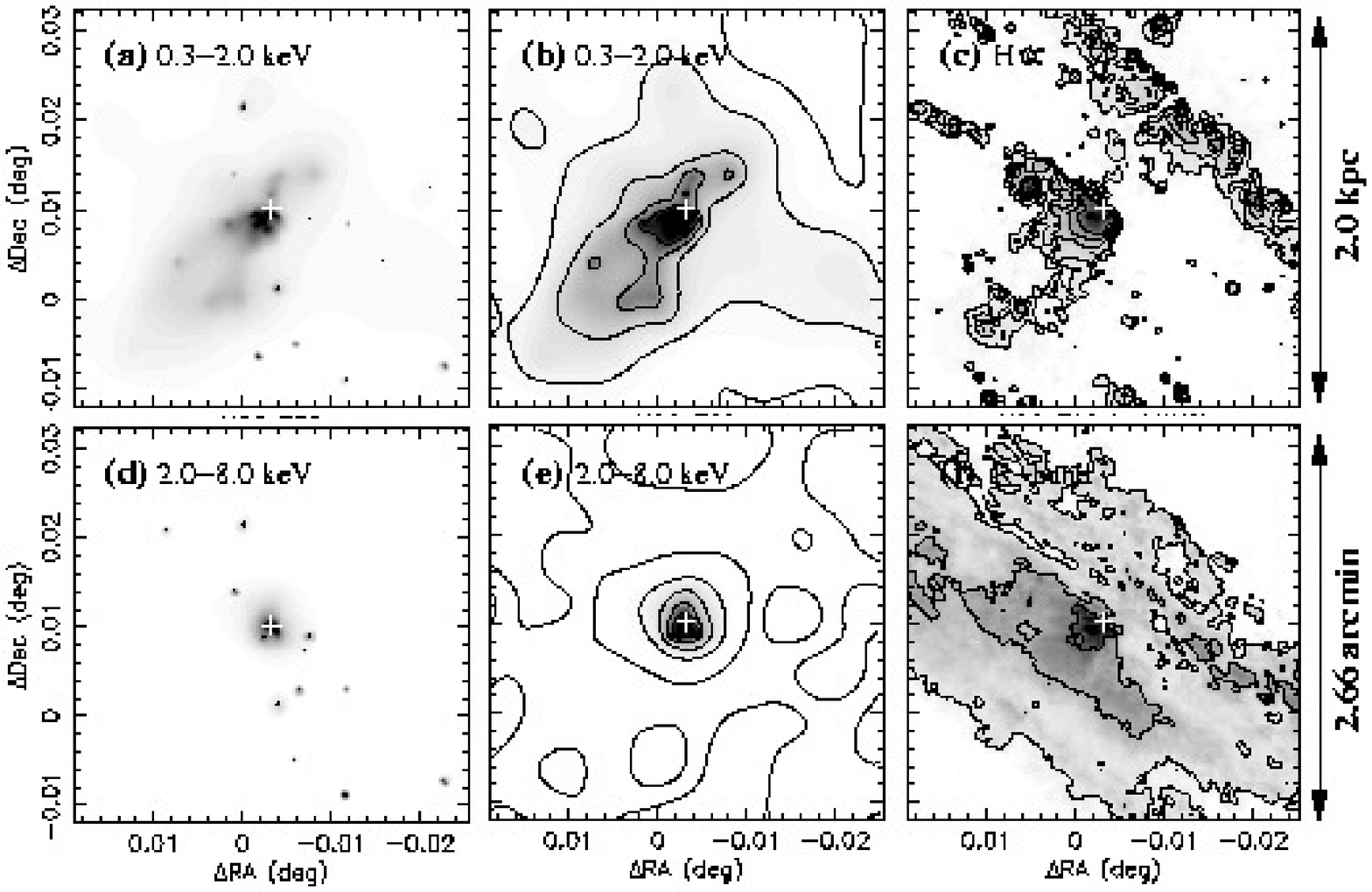}
\plotone{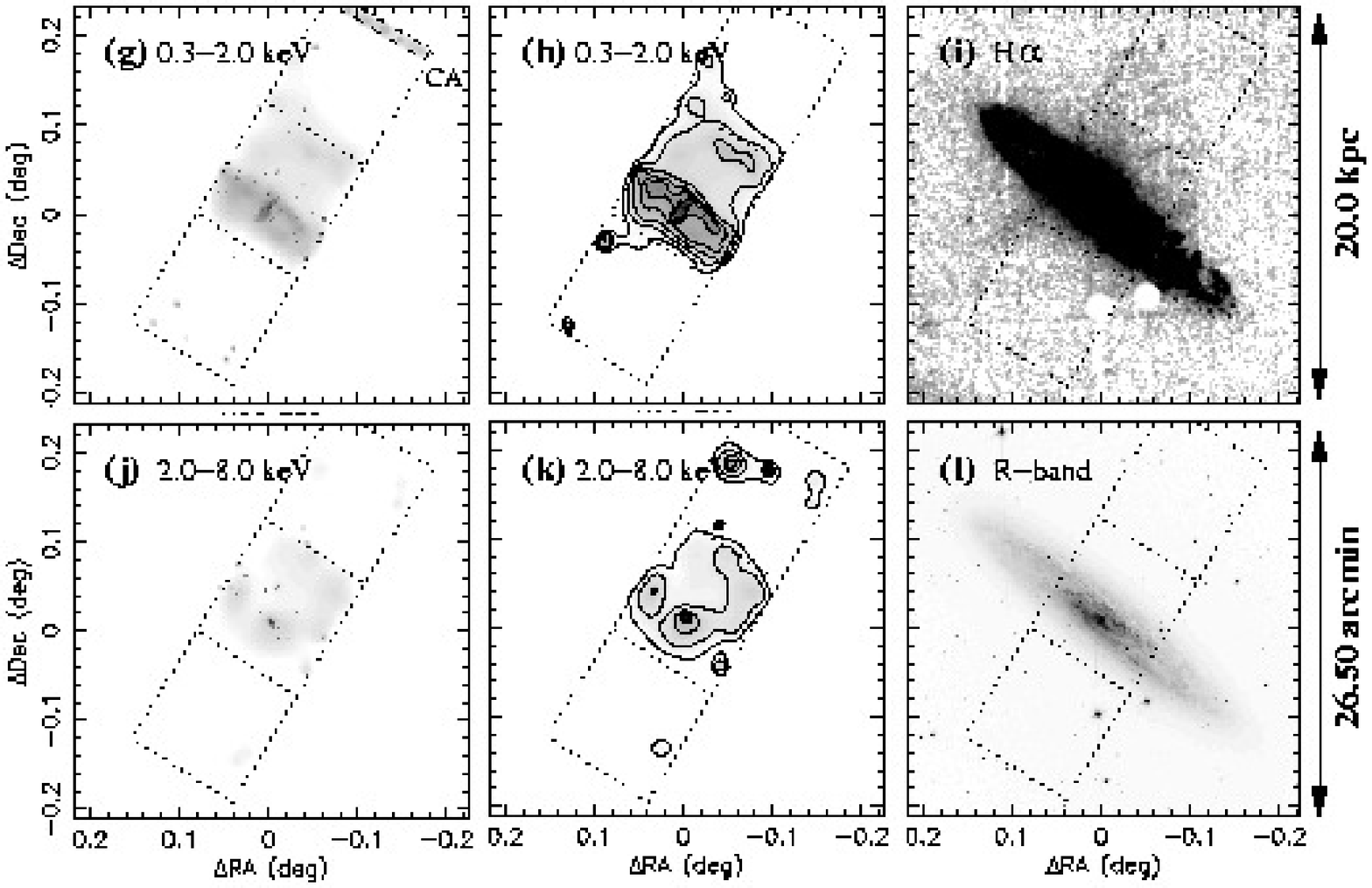}
\caption{{\it Chandra} X-ray and optical images of  NGC 253 (ObsID 969).
	The meaning of each panel, and the intensity scales and 
	contour levels used, are the same as those described 
	in Fig.~\ref{fig:images:m82}. The diffuse hard X-ray filling
	the majority of the S3 chip in panel k is
	an artifact (see \S~\ref{sec:results:gal_notes:n253}). 
	}
\label{fig:images:n253_15ks}
\end{figure*}

\begin{figure*}
\epsscale{2.0}

\plotone{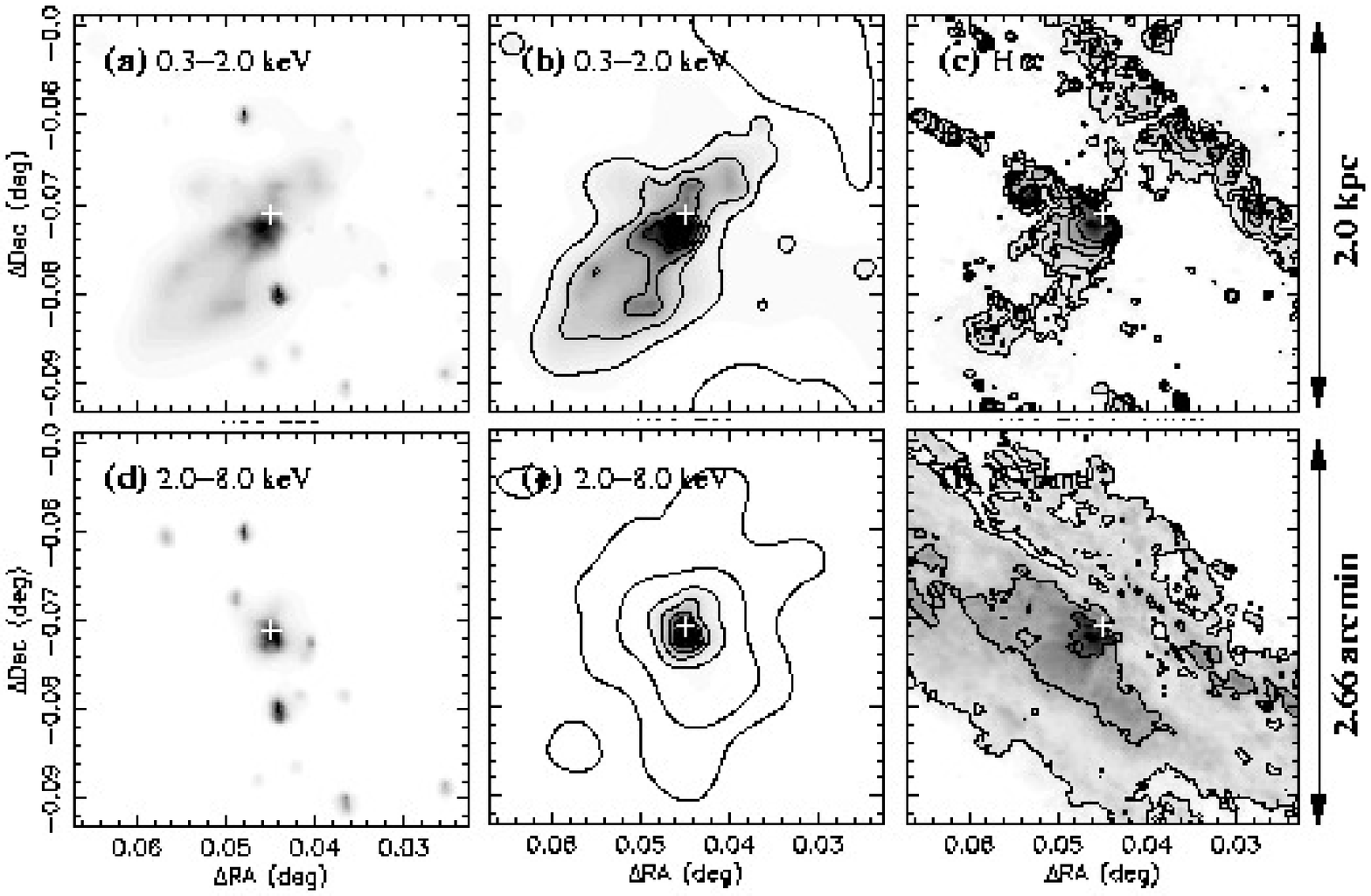}
\plotone{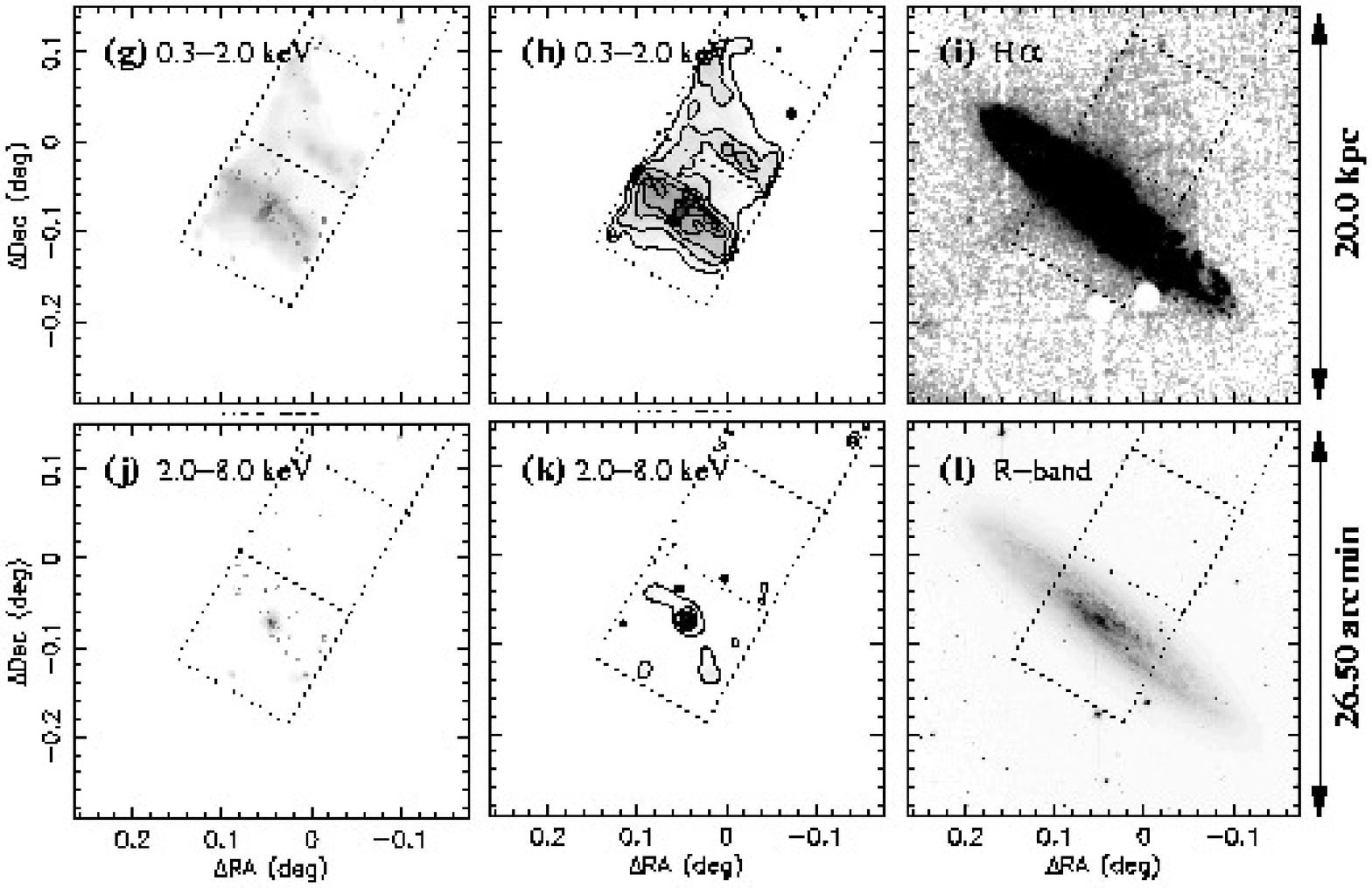}
\caption{{\it Chandra} X-ray and optical images of  NGC 253 (ObsID 790).
	The meaning of each panel, and the intensity scales and 
	contour levels used, are the same as those described 
	in Fig.~\ref{fig:images:m82}.
	}
\label{fig:images:n253_45ks}
\end{figure*}

\subsubsection{NGC 253}
\label{sec:results:gal_notes:n253}
NGC 253, the most massive galaxy in the Sculptor group \citep{puche88}, 
hosts a compact nuclear starburst, although recent star-formation is
also widespread throughout the 
disk \citep{ua97,engelbracht98,sofue,ulvestad00}. At a distance of 
only $\sim 2.6$ Mpc, it is the closest massive galaxy undergoing a starburst.
It has been heavily studied at X-ray wavelengths, originally with
{\it Einstein} \citep{fabbiano84}. Lacking any obvious nearby
companion galaxy, the
original cause of the starburst activity is a mystery.

Figs.~\ref{fig:images:n253_15ks} and \ref{fig:images:n253_45ks} display
both {\it Chandra} X-ray and optical images of NGC 253.
In general the morphology and surface brightness of the 
diffuse emission is consistent between the two observations.
The only significant exception to this is the apparently diffuse
hard X-ray emission filling the S3 chip in the shorter observation
(Fig.~\ref{fig:images:n253_15ks}k). The is undoubtedly an artifact,
as such hard diffuse emission
is not detected in the S2 chip which covers this location in the longer
observation (see Fig.\ref{fig:images:n253_45ks}k), and is due
to the excess background emission in the S3 chip
discussed in \citet{strickland02}.

We refer the reader to \citet{strickland00}, \citet{strickland02}
and \citet{weaver02} for detailed presentation and discussion of the
{\it Chandra} observations of NGC 253.

\begin{figure*}
\epsscale{2.0}

\plotone{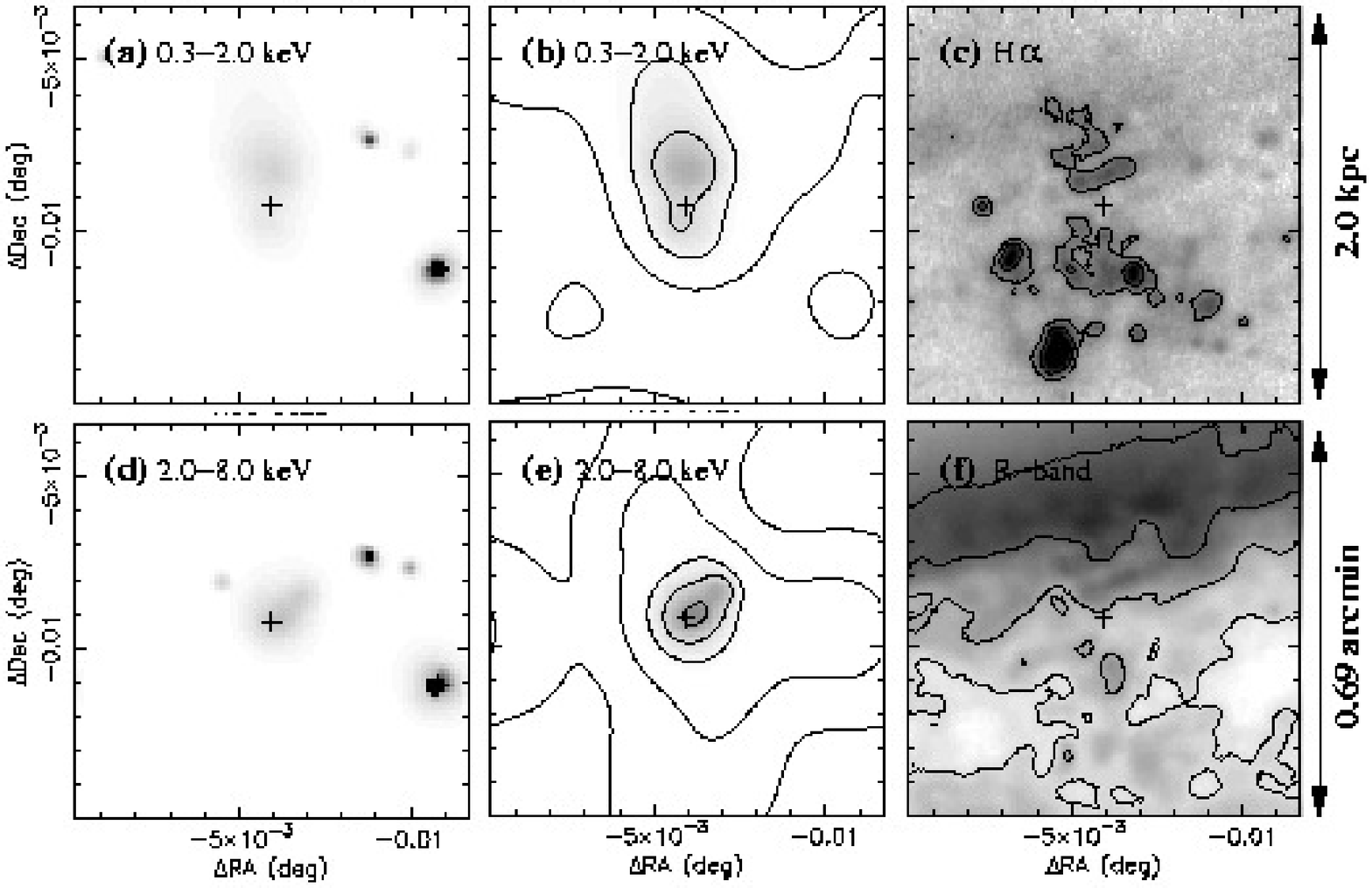}
\plotone{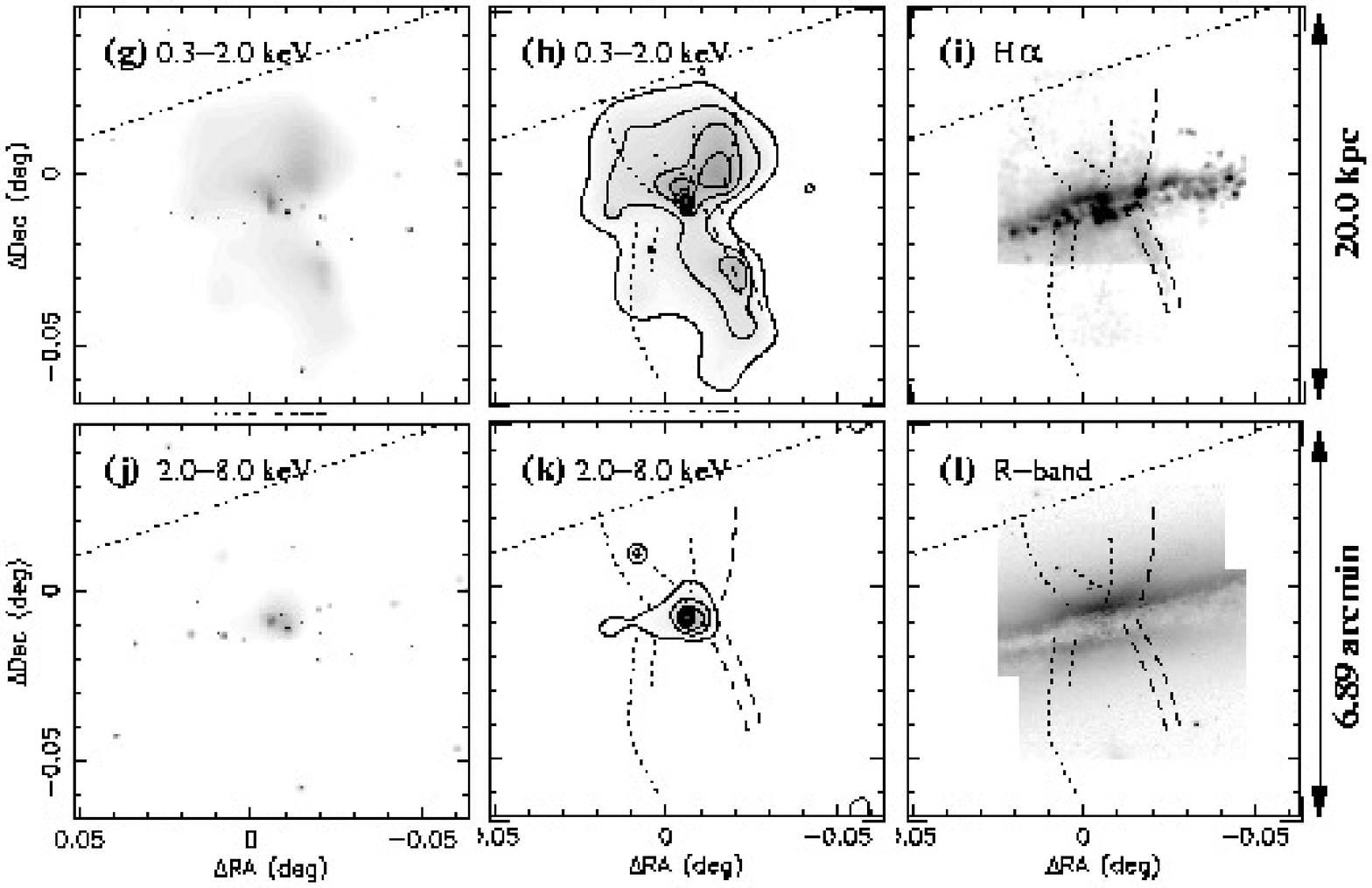}
\caption{{\it Chandra} X-ray and optical images of  NGC 3628.
	The meaning of each panel, and the intensity scales and 
	contour levels used, are the same as those described 
	in Fig.~\ref{fig:images:m82}.
	The dashed and dotted arcs marked on panels h, i, k, and l
	are intended to draw the reader's eye to the features discussed
	in \S~\ref{sec:results:gal_notes:n3628}.
	}
\label{fig:images:n3628}
\end{figure*}

\begin{figure*}
\epsscale{2.0}

\plotone{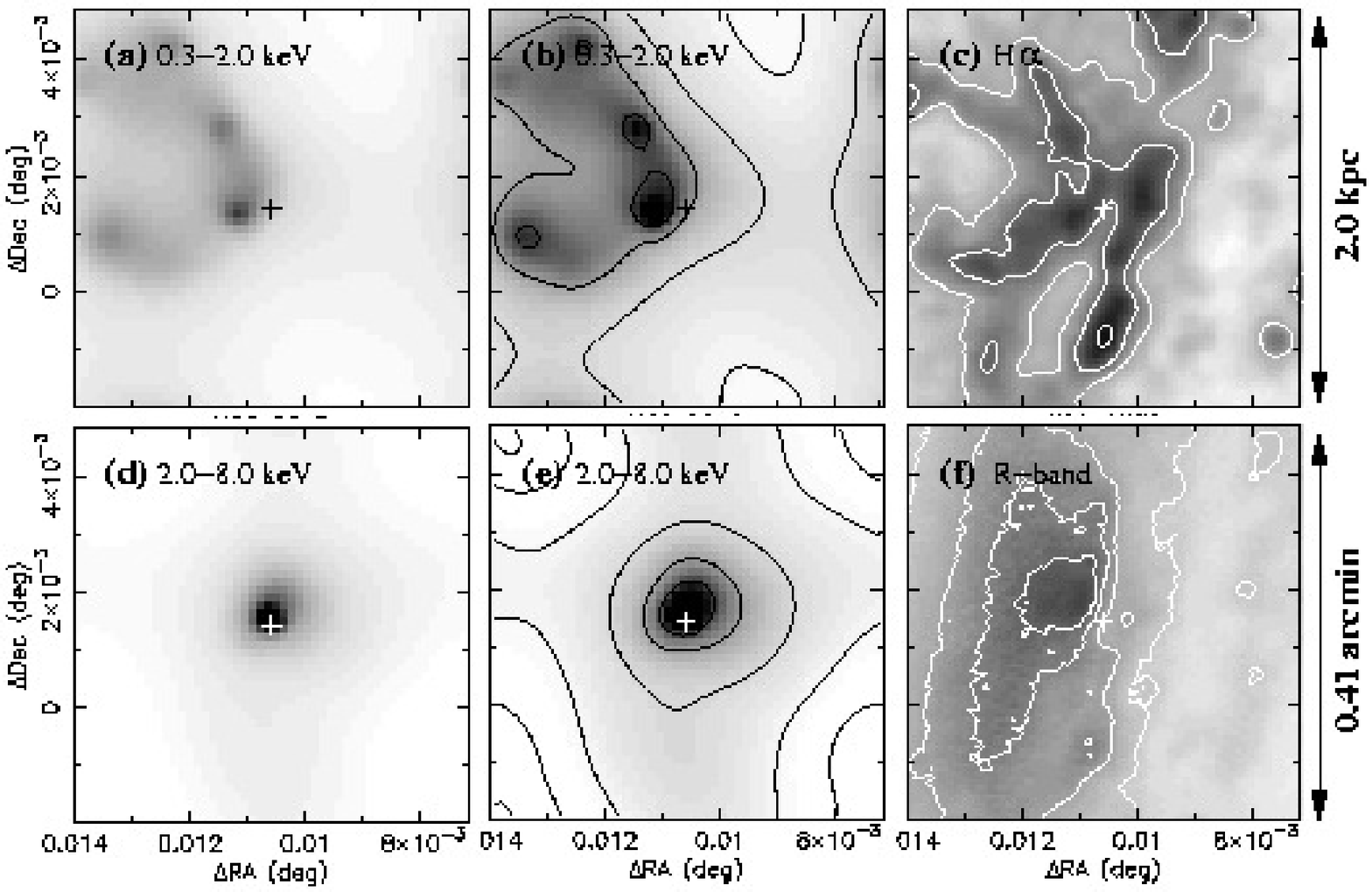}
\plotone{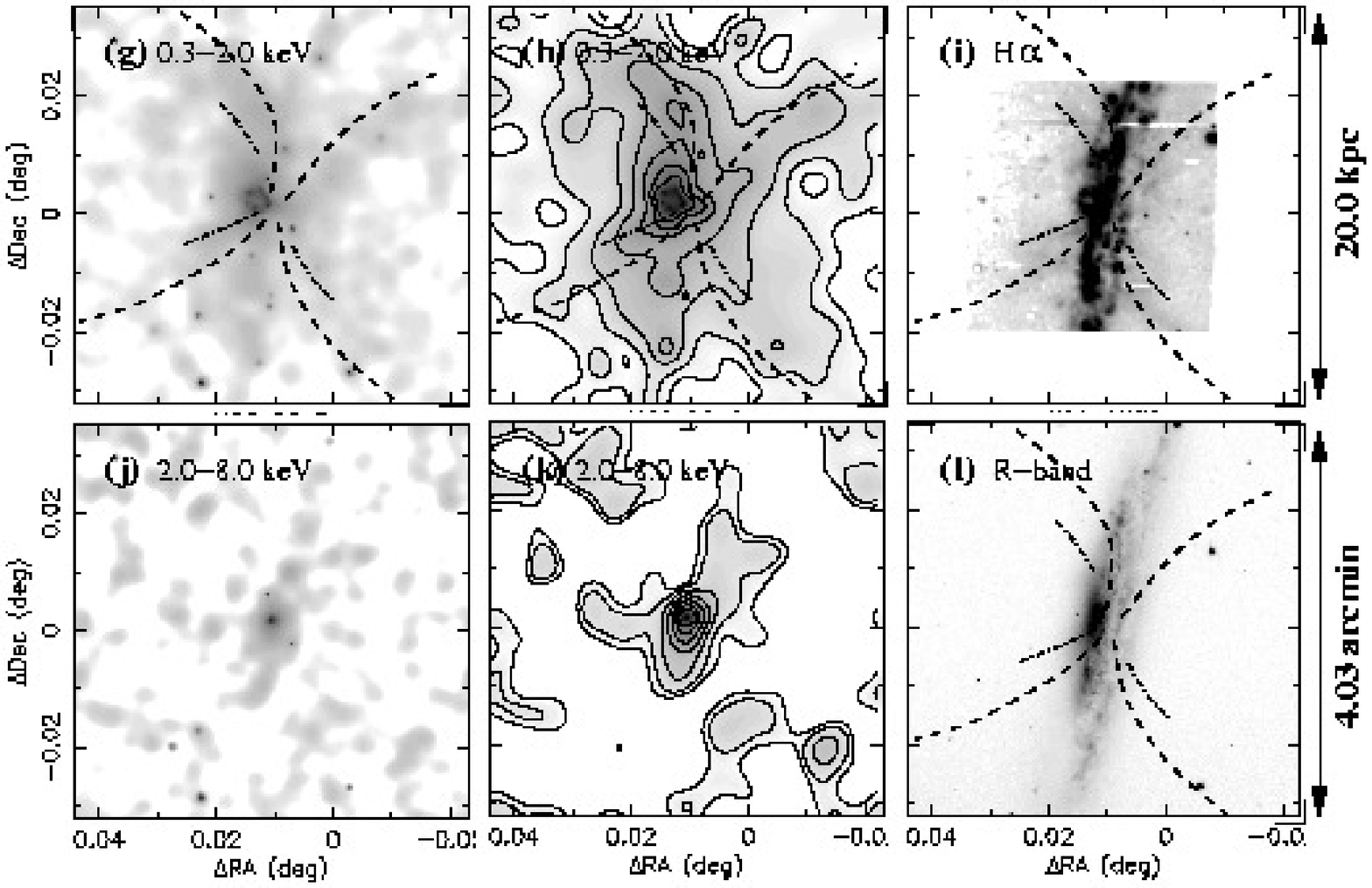}
\caption{{\it Chandra} X-ray and optical images of  NGC 3079.
	The meaning of each panel, and the intensity scales and 
	contour levels used, are the same as those described 
	in Fig.~\ref{fig:images:m82}.
	The arced dashed lines marked on panels h, i, k, and l
	are intended to draw the reader's eye to the features discussed
	in \S~\ref{sec:results:gal_notes:n3079}.
	}
\label{fig:images:n3079}
\end{figure*}

\subsubsection{NGC 3628}
\label{sec:results:gal_notes:n3628}
NGC 3628, a member of the 
Leo triplet, hosts a compact nuclear
star-forming disk, approximately
300 pc in diameter based on the non-thermal radio 
emission from young supernovae \citep*{carral90,irwin96,zhao97,cole98}, 
although the star-formation rate is high within the 
entire central kiloparsec (\eg \citealt{irwin96}).
An optical and \ion{H}{1} tidal arm extends $\sim 100$ kpc 
to the east of NGC 3628 \citep*{kormendy74,rots78,haynes84}, 
evidence of a close interaction between
NGC 3628 and NGC 3627 approximately $800$ Myr ago, based on the 
modeling of \citet{rots78}.
A very distinct boxy/peanut-shaped bulge suggests 
a bar may be present.

The {\it Chandra} X-ray observations (Fig.~\ref{fig:images:n3628}) 
reveal a wealth of detail, generally in agreement with 
the {\it Einstein} and {\it ROSAT} observations of \citet{fhk90},
\citet{rps97} and \citet{dahlem96}.

We find extended hard X-ray emission associated with the 
optically-obscured nuclear starburst region 
(Fig.~\ref{fig:images:n3628}d \& \ref{fig:images:n3628}e).
The major-axis spatial extent of this emission ($\sim 8\arcsec$)
is similar to the $6\arcsec$-diameter ridge of 1.4 GHz \citep{cole98}
and 15 GHz radio emission \citep{carral90}. The nuclear hard X-ray emission
is most likely predominantly a blend of spatially-unresolved 
high-mass X-ray binaries, rather than genuinely diffuse emission.
The relative faintness of the nuclear emission, as compared to other binaries
seen in projection within the central 2 kpc, is no doubt related to the
high absorbing column density toward the nucleus.

Relatively bright, but patchy, \halpha~emission extends $\sim 10\arcsec$ 
($\sim 450$ pc) to both the north and south of the location of 
this nuclear disk (Fig.~\ref{fig:images:n3628}c). We speculate that this
is emission from a well-collimated nuclear outflow, seen
along sight-lines of lower than normal obscuration. The {\it Chandra}
data supports this hypothesis, as we also see diffuse soft X-ray emission
in the same region (Fig.~\ref{fig:images:n3628}b), again
preferentially extended along the minor axis. The adaptive smoothing
applied to Fig.~\ref{fig:images:n3628}b tends to destroy details of the
structure of the diffuse emission. In the un-smoothed pixel maps
the diffuse soft X-ray emission clearly separates into two approximately
rectangular regions, separated from each other by a $\sim 3\arcsec$
thick soft-X-ray-dark 
region, corresponding spatially to the starburst region.
The region to the north the nucleus is $\sim8\arcsec$ wide 
($\sim 390$ pc), and extends $\sim12\arcsec$ ($\sim 580$ pc) to
the NNE along the minor axis. The southern region has a similar major-axis
extent, but only extends $\sim5\arcsec$ ($\sim 250$ pc) to the SSW.
The northern rectangle or cylinder is the brighter of the two,
consistent with the commonly-assumed geometry of NGC 3628, which is
that the disk of NGC 3628 is
inclined so we see its northern face. These structures are likely
to be analogous to the nuclear outflow cones seen in NGC 253 and 
NGC 4945.

On the larger scale (panels g to l in Fig.~\ref{fig:images:n3628}), 
diffuse soft X-ray emission is again found in
association with \halpha~emission, within the limited
field of view of the \halpha~image. To both the north and south of the
disk, the western-most edge of the diffuse emission is the brightest.
There is suggestive evidence for a eastern limb to the X-ray emission
in both the north and south, \ie
a bi-polar, limb-brightened, cavity somewhat similar to an open-ended
hourglass in morphology (dashed and dotted lines have added to 
Fig.~\ref{fig:images:n3628} to guide the reader these faint
features, and the others we describe below). 

The X-ray emission to the south of the optical
disk of the galaxy is brightest in the region bounded by the 
$\sim 4$ kpc-long \halpha~filament discovered by \citet{fhk90}
(for a clearer presentation of this, see Fig.~3 in \citealt{strickland_vulcano}).
This X-ray-plus-\halpha~filament is very 
similar to the $\sim 650$ pc-long filament 6 in the dwarf starburst
NGC 1569 \citep*{martin02}.
This filament appears to join the disk of the galaxy $20\arcsec$
($1.0$ kpc) to the west of the nucleus, although the inclination 
of the galaxy makes the location of the filament 
at heights $z \lesssim 1$ kpc 
of the plane uncertain.
To the east of the nucleus, extra-planar \halpha~filamentation 
is less pronounced,
and can only be traced with confidence to $\sim 2$ kpc south of the plane.
Diffuse soft X-ray is present within this region, but lacks any clearly
developed limb/filament of the sort found on the western limb.

On the northern side of the galaxy the \halpha~emission appears in the form
of a series of faint $\le 2$ kpc-long filaments, primarily above the nucleus
and to the west of the nucleus. The westernmost extent of the \halpha~emission
coincides with a bright \ion{H}{2} complex $\approx 40\arcsec$ 
($\approx 2$ kpc) to the west
of the nucleus. Extra-planar diffuse X-ray emission is relatively 
bright above this region. East of the nucleus, the lack of significant 
extra-planar \halpha~emission is matched by a similar lack in 
diffuse X-ray emission.

It is interesting to note that the base of the eastern and 
western edges of the extra-planar \halpha~and X-ray emission is
similar to the maximum extent of the box-shaped bulge seen in optical 
continuum images, which may be related to 
a bar (see \eg \citealt{irwin96,cole98}).

What is the cause of the pronounced east/west emission asymmetry?
Jonathan Bland-Hawthorn (2000, private communication)
has suggested that some of the asymmetry in the M82 wind, in particular
the apparent $\sim10\degr$ bending of the flow away from the minor axis,
might be due to ram pressure acting on the wind as M82 moves through
the inter-group medium. {\em If} NGC 3628 is moving primarily to the west,
then ram pressure confinement or compression would lead to higher gas
densities on the western side of the wind, and hence produce the higher
\halpha~and X-ray surface brightness that is seen. Unfortunately, 
NGC 3628's motion within the Leo triplet is unknown, as only
kinematic model of the Leo triplet \citep{rots78} deals with the
relative motion of NGC 3627 with respect to NGC 3628. 

With the higher sensitivity, and better point source rejection
afforded by the use of {\it Chandra}, we {\em robustly} trace the diffuse
emission out to $\sim 2\farcm4$ ($\sim 7.0$ kpc) to the north and
$\sim 3\farcm7$ ($\sim 10.8$ kpc) to the south of the disk.
The maximum major-axis width of the emission is $\sim 3\farcm7$
in the north, and slightly less to the south.

The diffuse X-ray emission 
to the north of the disk may extend even further from galaxy 
(beyond the outermost contour in Fig.~\ref{fig:images:n3628}h), 
out to $\sim 5\farcm0$ ($\sim 14.6$ kpc),
at low surface brightness, but the signal-to-noise is low, given 
that this region lies on the
lower sensitivity front-illuminated S2 chip.

The ability to robustly remove unrelated X-ray point sources
makes much more of a difference to determining the morphology and
extent of the diffuse X-ray emission in NGC 3628, than it does
in nearer objects such as M82 and NGC 253. Our large-scale diffuse emission
image is closest in resemblance to the smoothed point-source-subtracted
{\it ROSAT} PSPC image of \citet{rps97}. Many of the features in the
images of \citet{fhk90}, and the ostensibly point-source-subtracted
images of \citet{dahlem96}, are due to background point 
sources\footnote{We note in passing that the {\it Chandra} data
shows no connection between the diffuse emission and a
$z=2.15$ AGN, claimed by \citet{arp02} based on 
heavily smoothed {\it ROSAT} PSPC data (see \citealt{hardcastle00}
for a discussion of statistical inference from smoothed images).
Note that the z=2.15 AGN lies outside the field of view of 
Fig.~\ref{fig:images:n3628} panels g to l, which contain the full
extent of the diffuse X-ray emission. 
Nor does the diffuse X-ray emission
correlate with the ``chain of optical objects'' discussed 
in that paper. Only one object from the region of the 
optical chain that lies within the projected region of 
diffuse X-ray emission, the $z=0.995$ AGN, 
has an X-ray counterpart. The chance of finding an X-ray source 
this bright ($f_{\rm X} \sim 3 \times 10^{-14}$ 
erg s$^{-1}$ cm$^{-2}$ in each of the
soft and hard energy bands), 
within the $220\arcsec \times 320\arcsec$ area covered 
by the diffuse X-ray emission, is 20 to 40\% (based on the soft and
hard-band point source luminosity functions published 
in \citealt{giacconni2001}), which makes it most likely that this
is a chance superposition.}. Consequently, the spatial extent of the
diffuse emission we report is somewhat smaller than previously reported
values.

\subsubsection{NGC 3079 (and NGC 3073)}
\label{sec:results:gal_notes:n3079}
NGC 3079 is a massive spiral galaxy with a LINER nucleus 
\citep{heckman80}, and is home to both vigorous disk-wide 
star-formation (as evidenced by the high values of the IRAS
$60\micron$ to $100\micron$ flux ratio throughout the optical disk, 
see \citealt{mayya97}, in addition 
to \citealt{seaquist78,veilleux94,braine97}) 
and a weak AGN with a pc-scale radio jet 
(\eg \citealt{irwin88,trotter98,sawada00}). 
Direct radio detection of young SN remnants in the nuclear region
(as found in other galaxies with nuclear starbursts such as
M82, NGC 253 and NGC 3628) is currently lacking, although this
may be mainly due to a combination of the greater distance of 
NGC 3079 as compared to these other starbursts, and the general
focus of radio studies on the nuclear radio source and the probable
radio jet. Nevertheless, there is good observational evidence
pointing to recent star-formation within the molecular-gas-rich 
central few hundred pc \citep{armus95,israel98}.

We originally targeted NGC 3079 as part of a program seeking to
ascertain whether the close spatial correlation between X-ray and
\halpha~emission found in the southern NGC 253 nuclear outflow
cone \citep{strickland00} is a general feature of all superwinds.
NGC 3079 was known to have an outflowing, kiloparsec-scale,
limb-brightened optical emission line 
bubble \citep{ford86,ham90,filippenko92,veilleux94}, 
subsequently dramatically 
imaged by the HST WFPC2 \citep{cecil01},
and there was suggestive evidence for
bright structured diffuse X-ray emission in the same region
from ROSAT HRI observations \citep{dwh98,pietsch98}.

The {\it Chandra} observations of NGC 3079 show that the diffuse
X-ray emission is very closely associated with the optical 
filaments forming the bubble -- the lower half of which can be
seen in the left hand side of Fig.~\ref{fig:images:n3079} panels
a and c (see \citet{strickland_iau} for a direct overlay of the X-ray emission
on the \halpha~emission, or \citet*{cecil02}\footnote{The X-ray 
fluxes, emission integrals, luminosities, and other
derived physical properties of the X-ray-emitting gas quoted 
in \citet{cecil02} are incorrect.  
For example, in the region they define as the bubble, the
absorption-corrected intrinsic 0.1-6.5 keV energy band X-ray flux 
is $f_{\rm X} \approx 2\times10^{-13}$, a factor 400 lower than the
value \citeauthor{cecil02} quote, but consistent with the value
expected based on
the count rate within the older ROSAT HRI observations 
\citep[\eg][]{pietsch98}.
We find a best-fit emission integral 
for this bubble region $EI = n_{\rm e} n_{\rm H} \eta_{\rm X} V = 
5\times10^{62} (D/17.1 {\rm Mpc})^{2} \pcc$, again much
lower than the $10^{66} \pcc$ given in \citet{cecil02}, which
implies the electron densities, gas masses, pressures and energy content
are $\sim 50$ times lower than their quoted values.
The flaw in their analysis can not be explained as a simple error in
recording the value of a model normalization, as the
spectral model they describe fails to match the shape of the
observed spectrum. This pattern is typical of the use of an 
inconsistent set of spectral response files.
A further problem is that many of the physical properties 
quoted in \citet{cecil02}
also incorrectly state the dependence on the volume filling factor of the
hot gas $\eta_{\rm X}$.
} for a comparison to both the
HST images and the radio emission). Note that the 1.5 kpc-scale
radio lobes, the only large-scale-feature unambiguously associated with
the AGN, are offset from the X-ray/\halpha~bubble (see Fig.~3
in \citealt{cecil02}). The lack of any
clear correlation, or anti-correlation, between the bubble and the
radio lobes strongly suggests that they are unrelated and spatially distinct
phenomena, the partial overlap being due to projection.

On the larger scale, the diffuse X-ray emission is brightest
along a set of 10-kpc-scale arcs that define a roughly X-shaped
structure centered on the nucleus (see Fig~\ref{fig:images:n3079}h)
--- a structure visible in the lower signal-to-noise {\it ROSAT}
PSPC data \citep{rps97,dwh98,pietsch98}. Comparison with the
continuum-subtracted \halpha~image taken by 
\citet[-- see also Fig~\ref{fig:images:n3079}i]{ham90} 
 demonstrates that the X-ray arcs or
filaments correspond closely to the location of the multi-kpc-scale
X-shaped optical filaments discovered by \citeauthor{ham90}
The NGC 3079 observations were one of the shorter exposures in the
sample, and consequently the higher noise level is visible in the
hard X-ray images (Fig.~\ref{fig:images:n3079}j and k)

Diffuse soft X-ray emission almost completely fills the 
$20 \times 20$ kpc-scale boxes shown in Fig~\ref{fig:images:n3079}
panels g through l. The diffuse
X-ray emission is easily traced to $\sim 140\arcsec$ ($\sim 11.6$ kpc) above 
the disk (to the east) and $\sim 130\arcsec$ ($\sim10.8$ kpc) 
below the plane (to the west), with the tips of the arc-like X-features 
extending even further. The northern-most of the eastern arc
appears to extend out to $210\arcsec$ (17.4 kpc) above the plane.

Tracing the wind out to such large distances strengthens the case that
the peculiar \hi~distribution of the 
dwarf galaxy NGC 3073 ($10.0\arcmin$ away from NGC 3079, corresponding to a
minimum physical separation of $49.7$ kpc) is due to a wind
from NGC 3079 \citep{irwin87} ram-pressure stripping the ISM 
of this dwarf galaxy\footnote{\citet{irwin87} originally suggested
ram pressure stripping of NGC 3073 
when evidence for outflow was limited to the central
kiloparsec of NGC 3079.}. If this is occurring, there should be 
X-ray and \halpha~emission associated with a bow shock upstream 
from NGC 3073 (similar to the \citet{lhw99} 
model for the cloud found 11 kpc to the 
north of M82). We find no statistically-significant
level of X-ray emission associated with NGC 3073 itself. The upper limit 
on the soft X-ray emission within 
a 2\arcmin-diameter aperture centered on NGC 3073 
(this extends $\sim1\arcmin \sim5$ kpc further upstream than the \hi)
is $<1.3\times10^{-3}$ counts s$^{-1}$ ($3\sigma$, 0.3--2.0 keV energy band). 
Assuming a total
absorbing column equal to the foreground Galactic column
of $\nH = 2\times10^{20}\pcmsq$, emission from a hot plasma with
a temperature similar to what \citet{lhw99} 
found for the M82 northern cloud ($kT \approx 0.8$ keV), and using the
appropriate spectral responses, this corresponds to a flux limit
of $f_{\rm X} <6\times10^{-15} \ergps \pcmsq$, and $L_{\rm X} < 
2\times10^{38} \ergps$.  By way of comparison, 
the M82 northern cloud
has a soft X-ray luminosity of $2\times10^{38} \ergps$ and a
very similar \halpha~luminosity.
Were the M82 Northern cloud at a distance of 50 kpc from M82, we might expect
its X-ray and \halpha~luminosities to be a factor $(50/11)^{2} 
\sim 20$ times fainter, given that it would intercept a significantly
smaller fraction of the superwind's mechanical energy.
Deep optical~imaging 
may be a promising way of determining whether there really is a bow shock
with luminosity $L_{\rm H\alpha} \sim L_{\rm X} \sim 10^{37} \ergps$ 
upwind of NGC 3073 (to our knowledge this has not been attempted, despite
it being suggested by \citealt{irwin87}).

\begin{figure*}[!ht]
\epsscale{2.0}

\plotone{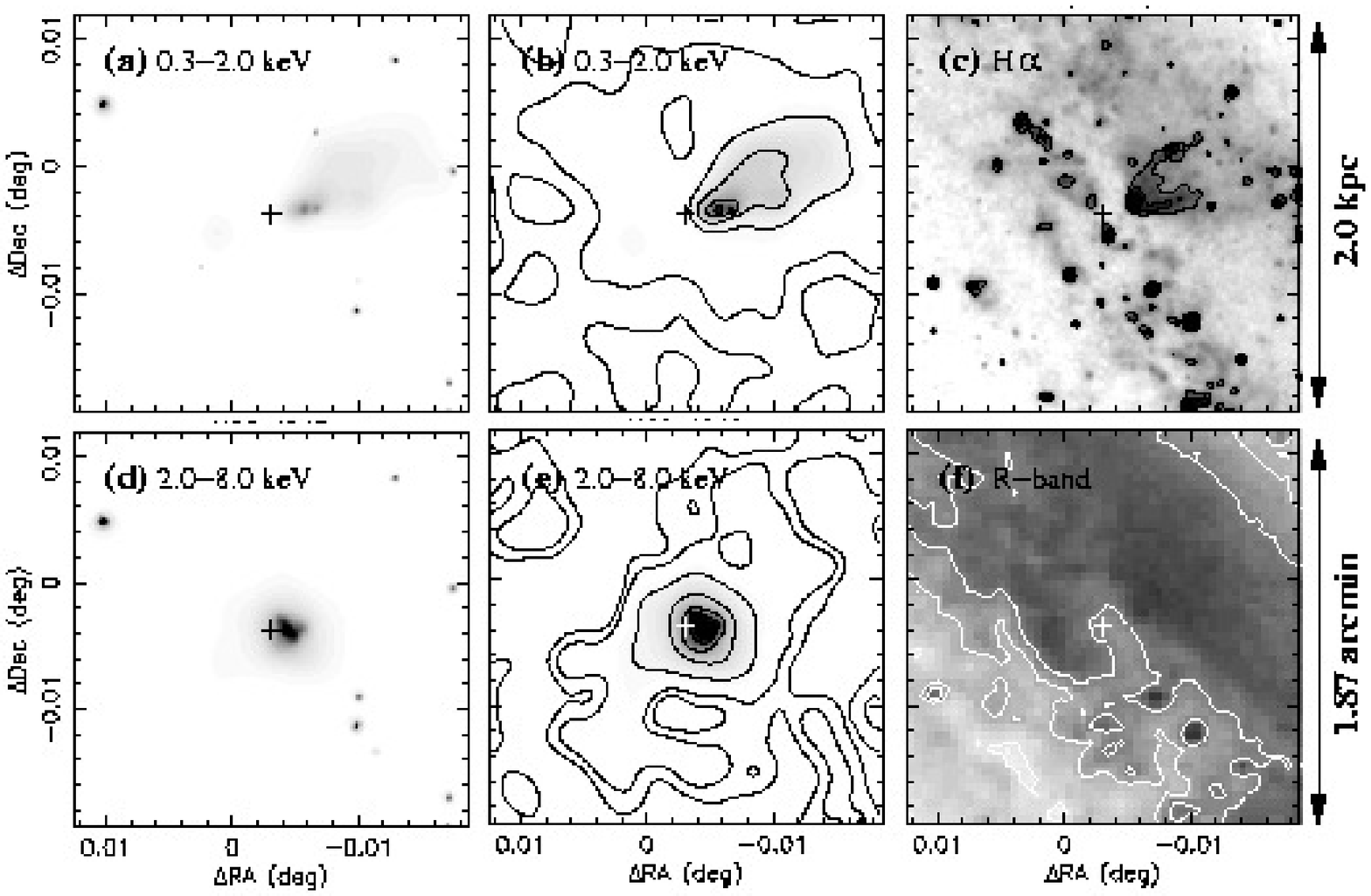}
\plotone{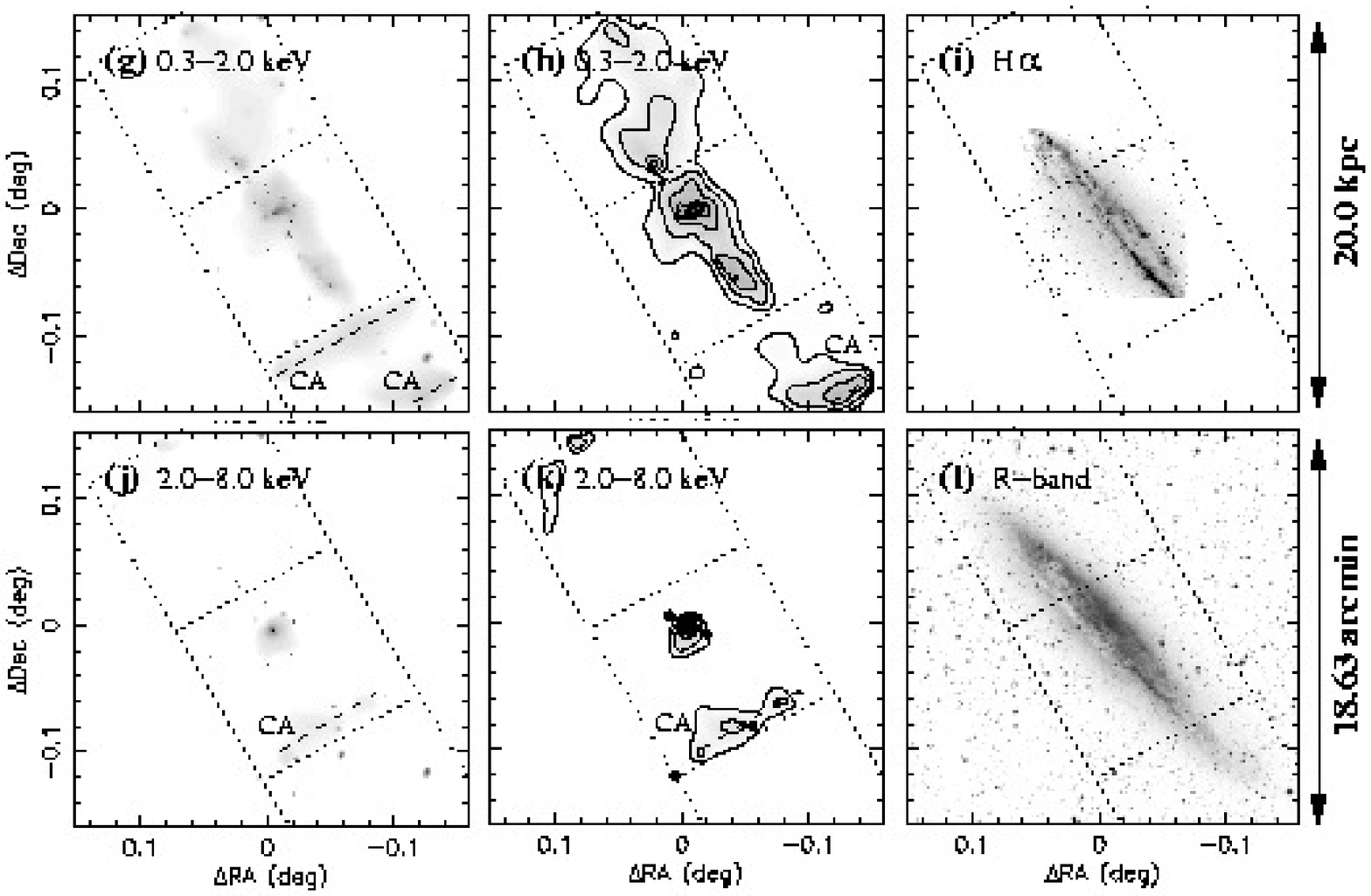}
\caption{{\it Chandra} X-ray and optical images of  NGC 4945.
	The meaning of each panel, and the intensity scales and 
	contour levels used, are the same as those described 
	in Fig.~\ref{fig:images:m82}.
	}
\label{fig:images:n4945}
\end{figure*}

\begin{figure*}
\epsscale{2.0}

\plotone{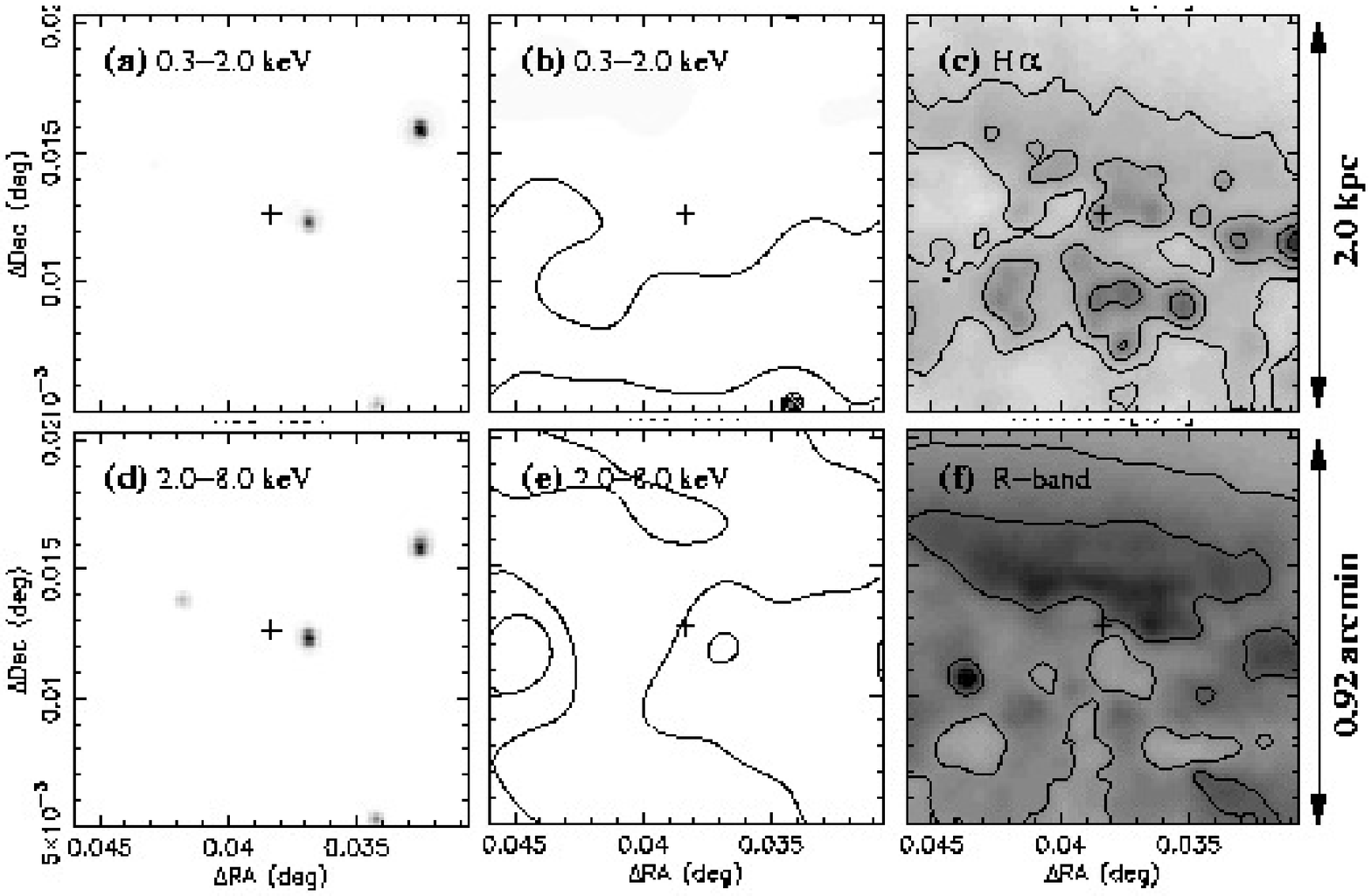}
\plotone{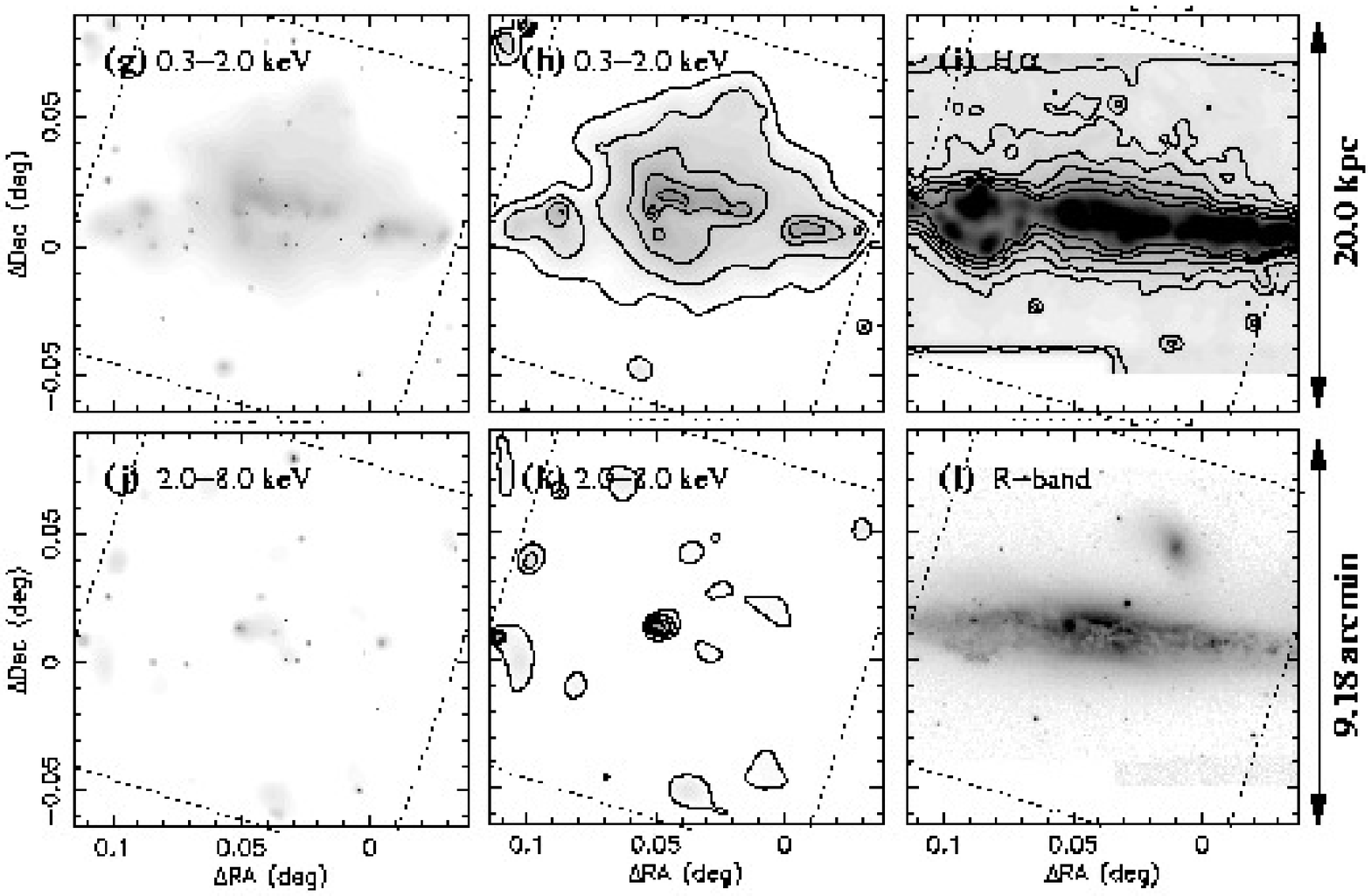}
\caption{{\it Chandra} X-ray and optical images of  NGC 4631.
	The meaning of each panel, and the intensity scales and 
	contour levels used, are the same as those described 
	in Fig.~\ref{fig:images:m82}.
	}
\label{fig:images:n4631}
\end{figure*}

\subsubsection{NGC 4945}
\label{sec:results:gal_notes:n4945}
NGC 4945 is one of the four brightest FIR sources in the sky outside
the local group, along with the starburst galaxies NGC 253, M82 and M83
\citep{rice88,soifer89}, and hosts both vigorous nuclear 
star-formation \citep{moorwood94,spoon00} 
and a peculiar AGN\footnote{
The obscured AGN, although often described as a Seyfert 2, is unusual
in several ways. NGC 4945 is the second brightest source in the sky at
E = 100 keV \citep*{done96}, yet appears completely obscured along our line
of sight at any energy less than $\sim 10$ keV 
(\citealt{marconi00,guianazzi00}, although \cf \citealt{krabbe01}).
Unless the nuclear AGN is completely obscured over 4$\pi$ steradian,
it can not have a typical Seyfert SED and must be strongly deficient
in UV photons given the low mid-IR-excitation and
lack of [\ion{O}{3}] in the outflow cone \citep{spoon00,marconi00}. 
\citet{madejski00} argue that total obscuration is
impossible if the absorber is larger in size than a light day or 
so, based on the X-ray variability. In addition, the models 
of \citet{levenson02} would predict a  iron K$\alpha$ equivalent 
width much higher than the value observed.
If the SED of the AGN is unusual, this in turn raises the uncertainty
associated with estimates of its bolometric luminosity 
based on extrapolations
from the X-ray flux \citep[\eg][]{guianazzi00,madejski00}.}
 with an estimated bolometric luminosity
of $\sim60$\% of the nuclear FIR luminosity \citep{madejski00,brock88}.

The star-formation is highly
 concentrated -- HST NICMOS Pa$\alpha$ observations
\citep{marconi00} reveal a 11\arcsec-diameter ($\sim 200$ pc)
 star-forming disk or ring, which corresponds closely in size
to the FWHM = 12\arcsec~FIR source that emits $\sim 50$\% of the 
100\micron~flux of the entire galaxy \citep{brock88}. \citet{moorwood94}
find Br$\gamma$~emission confined to the central 20\arcsec. This implies
that the current star-formation lies within the molecular gas torus
found by \citeauthor{bergman92} (1992, who argue that the 
CO emission comes from a ring of inner radius $\sim7\arcsec$ 
and outer radius $\sim16\arcsec$). Based on ISO-spectroscopy,
\citet{spoon00} argue that a starburst of age $\ge 5\times10^{6}$ years
is responsible for between 50 -- 100\% of the nuclear region's bolometric 
luminosity.

\citet{ham90} discovered a $\sim 500 \kmps$ conical outflow from the nucleus,
extending at least $\sim 600$ pc to the NW along the minor axis.
This is very similar to the starburst-driven outflow cone in NGC 253, 
and can {\em not} be a conventional AGN-lit ionization cone given the
lack of [\ion{O}{3}] emission \citep{moorwood96}. 
This \halpha~outflow cone (Fig.~\ref{fig:images:n4945}c) is matched by a
 co-spatial limb-brightened
X-ray outflow cone (Fig.~\ref{fig:images:n4945}b, 
see also \citealt{strickland_vulcano,schurch02}). Note that 
the local signal-to-noise maximization used in the adaptive smoothing of
the X-ray images tends to artificially reduce the apparent 
limb-brightening.

On larger scales, diffuse X-ray emission associated with the star-forming
disk of NGC 4945 is apparent in Fig.~\ref{fig:images:n4945}h. 
Indeed, the spatial extent of diffuse X-ray emission is larger along
the major axis of the galaxy than the
minor axis. There is little or no evidence for significant extra-planar
emission in other wave-bands. Optical filaments have only been traced
out to $\sim 2$ kpc from the plane \citep{nakai89}, and no radio emission
has been discovered \citep{elmouttie97}.

NGC 4945 lies at low Galactic latitude, and consequently, the foreground
\hi~column is high. The current lack of evidence for any larger, 10-kpc-scale,
extra-planar emission, \ie emission
that might be expected from a starburst-driven superwind, is possibly
due to the resulting high foreground optical depth of
$\tau \sim 1$ for \halpha~and soft X-ray radiation 
(see Table~\ref{tab:xray_bgprops}). The X-ray data from {\it Chandra}
is further biased against a detection of diffuse emission due to the
significantly higher than normal background experienced during this
observation.

\subsubsection{NGC 4631}
\label{sec:results:gal_notes:n4631}
NGC 4631 is often erroneously
presented as an example of a
 ``normal'' spiral galaxy with a X-ray-emitting halo 
(\eg see \citealt{wang2001}). However, its
large-scale-height non-thermal radio halo, warm IRAS 60 to $100\mu$m flux
ratio, high \halpha~luminosity
and optically disturbed morphology show  it to be a
highly atypical spiral galaxy. 
It is a good example of a galaxy experiencing a mild, disk-wide,
starburst event \citep{golla94b}.

In the optical and UV, NGC 4631 is a large, distorted, edge-on disk galaxy, 
with many \ion{H}{2} regions, dust filaments and young stellar clusters
spread throughout the disk \citep*{rand92,smith01}. Four $\sim35$ kpc-long
\hi~spurs \citep{rand94} point to strong tidal interactions with 
a dwarf elliptical companion, NGC 4627 (2\farcm5 to the NW, $\sim5$ kpc 
projected separation), and the edge-on spiral
galaxy NGC 4656 (32\farcm4 to the SE, projected separation $\sim71$ kpc).
This interaction is thought to be the cause for the enhanced star-formation
within the disk of NGC 4631. Molecular gas, and the recent star-formation,
is less-centrally concentrated in NGC 4631 than in the other starburst
galaxies of this sample \citep{golla94b,golla96,golla99,rand00}, 
and \citeauthor{golla94b}
describe the nuclear star-formation as an optically-obscured
``mild central starburst.''  
 NGC 4631 was one of the first spiral galaxies
discovered to have a non-thermal radio halo \citep{ekers77},
which has been traced out to $\sim 4\arcmin$ ($z_{\rm max} \sim9$ kpc)
above and below the plane of the galaxy \citep{golla94a}. 
The observed alignment of the halo magnetic field strongly
suggests a large-scale outflow from the central 5 kpc of the disk
\citep{golla94a}, 
and although kinematic evidence for outflow exists
\citep[\cf \citealt{martin01}]{golla96,rand00} 
it is less compelling than in the more
powerful starbursts discussed earlier.
A faint V-shaped pair of
$\sim 1.5$ kpc-long \halpha~filaments is seen directly 
above the optically-obscured nucleus,  reminiscent of the outflow cones
seen in NGC 253 and NGC 4945 
\citep[see also the HST WFPC2 images in \citealt{wang2001}]{rand92}.
The extra-planar \halpha~emission is somewhat less vertically
extended than the radio emission ($z_{\rm max} \sim 5$ kpc, 
\citealt{martin01}), and is less-well organized
into distinct multi-kpc filaments and arcs than it
is in the other starburst galaxies \citep{rand92,hoopes99}. 
Vertically extended diffuse X-ray emission was first robustly detected
by {\it ROSAT} PSPC observations (\eg \citealt{wang95,vogler96,rps97,dwh98}).
Results from the  {\it Chandra} ACIS-S observation of NGC 4631 were
published by \citet{wang2001}.

The lack of central concentration in the star-formation is 
dramatically illustrated by the weakness of both soft and hard
X-ray emission from the central
2 kpc (see Fig.~\ref{fig:images:n4631}a, b, d and e), 
both diffuse and point-like, in comparison to the starburst galaxies
previously considered. There is diffuse emission in the central
2 kpc, but at low surface brightness \citep[\eg][]{wang2001}.

There is no evidence of AGN-like activity in NGC 4631. 
The aligned triple radio source $\sim6\arcsec$ to the SW
of the dynamical center of the galaxy, identified by
\citet{golla99} as a low luminosity AGN candidate,
has no X-ray counterpart in the {\it Chandra} 
observations.

The diffuse X-ray emission is brightest, and has its greatest vertical extent,
above the central $\sim 4$ kpc of NGC 4631 
(Fig.~\ref{fig:images:n4631}g and h). This region corresponds
to the CO ring or bar \citep{rand00}, is also the location 
of the brightest radio emission \citep{golla94a,golla99}, 
and is bounded on the east by the most prominent giant \ion{H}{2} 
region visible within NGC 4631, CM67\footnote{\citet{golla96}
present long-slit \halpha~spectra with features consistent 
with outflow at velocities of several hundred km/s to the south of CM67.
\citet{rand00} presents evidence for outflow of molecular gas to
the north of CM67.} \citep{crillon69}. Diffuse X-ray emission definitely
extends at least $180\arcsec$ (6.5 kpc) to the north and 
$130\arcsec$ (4.7 kpc) to the south of the disk, and appears to extend
out to the edge of the S3 chip at $z\sim \pm8$ kpc at lower surface
brightness. 
Given the low diffuse X-ray flux, and its lack of concentration
into specific filaments or arcs, the S/N of the diffuse X-ray images
is low. This hinders a meaningful feature-by-feature comparison
with the \halpha~emission, beyond the general observation that the
diffuse X-ray emission is brightest above the regions of the disk 
with the most prominent extra-planar \halpha~emission.

Based on the older {\it ROSAT} PSPC observations 
of NGC 4631 it was already known
that in terms of overall spatial extent the non-thermal radio emission
\citep*[at either 1.4 GHz or 4.86 Ghz, see][]{golla94a,dahlem95}
resembles the soft X-ray emission, the radio emission
being somewhat more extended and/or
brighter to the north of the 
disk, and most vertically extended above the same region of the disk 
\citep[see the X-ray/radio overlays presented in][]{wang95,vogler96}. 
It is difficult to make a more detailed morphological comparison
than this using either the {\it ROSAT} data or the {\it Chandra} 
observations, in part due to the spatial filtering inherent in
radio imaging. \citet{vogler96} suggest, rather convincingly in our
opinion, that ``the north-eastern and south-eastern radio spurs seem
to be avoided by the X-ray emission'', although both the radio and X-ray
emission are faint in these regions.
Nevertheless, when comparing the brightness of the
soft X-ray emission to that of the nonthermal radio emission on
a location by location basis it is as easy to find regions of apparent
spatial correlation as to find regions with X-ray/radio anti-correlation 
(based on the images presented in \citealt{wang95,vogler96} or when 
overlaying the 1.4 GHz emission from the NVSS survey on the Chandra 
X-ray images), leading us to speculate that the halo
magnetic fields and cosmic rays may be somewhat decoupled from the hot
plasma.  Perhaps the only robust conclusion that can be drawn is that there
are magnetic fields and cosmic rays in the halo of NGC 4631, in addition to
hot plasma.

\begin{figure*}[!ht]
\epsscale{2.0}

\plotone{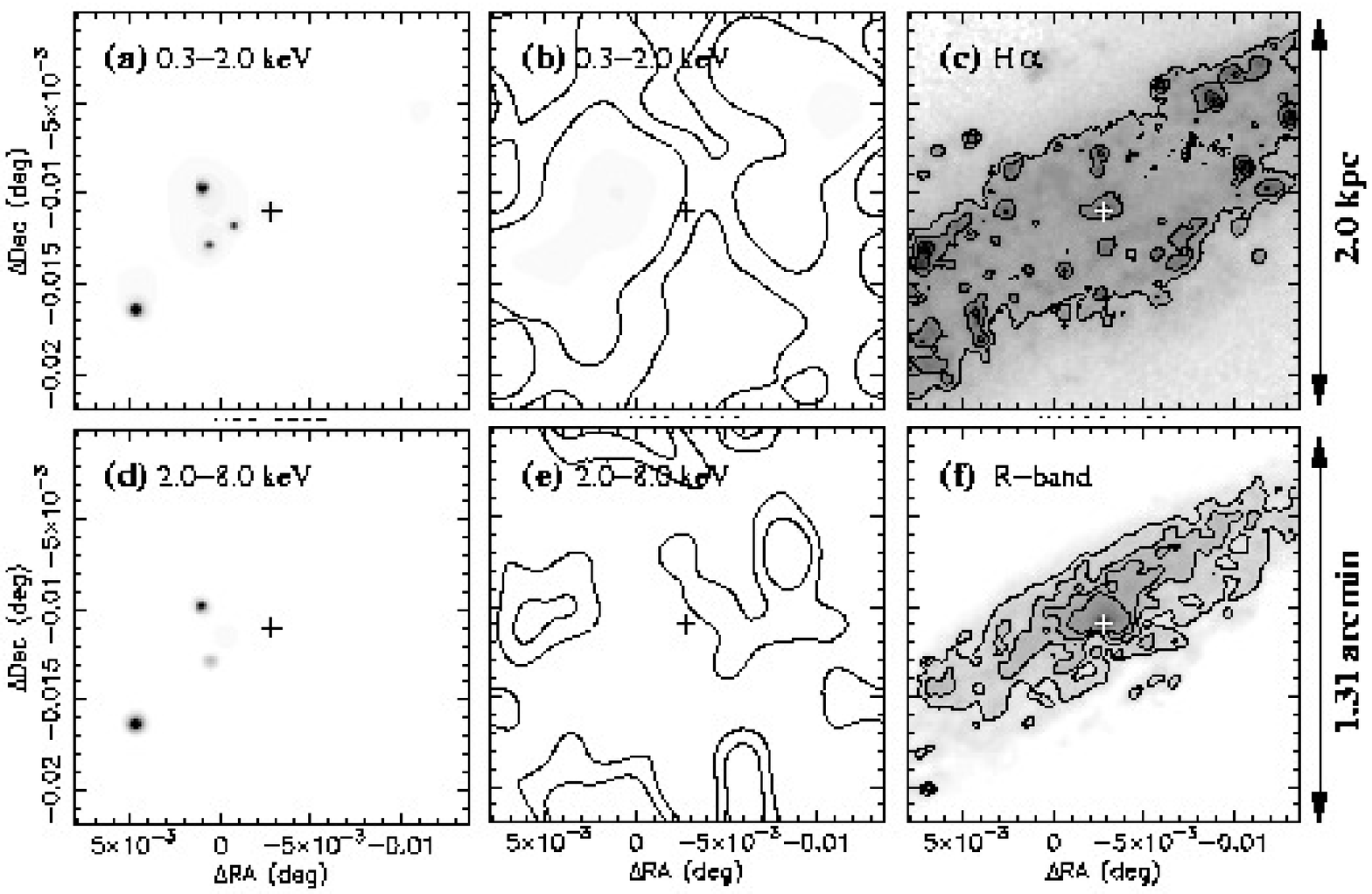}
\plotone{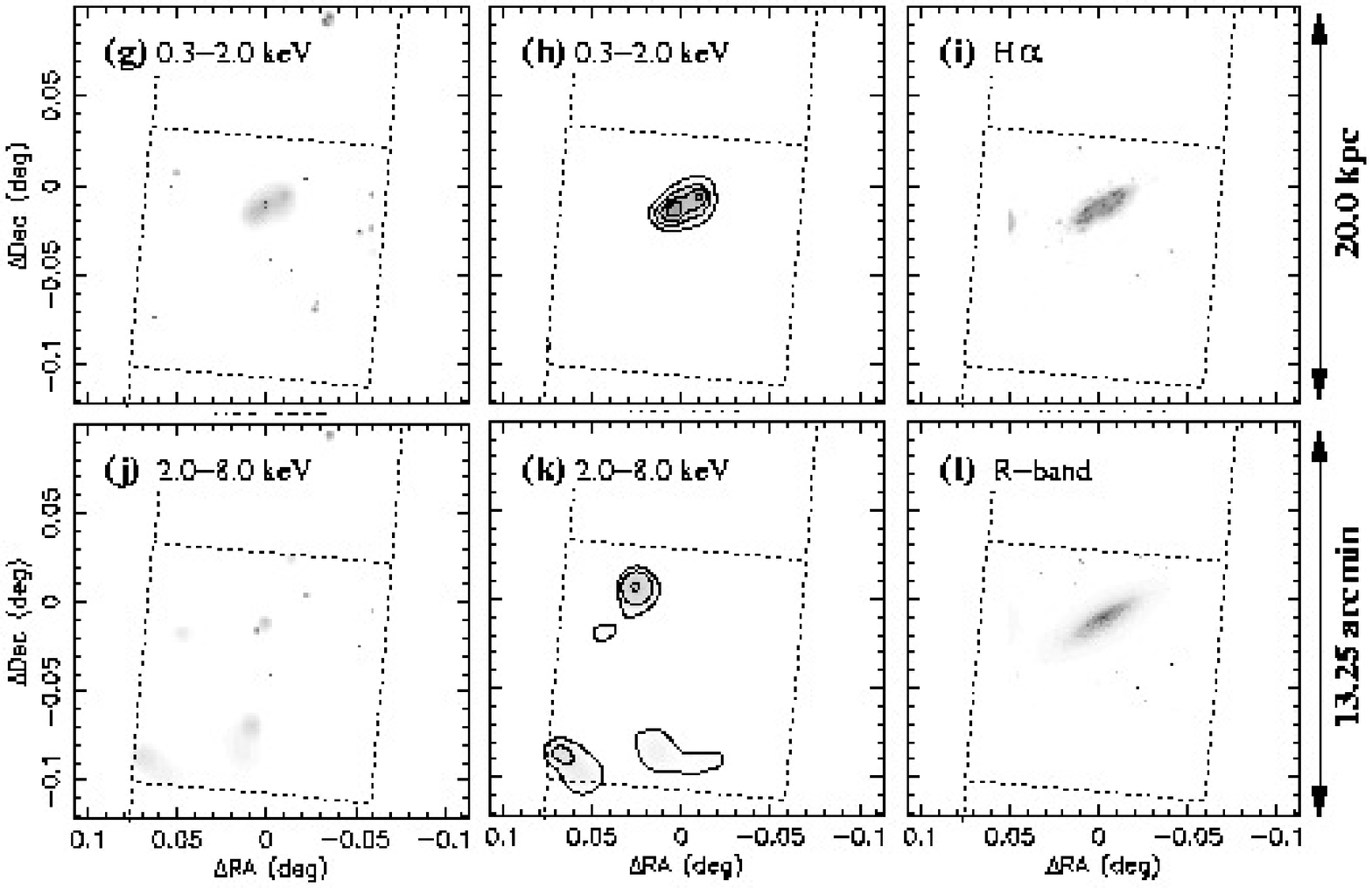}
\caption{{\it Chandra} X-ray and optical images of  NGC 6503.
	The meaning of each panel, and the intensity scales and 
	contour levels used, are the same as those described 
	in Fig.~\ref{fig:images:m82}.
	}
\label{fig:images:n6503}
\end{figure*}

\subsubsection{NGC 6503}
\label{sec:results:gal_notes:n6503}
NGC 6503 is a low mass Sc galaxy, similar in size, mass, blue luminosity
and infrared warmth to M33 \citep{devaucouleurs82,rice88}.
While otherwise an unremarkable galaxy, it has a compact, low velocity 
dispersion nucleus \citep{bottema97}, best classified as a 
borderline starburst-LINER \citep{lira02}. 

The moderate values of the IRAS $f_{60}/f_{100}$ flux ratio
and star-formation rate given in Table~\ref{tab:galaxies} imply that 
NGC 6503 as a whole should not be considered as a starburst galaxy.

Diffuse X-ray emission, first reported by \citet{lira02}, is weak,
and confined exclusively within the optical confines of the disk
of NGC 6503 (see Fig.~\ref{fig:images:n6503}). We can only place upper
limits on diffuse X-ray emission from the halo. The nucleus of the 
galaxy is marked by a very weak X-ray point source (see \citet{lira02} for 
more discussion of the point sources), and there is no obvious 
concentration of the diffuse X-ray emission toward the nucleus.

\begin{figure*}
\epsscale{2.0}

\plotone{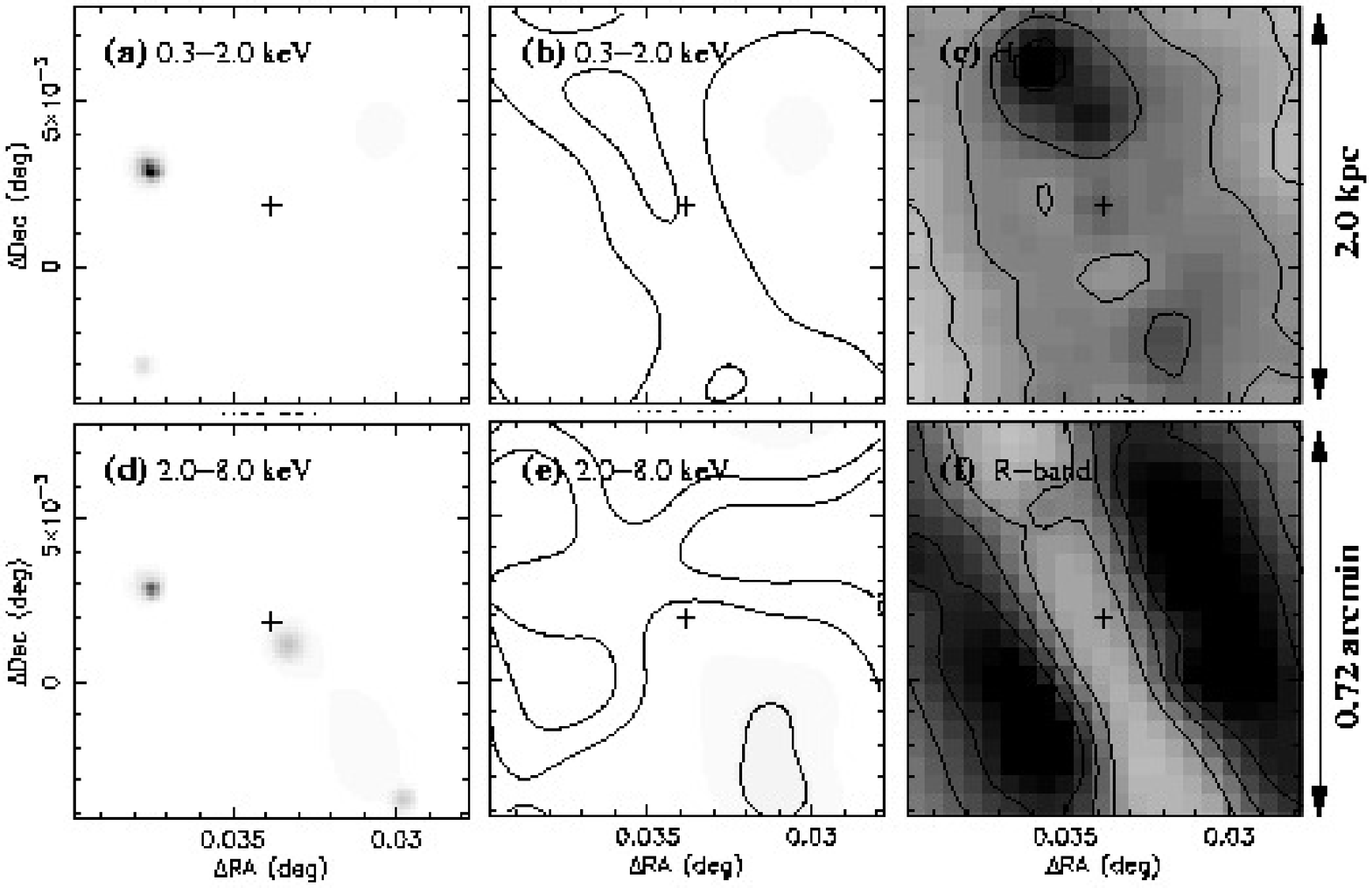}
\plotone{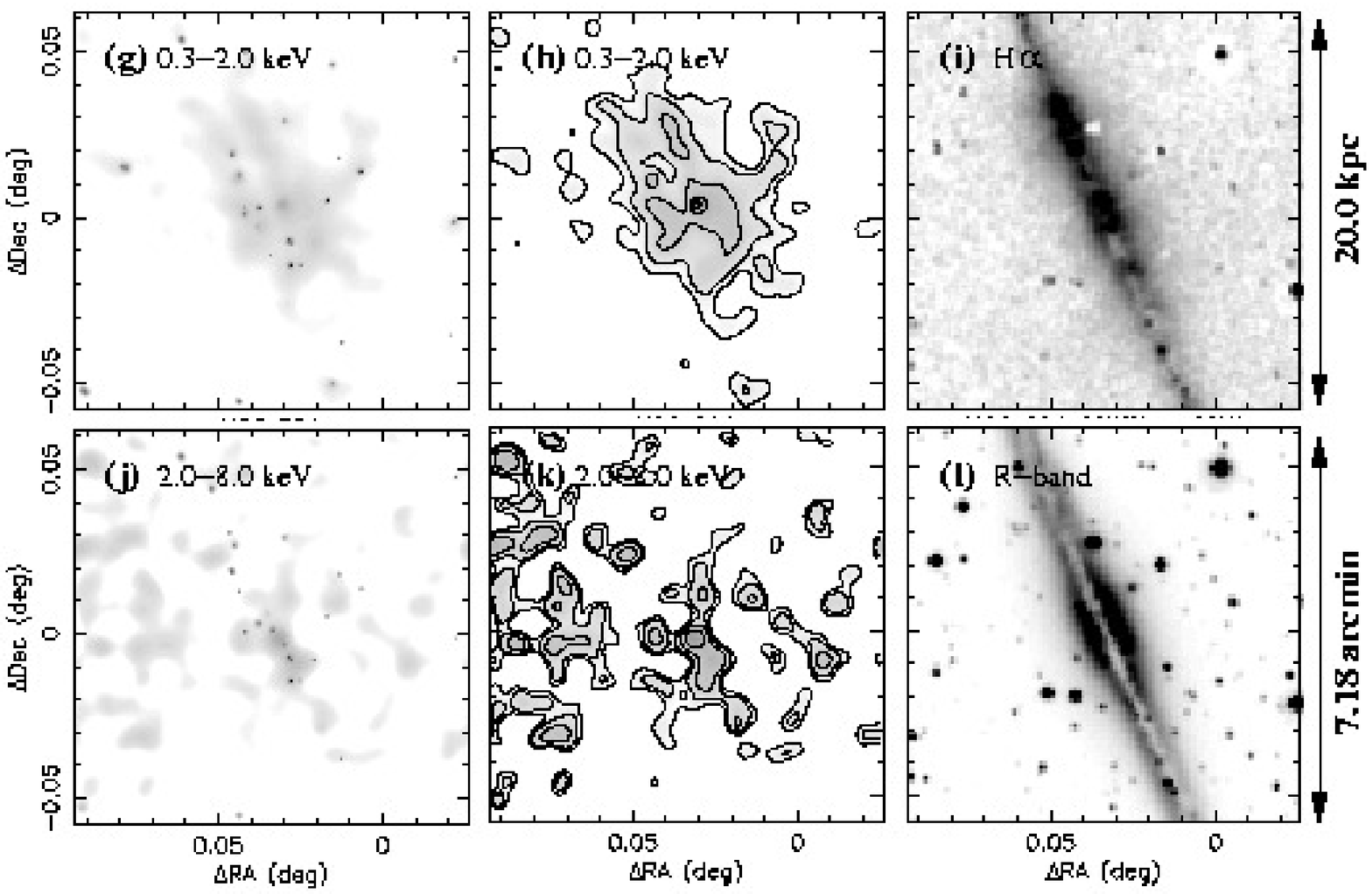}
\caption{{\it Chandra} X-ray and optical images of  NGC 891.
	The meaning of each panel, and the intensity scales and 
	contour levels used, are the same as those described 
	in Fig.~\ref{fig:images:m82}.
	}
\label{fig:images:n891}
\end{figure*}

\subsubsection{NGC 891}
\label{sec:results:gal_notes:n891}
NGC 891 is a spiral galaxy very similar in size, total luminosity, 
spiral-type and molecular gas spatial distribution
to the Milky Way, seen almost exactly edge-on \citep{kruit84,scoville93}.
Despite these similarities, {\em NGC 891 is forming stars 
more intensely than the Milky Way}, 
as it is twice as rich in molecular gas \citep{scoville93}, twice
as IR-luminous as the Milky Way \citep*{wainscoat87}, and several times more
radio-luminous at 1.4 GHz \citep*{allen78}. The IRAS $f_{60}/f_{100}$
ratio for NGC 891 is higher than that of the Milky Way (0.31 for NGC 891
vs. $\sim0.2$ for the Milky Way, see \citealt{rice88} and 
\citealt{boulanger88}).

There is no direct evidence for a weak AGN in NGC 891, although
the {\it Chandra} observations do show a weak hard X-ray source 
(see Fig.~\ref{fig:images:n891}a and d) $\sim3\arcsec$ to SW of the
position of the radio continuum point source we have adopted as
the nucleus \citep{rupen91}. Given the $\sim2\arcsec$ positional
uncertainty associated with the radio source, the X-ray source may
be associated with it.

Little is known about the star formation rate in the
nucleus of NGC 891 --- optically the region is highly obscured by the
prominent dust lane bisecting much of the disk. 
Diffuse X-ray emission within the central few kiloparsecs is faint,
with a surface brightness 1 to 1.5 orders of
magnitude fainter than in the starburst galaxies such as M82, NGC 1482
and NGC 3079. In the 0.3--1.0 keV energy band the presence of the
dust lane is obvious (although not immediately apparent 
in Fig.~\ref{fig:images:n891}b), and forms the basis of the X-ray
shadowing experiment performed by \citet{bregman_and_irwin}.

NGC 891 has a radio \citep{allen78}, 
\halpha~\citep{rand90,dettmar90,hoopes99}, \hi~\citep*{swaters97}
and X-ray halo, all of which extend to heights $\sim 5$ kpc above
the plane of the galaxy. Dense dust clouds and filaments have
been traced to heights of $|z| \sim 2$ kpc \citep{howk97,howk00}.
Diffuse X-ray emission in the halo of NGC 891 was discovered
in {\it ROSAT} PSPC observations \citep{bregman_and_pildis,bregman97},
and it is currently the only normal spiral galaxy around which extra-planar
diffuse X-ray emission has been detected.

Star-formation, as traced by 50\micron~IR emission \citep{wainscoat87},
non-thermal radio emission \citep{rupen87}, 
and \halpha~emission \citep[\eg][]{hoopes99},
is not uniformly distributed throughout the disk of NGC 891.
Star formation activity appears stronger along the major axis to the
north-east of the nucleus than along the major axis to the south-west of
nucleus.
The distribution of extra-planar \halpha~and soft X-ray
emission is similarly asymmetric with respect to the nucleus, emission being
strongest above the region of the disk to the north-east of
nucleus (Fig.~\ref{fig:images:n891}h and i). 
This strongly suggests that
the extra-planar \halpha~and soft X-ray is created by star-formation
processes within the disk, as opposed to some form of accretion
scenario \citep{toft02,sommerlarsen02}. 
Diffuse {\em halo} X-ray emission extends along the plane of the
galaxy $\sim150\arcsec$ to the NE, and $\sim 90\arcsec$ to the SW,
of the nucleus ($r\sim 7$ kpc and $r\sim4.2$ kpc respectively), and appears
to reach a height of $\sim 90\arcsec$ above and below the plane of
galaxy over most of this region.

As in the case of NGC 4631,
both the X-ray and \halpha~emission appear relatively smooth and
spatially uniform, although \halpha~observations do show a variety
of vertical filaments reaching heights of $\sim1.5$ kpc above the 
plane \citep[\eg][]{howk00}.

Note also that the diffuse X-ray emission is particularly
intense, and has its greatest vertical extent,
above (to the NW of) the nucleus (Fig.~\ref{fig:images:n891}h),
extending  $\sim140\arcsec$ ($z\sim6.5$ kpc) from the nucleus.
This suggests significant nuclear star formation is
occurring in this region (perhaps associated with
the $\sim15\arcsec$-radius [$\sim 700$ pc] bar seen in CO emission, see
\citealt{garciaburillo95}).

\begin{figure*}
\epsscale{2.0}

\plotone{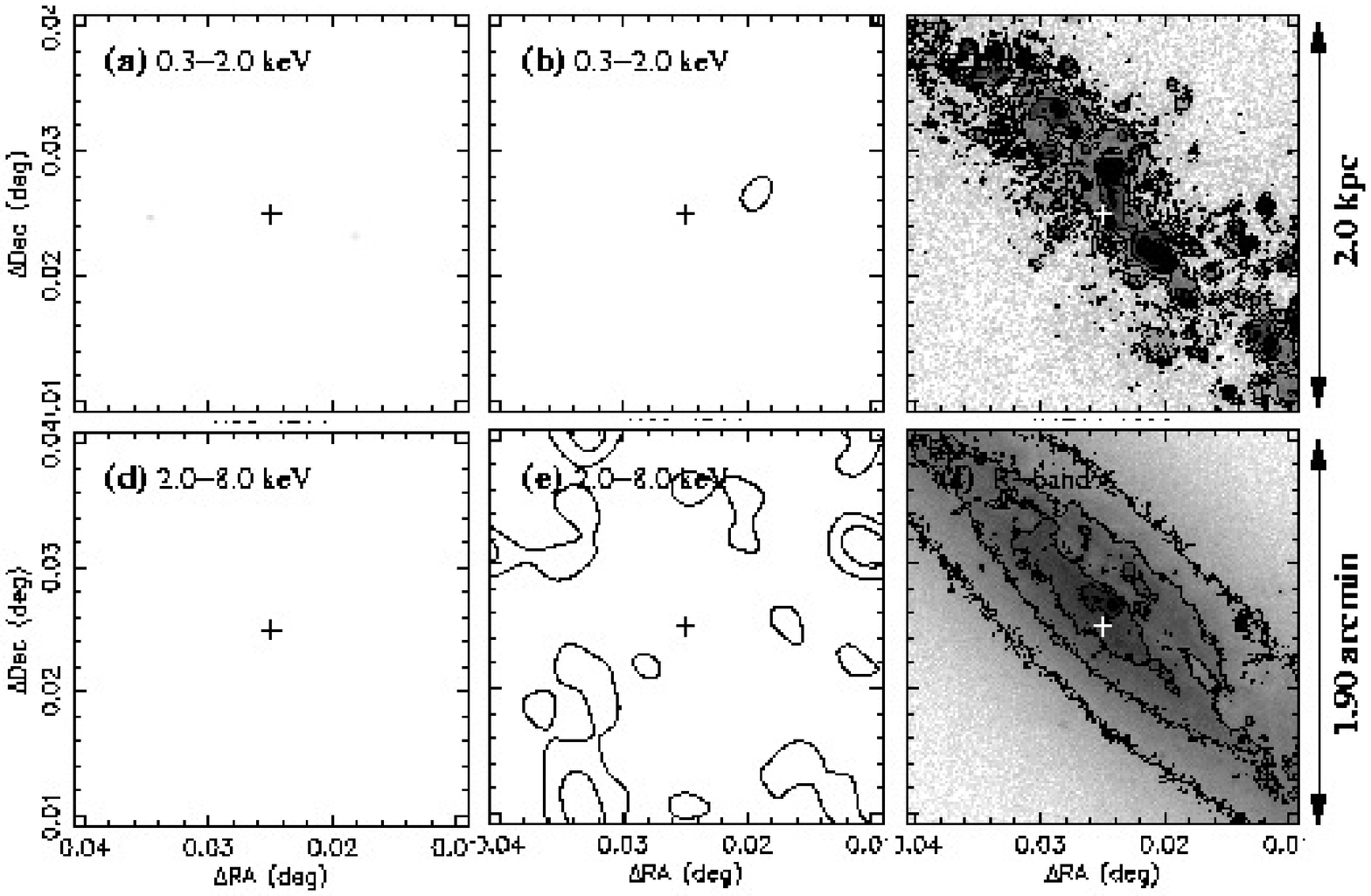}
\plotone{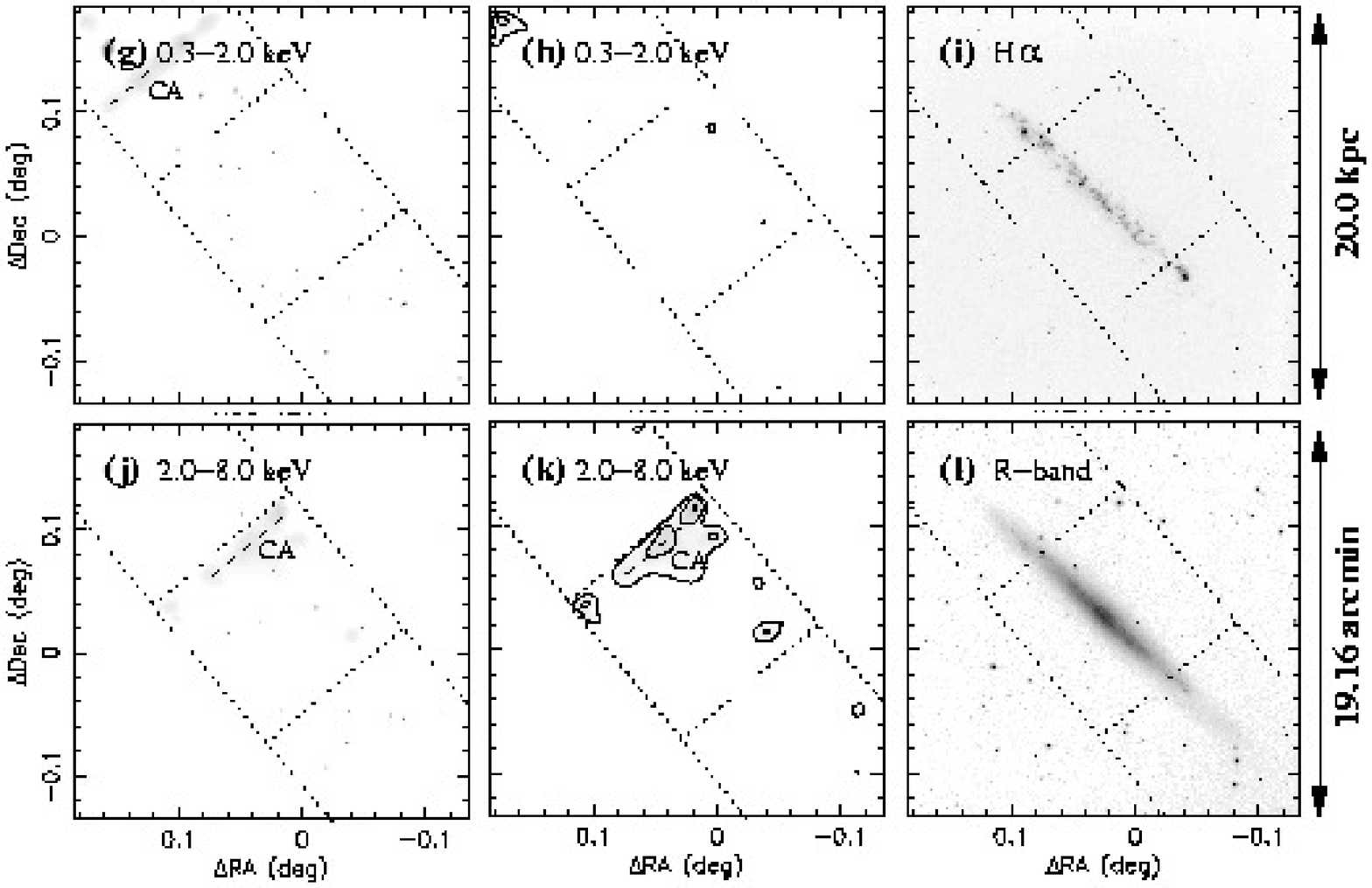}
\caption{{\it Chandra} X-ray and optical images of NGC 4244.
	The meaning of each panel, and the intensity scales and 
	contour levels used, are the same as those described 
	in Fig.~\ref{fig:images:m82}.
	}
\label{fig:images:n4244}
\end{figure*}

\subsubsection{NGC 4244}
\label{sec:results:gal_notes:n4244}
NGC 4244 is a nearby, low-mass, 
Scd galaxy ($M_{\rm tot} \approx 10^{10} \Msol$: \citealt{fry99})
at a distance of 3.6 Mpc. 
The current star formation rate is low, based on the low
IR \citep{rice88} and \halpha~luminosities \citep{hoopes99}.
No evidence for a radio, UV or X-ray-emitting halo has
been found \citep[this paper]{bregman82,schlickeiser84,deharveng86}. 
Deep \halpha~imaging \citep{hoopes99} demonstrates
that warm ionized gas at $T \sim 10^{4}$ K is 
confined to within $40\arcsec$ (700 pc) of the disk of the galaxy.
No nuclear activity is known in NGC 4244. 

Despite the relatively long {\it Chandra} observation,
there is a remarkable lack of X-ray emission from NGC 4244, with 
little or no concentration of X-ray point sources within the 
optical confines of the galaxy (see Fig.~\ref{fig:images:n4244}). 
There is only a weak, $\sim 3\sigma$,
detection of residual emission within the disk after point source
removal, and only an upper limit of $f_{\rm X} < 1.6\times10^{-14} \ergps
\pcmsq$ ($3\sigma$ upper limits in the 0.3--2.0 keV energy band, 
$L_{\rm X} < 2.5 \times 10^{37} \ergps$) 
could be obtained on diffuse emission from the halo.

\subsection{Diffuse X-ray surface brightness and spectral hardness
distributions}
\label{sec:results:extent}

Quantitative measures of the spatial
extent of the diffuse X-ray emission that can be used for comparative
purposes are rare to non-existent 
within the extensive {\it ROSAT}-based literature. 
Typically individual papers, even those of samples of star-forming
galaxies, only quote the size at which the diffuse emission disappears into
the noise, \ie observation-quality-dependent,
rather than a size characteristic of the emitting galaxy.

To rectify this problem, we extracted one-dimensional X-ray surface 
brightness profiles, along both the major 
and minor axis, from the
background-and-point-source-subtracted, exposure-map-corrected,
diffuse X-ray images. 

\begin{figure*}
\epsscale{2.0}
\plotone{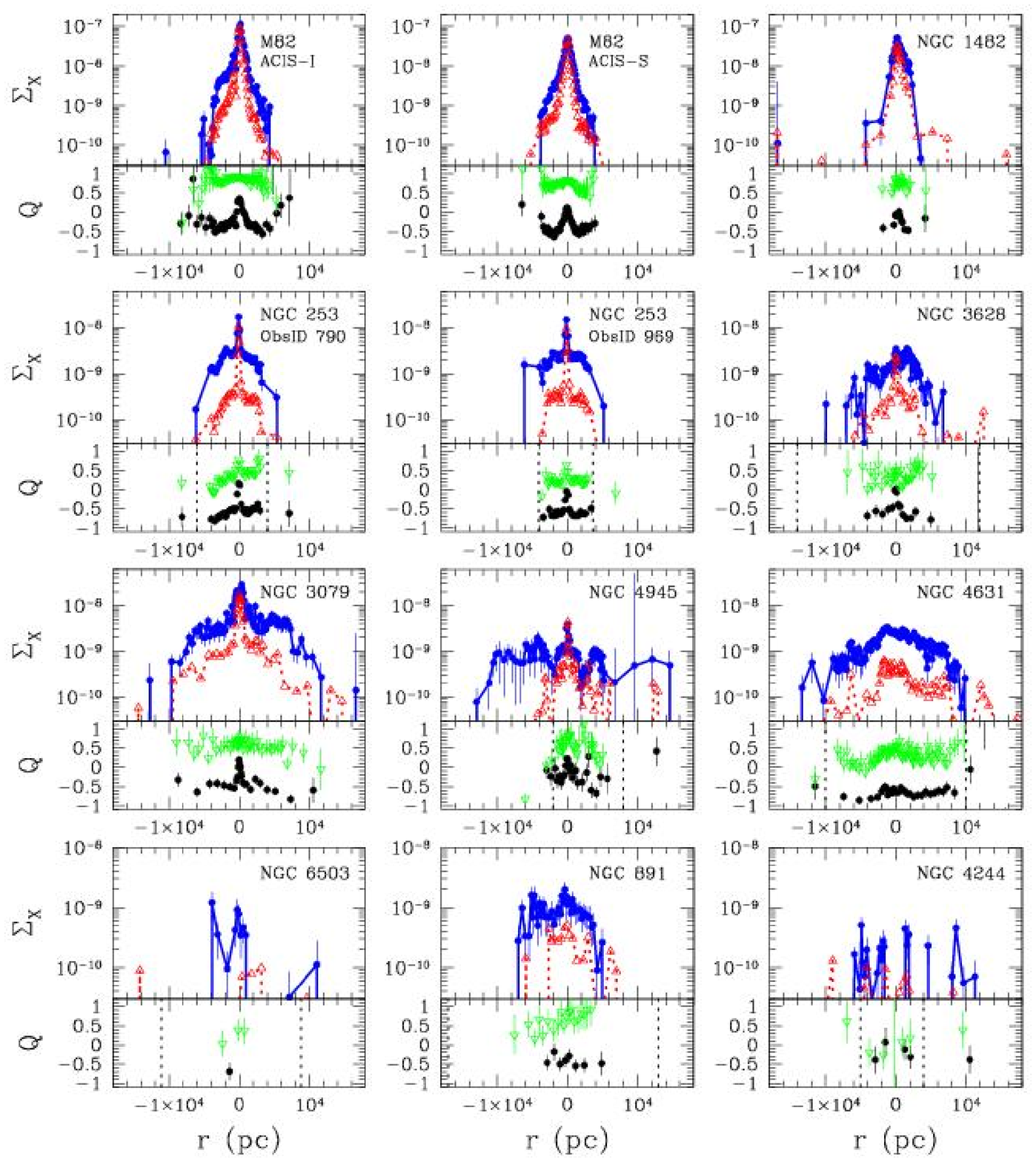}
\caption{Major axis diffuse X-ray surface brightness and spectral hardness
	profiles. The soft band (0.3-1.0 keV) and medium 
	band (1.0--2.0 keV)  surface 
	brightness $\Sigma_{\rm X}$ (photons s$^{-1}$ 
	cm$^{-2}$ arcsec$^{-2}$) are shown as blue filled circles and
	red open triangles respectively. The spectral hardness ratios
	$Q_{\rm A}$ and $Q_{\rm B}$ (defined in \S~\ref{sec:results:extent})
	for each galaxy are shown in the panel below the surface 
	brightnesses, as green 
	inverted open triangles and filled black circles respectively. 
	Error bars show 68\% confidence regions, 
	based on the \protect\citet{gehrels86}
	approximation to Poisson statistics (error bars are not shown
	for surface brightness values in the 1.0--2.0 keV energy band).
	Vertical dashed lines show the approximate location
	of the edges of the back-illuminated S3 chip. Negative $r$ is defined
	as the vector running from the nucleus along 
	the major axis that has a 
	position angle in the range $0\le$ PA (deg) $<180$.
	Data was accumulated within $|z| \le 5$ kpc of the
	midplane. Note that the axis scale on the ordinate is constant on
	a given row of figures, but varies between different rows.	
	}
\label{fig:major_axis_profiles}
\end{figure*}

The major axis surface brightness profiles 
(Fig.~\ref{fig:major_axis_profiles}) were extracted in slices
10 kpc high (\ie extending from 5 kpc above and below the plane of the
galaxy along the minor axis). The major-axis
width of each bin was chosen to achieve $\ge 20$ diffuse counts 
per slice. Errors were calculated taking into account both the Poisson
uncertainties in the diffuse emission and the background. We also use 
the more conservative \citet{gehrels86} approximation to Poisson statistics,
rather than using $\surd N$ errors.
Regions of sky exterior to any of the CCD chips were removed to avoid
biasing the surface brightness calculations.

\begin{figure*}
\epsscale{2.0}
\plotone{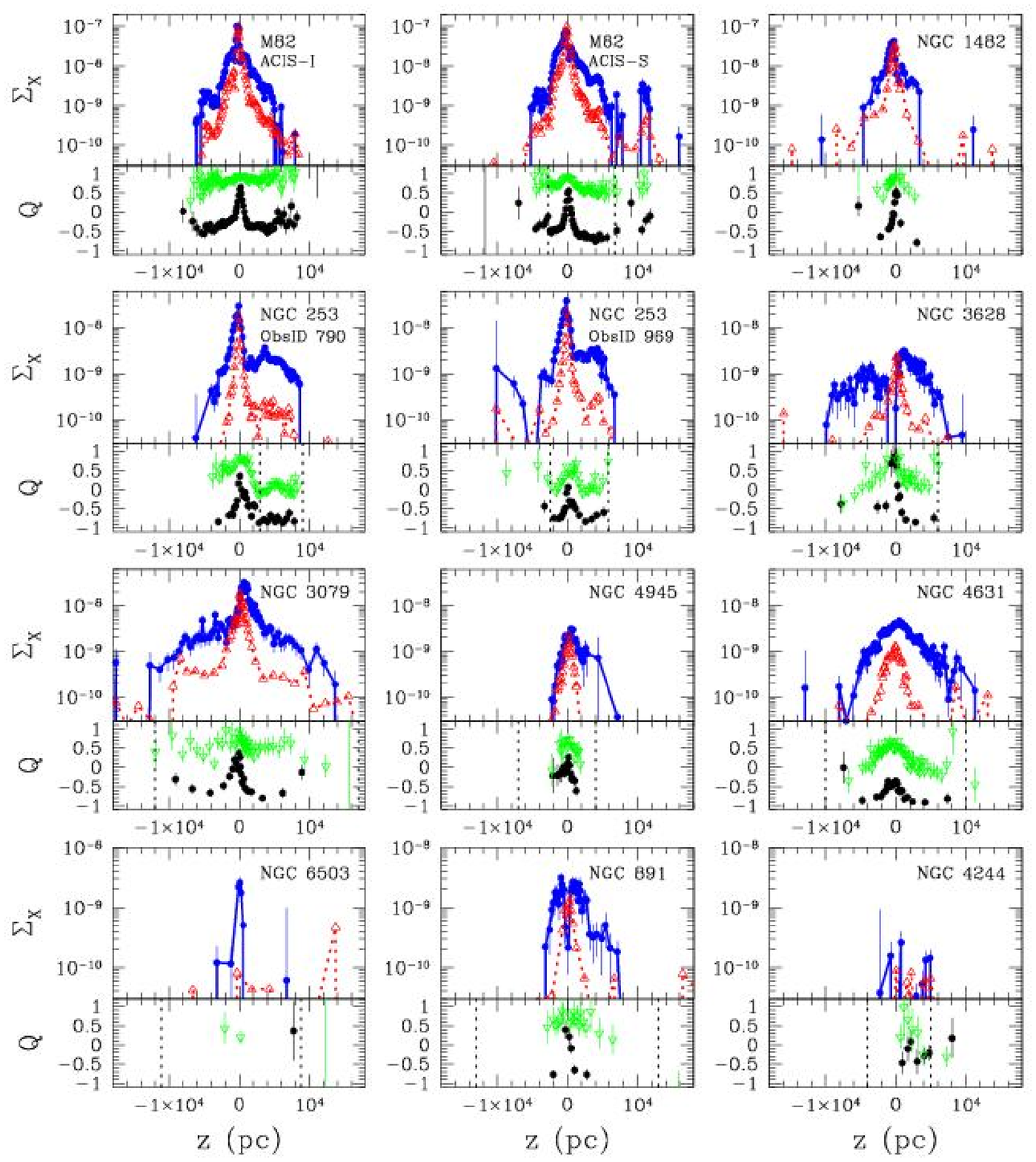}
\caption{Minor axis diffuse X-ray surface brightness and spectral
	hardness profiles. The meaning of the symbols is the same
	as that used in Fig.~\ref{fig:major_axis_profiles}.
	Positive $z$ is defined
	as the vector running from the nucleus along 
	the minor axis that has a position angle 
	in the range $0\le$ PA (deg) $<90$
	or $270\le$ PA (deg) $<360$. Data was accumulated 
	within $|r| \le 5$ kpc of the
	minor axis.  Note that the axis scale on the ordinate is constant on
	a given row of figures, but varies between different rows.}
\label{fig:minor_axis_profiles}
\end{figure*}

The minor axis profiles (Fig.~\ref{fig:minor_axis_profiles})
were created using the same method, with slices of total width
10 kpc parallel to the major axis (\ie within 5 kpc of the minor
axis of the galaxy).

Profiles were extracted in two energy bands - a soft 0.3--1.0 keV band
and a medium 1.0--2.0 keV energy band. The medium band contains
fewer counts than the soft energy band, but is significantly less affected by
absorption. 
The effect of absorption of the X-ray emission by the gas
within $\sim 1$ kpc of the planes of these galaxies 
is particularly noticeable in the minor
axis surface brightness profiles of NGC 3628 and NGC 891, 
where the surface brightness
in the 0.3--1.0 keV band drops to zero.

Two further sets of profiles, in 0.3--0.6 keV and 0.6--1.0 keV
energy bands were also extracted for the purposes of mapping the variation in
X-ray spectral hardness along the major and minor axes. If the surface
brightness, or alternatively the count rates, in these four bands are
$S_{0.3-1.0}, S_{1.0-2.0}, S_{0.3-0.6}$ and $S_{0.6-1.0}$ 
respectively, then we can define the following normalized hardness ratios:
\begin{displaymath}
Q_{A} = \frac{S_{0.6-1.0}-S_{0.3-0.6}}{S_{0.3-0.6}+S_{0.6-1.0}}, \, {\rm and}
\end{displaymath}
\begin{equation}
 Q_{B} =  \frac{S_{1.0-2.0}-S_{0.3-1.0}}{S_{0.3-1.0}+S_{1.0-2.0}}.
\label{equ:qa_qb}
\end{equation}
For our present purposes, it should be noted that $Q_{A}$ is 
sensitive to gas temperature, absorption and metal abundance
(increased values of which lead to higher values of $Q_{A}$),
while $Q_{B}$ is most sensitive to the temperature 
of the X-ray emitting plasma, and then to absorption.
See Fig.~7 in \citet{strickland02}.
To enhance the signal-to-noise of the hardness profiles, 
we re-binned the surface
brightness profiles to achieve $\ge 40$ counts in the harder 
of each of the two
bands used (this is often, but not always, the band with the 
fewest counts). Finally, points
with hardness ratio uncertainties $\sigma_{Q} \ge 1$ are not plotted
in Figs.~\ref{fig:major_axis_profiles} and \ref{fig:minor_axis_profiles}, 
given that the poor constraint on their 
spectral hardness makes them meaningless. 

We explored three independent methods to quantify the spatial extent of the
diffuse emission along both the minor and major axes: 
sizes based on enclosing a fixed fraction of the total diffuse
emission, sizes at fixed surface brightness levels 
(see Tables~\ref{tab:enclosed_flux_fractions:minor} 
and \ref{tab:enclosed_flux_fractions:major}), and parameterized
exponential, Gaussian and power-law model fits to the data 
(Table~\ref{tab:exponential_models}).

The galaxies within this sample are among those few edge-on galaxies
that are X-ray bright enough, and close enough,
to allow robust surface brightness profile fitting.
In contrast, diffuse emission sizes based 
on a fixed flux fraction, or a fixed isophotal size,
are more easily computed, even for distant and/or faint galaxies.
We present the results from all three methods to allow both a 
cross-comparison between the various methods, and to allow for a later direct
comparison of these galaxies against larger samples of more distant
spiral and starburst galaxies.

\begin{deluxetable}{lrcrrrrrrrrr}
 \tabletypesize{\scriptsize}%
\tablecolumns{12} 
\tablewidth{0pc} 
\tablecaption{Minor-axis diffuse emission spatial extent at fixed enclosed flux fractions
	\label{tab:enclosed_flux_fractions:minor}} 
\tablehead{ 
\colhead{Galaxy} & \colhead{PA} & \colhead{Energy}  
	& \colhead{$z_{0.5}$}  
	& \colhead{$z_{0.75}$}  
	& \colhead{$z_{0.9}$}  
	& \colhead{$z_{0.95}$}  
	& \colhead{$\Sigma_{0.5}$}  & \colhead{$\Sigma_{0.75}$}  
	& \colhead{$\Sigma_{0.9}$}  & \colhead{$\Sigma_{0.95}$} 
	& \colhead{$z_{1e-9}$} \\
\colhead{(1)} & \colhead{(2)} & \colhead{(3)}
	&  \colhead{(4)} & \colhead{(5)} &  \colhead{(6)} &  \colhead{(7)} 
	& \colhead{(8)} &  \colhead{(9)} &  \colhead{(10)} &  \colhead{(11)}
	& \colhead{(12)}  
	}
\startdata
M82$^{a}$ & 335 & 0.3-1.0 &   0.990 & 2.23 & 3.85 & 4.60 &   1.83e-08 & 1.31e-08 & 8.86e-09 & 7.78e-09 & $ 4.95$ \\
M82$^{a}$ & 155 & 0.3-1.0 &   0.555 & 1.36 & 2.19 & 3.53 &   7.20e-08 & 3.68e-08 & 2.55e-08 & 1.59e-08 & $ 5.48$ \\

M82$^{b}$ & 335 & 0.3-1.0 &   1.11 & 2.42 & 4.08 & 4.73 &   2.19e-08 & 1.60e-08 & 1.16e-08 & 1.06e-08 & $ 11.7$ \\
M82$^{b}$ & 155 & 0.3-1.0 &   0.522 & 1.12 & 1.66 & 1.97 &   6.58e-08 & 4.95e-08 & 4.11e-08 & 3.69e-08 & $ 4.81$ \\

NGC 1482 & 13 & 0.3-1.0 &   1.05 & 1.90 & 2.48 & 2.70 &   5.92e-09 & 5.13e-09 & 4.92e-09 & 4.81e-09 & $ 4.13$ \\
NGC 1482 & 193 & 0.3-1.0 &   0.587 & 0.984 & 1.64 & 2.05 &   3.28e-08 & 3.06e-08 & 2.22e-08 & 1.89e-08 & $ 4.22$ \\

NGC 253$^{c}$ & 319 & 0.3-1.0 &   0.601 & 2.54 & 3.64 & 4.08 &   1.15e-08 & 4.55e-09 & 3.95e-09 & 3.75e-09 & $ 5.50$ \\
NGC 253$^{c}$ & 139 & 0.3-1.0 &   0.302 & 0.658 & 1.08 & 1.39 &   3.02e-08 & 2.44e-08 & 1.89e-08 & 1.59e-08 & $ 2.03$ \\

NGC 253$^{d}$ & 319 & 0.3-1.0 &   2.63 & 4.16 & 5.45 & 5.92 &   3.22e-09 & 2.71e-09 & 2.37e-09 & 2.27e-09 & $ 6.42$ \\
NGC 253$^{d}$ & 139 & 0.3-1.0 &   0.265 & 0.632 & 1.21 & 1.78 &   2.47e-08 & 1.96e-08 & 1.37e-08 & 1.01e-08 & $ 2.60$ \\

NGC 3628 & 14 & 0.3-1.0 &   1.76 & 2.92 & 3.87 & 4.27 &   2.36e-09 & 2.06e-09 & 1.85e-09 & 1.75e-09 & $ 4.05$ \\
NGC 3628 & 194 & 0.3-1.0 &   3.48 & 4.58 & 5.40 & 6.03 &   4.66e-10 & 5.54e-10 & 5.89e-10 & 5.54e-10 & $ 4.73$ \\

NGC 3079 & 75 & 0.3-1.0 &   1.15 & 2.66 & 5.69 & 7.37 &   2.25e-08 & 1.55e-08 & 8.88e-09 & 7.23e-09 & $ 11.3$ \\
NGC 3079 & 255 & 0.3-1.0 &   3.26 & 5.89 & 8.22 & 9.48 &   3.74e-09 & 3.16e-09 & 2.71e-09 & 2.44e-09 & $ 8.64$ \\

NGC 4945 & 313 & 0.3-1.0 &   0.623 & 1.08 & 1.54 & 1.84 &   2.45e-09 & 2.31e-09 & 2.03e-09 & 1.84e-09 & $ 2.72$ \\
NGC 4945 & 133 & 0.3-1.0 &   0.313 & 0.800 & 1.20 & 1.36 &   1.91e-09 & 1.36e-09 & 1.12e-09 & 1.03e-09 & $ 0.553$ \\

NGC 4631 & 356 & 0.3-1.0 &   1.61 & 3.42 & 5.42 & 6.65 &   3.74e-09 & 2.81e-09 & 2.18e-09 & 1.87e-09 & $ 7.02$ \\
NGC 4631 & 176 & 0.3-1.0 &   1.37 & 2.39 & 3.66 & 4.42 &   3.14e-09 & 2.77e-09 & 2.22e-09 & 1.97e-09 & $ 3.81$ \\

NGC 6503 & \nodata & 0.3-1.0 
	&   (1.0) & \nodata & (4.0) & \nodata 
	&   5.61e-9 & \nodata & 5.63e-10 & \nodata & \nodata \\

NGC 891 & 293 & 0.3-1.0 &   1.56 & 2.52 & 3.74 & 4.59 &   1.47e-09 & 1.46e-09 & 1.15e-09 & 9.94e-10 & $ 2.88$ \\
NGC 891 & 113 & 0.3-1.0 &   0.717 & 0.884 & 1.04 & 1.13 &   1.31e-09 & 1.65e-09 & 1.65e-09 & 1.61e-09 & $ 2.13$ \\
NGC 891 & \nodata & 0.3-1.0 
	&   (1.0) & \nodata & (4.0) & \nodata 
	&   3.37e-9 & \nodata & 2.34e-9 & \nodata & \nodata \\

NGC 4244 & \nodata & 0.3-1.0 
	&   (1.0) & \nodata & (4.0) & \nodata 
	&   6.97e-11 & \nodata & 5.61e-11 & \nodata & \nodata \\

\tableline

M82$^{a}$ & 335 & 1.0-2.0 &   0.256 & 0.631 & 1.91 & 3.03 &   3.60e-08 & 2.34e-08 & 1.07e-08 & 6.90e-09 & $ 2.60$ \\
M82$^{a}$ & 155 & 1.0-2.0 &   0.160 & 0.555 & 1.40 & 1.86 &   8.68e-08 & 5.55e-08 & 2.08e-08 & 1.55e-08 & $ 2.13$ \\

M82$^{b}$ & 335 & 1.0-2.0 &   0.274 & 0.645 & 1.84 & 3.35 &   4.94e-08 & 3.50e-08 & 1.56e-08 & 9.16e-09 & $ 2.73$ \\
M82$^{b}$ & 155 & 1.0-2.0 &   0.209 & 0.582 & 1.30 & 1.71 &   7.99e-08 & 5.21e-08 & 2.95e-08 & 2.39e-08 & $ 2.27$ \\

NGC 1482 & 13 & 1.0-2.0 &   0.182 & 0.649 & 1.26 & 1.59 &   1.43e-08 & 6.26e-09 & 4.09e-09 & 2.81e-09 & $ 1.48$ \\
NGC 1482 & 193 & 1.0-2.0 &   0.339 & 0.714 & 1.75 & 3.52 &   2.94e-08 & 2.26e-08 & 9.83e-09 & 5.31e-09 & $ 1.77$ \\

NGC 253$^{c}$ & 319 & 1.0-2.0 &   0.155 & 0.602 & 2.50 & 3.78 &   1.15e-08 & 5.90e-09 & 1.76e-09 & 1.32e-09 & $ 0.814$ \\
NGC 253$^{c}$ & 139 & 1.0-2.0 &   0.0642 & 0.291 & 0.673 & 1.02 &   2.25e-08 & 1.41e-08 & 8.44e-09 & 6.13e-09 & $ 0.953$ \\

NGC 253$^{d}$ & 319 & 1.0-2.0 &   0.119 & 0.575 & 3.56 & 4.84 &   1.11e-08 & 5.31e-09 & 1.14e-09 & 8.92e-10 & $ 0.676$ \\
NGC 253$^{d}$ & 139 & 1.0-2.0 &   0.0483 & 0.211 & 0.496 & 0.744 &   1.81e-08 & 1.41e-08 & 8.28e-09 & 6.27e-09 & $ 0.774$ \\

NGC 3628 & 14 & 1.0-2.0 &   0.355 & 1.02 & 2.55 & 3.55 &   1.77e-09 & 9.72e-10 & 4.38e-10 & 3.43e-10 & $ 0.322$ \\
NGC 3628 & 194 & 1.0-2.0 &   0.435 & 0.963 & 1.75 & 2.10 &   8.01e-10 & 5.68e-10 & 3.64e-10 & 3.28e-10 & $ 0.149$ \\

NGC 3079 & 75 & 1.0-2.0 &   0.628 & 1.37 & 5.63 & 8.75 &   1.11e-08 & 7.64e-09 & 2.13e-09 & 1.51e-09 & $ 2.03$ \\
NGC 3079 & 255 & 1.0-2.0 &   1.02 & 3.86 & 7.89 & 9.38 &   4.33e-09 & 1.60e-09 & 1.03e-09 & 8.96e-10 & $ 1.93$ \\

NGC 4945 & 313 & 1.0-2.0 &   0.220 & 0.392 & 0.520 & 0.576 &   1.81e-09 & 1.70e-09 & 1.58e-09 & 1.52e-09 & $ 0.590$ \\
NGC 4945 & 133 & 1.0-2.0 &   0.346 & 0.857 & 1.50 & 2.17 &   8.93e-10 & 5.95e-10 & 4.01e-10 & 2.80e-10 & $ 0.225$ \\

NGC 4631 & 356 & 1.0-2.0 &   0.657 & 1.48 & 3.70 & 7.49 &   8.47e-10 & 5.82e-10 & 2.65e-10 & 1.40e-10 & $ 0.221$ \\
NGC 4631 & 176 & 1.0-2.0 &   0.839 & 1.56 & 2.94 & 5.14 &   9.02e-10 & 7.53e-10 & 4.49e-10 & 2.58e-10 & $ 0.242$ \\

NGC 6503 & \nodata & 1.0-2.0 
	&   (1.0) & \nodata & (4.0) & \nodata 
	&   6.84e-10 & \nodata & 1.07e-10 & \nodata & \nodata \\

NGC 891 & 293 & 1.0-2.0 &   0.740 & 4.90 & 15.5 & 16.9 &   9.00e-10 & 1.76e-10 & 8.75e-11 & 8.56e-11 & $ 0.589$ \\
NGC 891 & 113 & 1.0-2.0 &   0.209 & 0.319 & 0.385 & 0.407 &   9.42e-10 & 9.27e-10 & 9.18e-10 & 9.15e-10 & $ 0.226$ \\
NGC 891 & \nodata & 1.0-2.0 
	&   (1.0) & \nodata & (4.0) & \nodata 
	&   2.72e-9 & \nodata & 6.39e-10 & \nodata & \nodata \\

NGC 4244 & \nodata & 1.0-2.0 
	&   (1.0) & \nodata & (4.0) & \nodata 
	&   1.18e-10 & \nodata & 2.80e-11 & \nodata & \nodata \\

\enddata
\tablecomments{Fixed enclosed flux and isophotal {\em heights} of
	the diffuse X-ray emission.
	Column 2: Position angle along which the surface brightness
	profile was extracted.
	Column 3: Energy band in keV.
	Column 4: Distance along minor axis (in kpc) at which 50\% of the total
	diffuse X-ray counts, {\em as measured from the major axis along this
	specific position angle}, 
	are contained. In other words, $z_{0.5}$ is 
	the minor axis X-ray half-light height.
	Each side of the disk is treated
	independently.
	Columns 5, 6 and 7: As column (4), but the 75, 90 and 95\% 
	X-ray-light-enclosed heights.
	Columns 8, 9, 10 and 11: The effective diffuse 
	X-ray surface brightness,
	in units of photons s$^{-1}$ cm$^{-2}$ arcsec$^{-2}$, between the
	major axis and $z_{0.5}$, $z_{0.75}$, $z_{0.90}$ and $z_{0.95}$ 
	respectively. This value is based on the exposure and 
	chip-sensitivity corrected images. Note that this is not the 
	same as the surface brightness at that height.
	Column 12: Minor axis isophotal height (in kpc), 
	at which the X-ray surface
	brightness drops to $10^{-9}$ photons s$^{-1}$ cm$^{-2}$ 
	arcsec$^{-2}$. Note that when a position angle is not given, 
	this denotes a measurement based on the count rate within a 
	circular aperture from the diffuse-emission-only images, and
	not from the 1-D surface brightness profiles. The radius chosen
	is shown in brackets in columns 4 and 6. See 
	\S~\ref{sec:results:extent} for 
	details.
	}
\tablenotetext{a}{ACIS-I observations of M82, ObsIDs 361 and 1302.}
\tablenotetext{b}{ACIS-S observation of M82, ObsID 2933.}
\tablenotetext{c}{NGC 253 observation ObsID 969, nucleus and disk on chip S3, northern halo on S4.}
\tablenotetext{d}{NGC 253 observation ObsID 790, northern halo on chip S3, disk and nucleus on S2.}
\end{deluxetable}


\begin{figure*}
\epsscale{2.0}
\plotone{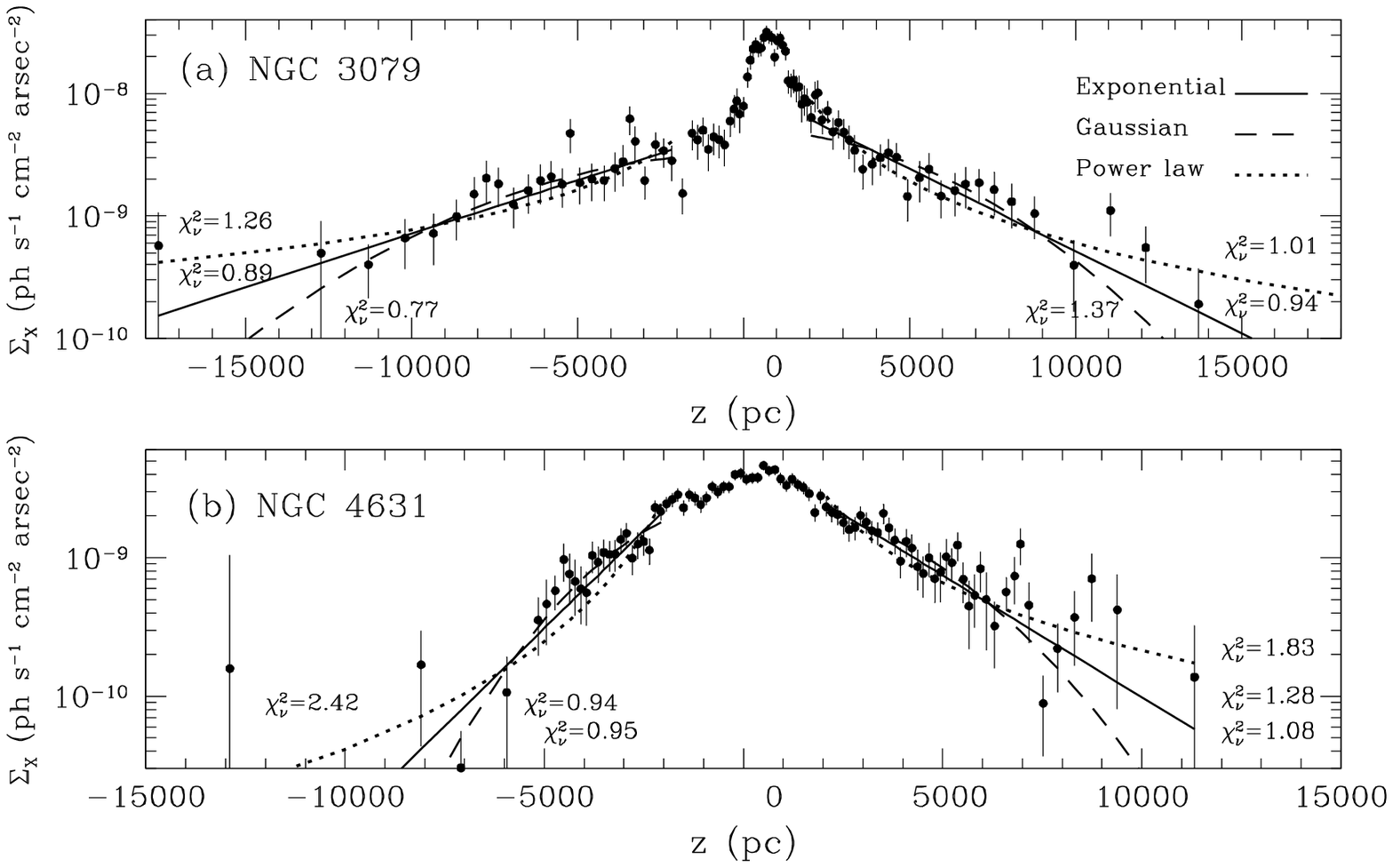}
\caption{Two examples of the observed 0.3--1.0 keV minor axis diffuse X-ray 
	surface brightness profiles, with the 
	best-fit exponential (solid line), Gaussian (dashed) and 
	power law (dotted) models. The reduced $\chi^{2}$ values of the
	fits are
	given alongside the appropriate curve. 
	Points within 2 kpc of the plane of the galaxy were not used
	in the fitting process 
	due to the high, and spatially non-uniform, absorption near the disk
	midplane.}
\label{fig:example_minor_axis_fits}
\end{figure*}

\begin{deluxetable}{lrcrrrrrrrrr}
 \tabletypesize{\scriptsize}%
\tablecolumns{12} 
\tablewidth{0pc} 
\tablecaption{Major-axis diffuse emission 
	spatial extent at fixed enclosed flux fractions
	\label{tab:enclosed_flux_fractions:major}} 
\tablehead{ 
\colhead{Galaxy} & \colhead{PA} & \colhead{Energy}  
	& \colhead{$r_{0.5}$}  
	& \colhead{$r_{0.75}$}  
	& \colhead{$r_{0.9}$}  
	& \colhead{$r_{0.95}$}  
	& \colhead{$\Sigma_{0.5}$}  & \colhead{$\Sigma_{0.75}$}  
	& \colhead{$\Sigma_{0.9}$}  & \colhead{$\Sigma_{0.95}$} 
	& \colhead{$r_{1e-9}$} \\
\colhead{(1)} & \colhead{(2)} & \colhead{(3)}
	& \colhead{(4)} & \colhead{(5)} &  \colhead{(6)} &  \colhead{(7)} 
	& \colhead{(8)} &  \colhead{(9)} &  \colhead{(10)} &  \colhead{(11)}
	& \colhead{(12)}  
	}
\startdata

M82$^{a}$ & 245 & 0.3-1.0 &   0.313 & 0.683 & 1.20 & 1.87 &   9.59e-08 & 6.70e-08 & 4.63e-08 & 3.13e-08 & $ 3.57$ \\
M82$^{a}$ & 65 & 0.3-1.0 &   0.416 & 1.35 & 2.43 & 2.94 &   4.59e-08 & 2.31e-08 & 1.45e-08 & 1.25e-08 & $ 3.65$ \\

M82$^{b}$ & 245 & 0.3-1.0 &   0.274 & 0.622 & 1.11 & 1.74 &   4.48e-08 & 3.31e-08 & 2.35e-08 & 1.63e-08 & $ 2.63$ \\
M82$^{b}$ & 65 & 0.3-1.0 &   0.321 & 0.876 & 1.85 & 2.48 &   3.54e-08 & 2.22e-08 & 1.32e-08 & 1.05e-08 & $ 3.17$ \\

NGC 1482 & 283 & 0.3-1.0 &   0.505 & 0.934 & 1.34 & 1.55 &   4.05e-08 & 3.40e-08 & 2.89e-08 & 2.60e-08 & $ 3.22$ \\
NGC 1482 & 103 & 0.3-1.0 &   0.139 & 0.278 & 0.394 & 0.429 &   2.65e-08 & 2.34e-08 & 2.10e-08 & 2.07e-08 & $ 1.97$ \\

n253$^{c}$ & 229 & 0.3-1.0 &   0.843 & 1.61 & 2.25 & 2.56 &   3.29e-09 & 2.85e-09 & 2.73e-09 & 2.64e-09 & $ 3.59$ \\
n253$^{c}$ & 49 & 0.3-1.0 &   0.452 & 1.59 & 2.30 & 2.79 &   8.77e-09 & 4.40e-09 & 3.89e-09 & 3.58e-09 & $ 7.59$ \\
n253$^{d}$ & 229 & 0.3-1.0 &   0.750 & 1.50 & 2.15 & 2.46 &   4.22e-09 & 3.33e-09 & 2.98e-09 & 2.88e-09 & $ 3.20$ \\
n253$^{d}$ & 49 & 0.3-1.0 &   0.847 & 2.04 & 2.97 & 3.48 &   6.95e-09 & 4.66e-09 & 3.84e-09 & 3.51e-09 & $ 4.51$ \\

NGC 3628 & 284 & 0.3-1.0 &   1.58 & 2.43 & 3.17 & 3.49 &   2.33e-09 & 2.27e-09 & 2.07e-09 & 1.96e-09 & $ 3.49$ \\
NGC 3628 & 104 & 0.3-1.0 &   1.70 & 3.31 & 4.05 & 4.71 &   1.42e-09 & 1.14e-09 & 1.12e-09 & 8.52e-10 & $ 4.10$ \\

NGC 3079 & 345 & 0.3-1.0 &   1.99 & 4.87 & 6.62 & 7.40 &   1.12e-08 & 7.13e-09 & 6.32e-09 & 5.98e-09 & $ 9.39$ \\
NGC 3079 & 165 & 0.3-1.0 &   1.43 & 3.82 & 5.63 & 6.23 &   1.05e-08 & 6.02e-09 & 4.98e-09 & 4.75e-09 & $ 7.52$ \\

NGC 4945 & 223 & 0.3-1.0 &   1.52 & 3.65 & 4.42 & 4.90 &   1.34e-09 & 8.14e-10 & 8.30e-10 & 7.99e-10 & $ 4.16$ \\
NGC 4945 & 43 & 0.3-1.0 &   1.67 & 4.64 & 7.80 & 9.59 &   1.06e-09 & 7.56e-10 & 6.08e-10 & 5.46e-10 & $ 6.09$ \\

NGC 4631& 266 & 0.3-1.0 &   2.59 & 4.73 & 6.37 & 6.94 &   2.22e-09 & 1.87e-09 & 1.69e-09 & 1.65e-09 & $ 7.17$ \\
NGC 4631& 86 & 0.3-1.0 &   1.94 & 3.82 & 6.17 & 6.74 &   2.75e-09 & 2.20e-09 & 1.70e-09 & 1.66e-09 & $ 6.73$ \\

NGC 891 & 203 & 0.3-1.0 &   0.929 & 1.60 & 2.04 & 2.18 &   1.10e-09 & 1.02e-09 & 9.67e-10 & 9.56e-10 & $ 1.05$ \\
NGC 891 & 23 & 0.3-1.0 &   3.04 & 4.81 & 6.33 & 9.67 &   9.15e-10 & 9.17e-10 & 8.26e-10 & 5.51e-10 & $ 5.31$ \\

\tableline
M82$^{a}$ & 245 & 1.0-2.0 &   0.184 & 0.401 & 0.736 & 1.00 &   8.41e-08 & 6.37e-08 & 4.13e-08 & 3.21e-08 & $ 1.48$ \\
M82$^{a}$ & 65 & 1.0-2.0 &   0.0995 & 0.329 & 1.32 & 2.10 &   6.58e-08 & 4.04e-08 & 1.37e-08 & 8.52e-09 & $ 2.03$ \\

M82$^{b}$ & 245 & 1.0-2.0 &   0.156 & 0.379 & 0.730 & 1.08 &   4.19e-08 & 3.10e-08 & 2.06e-08 & 1.52e-08 & $ 1.32$ \\
M82$^{b}$ & 65 & 1.0-2.0 &   0.153 & 0.416 & 0.971 & 1.82 &   3.72e-08 & 2.55e-08 & 1.42e-08 & 8.20e-09 & $ 1.48$ \\

NGC 1482 & 283 & 1.0-2.0 &   0.395 & 0.747 & 1.22 & 1.57 &   2.66e-08 & 2.23e-08 & 1.62e-08 & 1.30e-08 & $ 2.64$ \\
NGC 1482 & 103 & 1.0-2.0 &   0.131 & 0.348 & 0.637 & 0.746 &   1.42e-08 & 9.20e-09 & 6.29e-09 & 5.33e-09 & $ 1.40$ \\

n253$^{c}$ & 229 & 1.0-2.0 &   0.149 & 1.21 & 2.05 & 2.47 &   2.90e-09 & 7.72e-10 & 6.05e-10 & 5.61e-10 & $ 0.266$ \\
n253$^{c}$ & 49 & 1.0-2.0 &   0.0682 & 0.303 & 1.63 & 2.21 &   9.50e-09 & 5.95e-09 & 1.65e-09 & 1.35e-09 & $ 0.450$ \\
n253$^{d}$ & 229 & 1.0-2.0 &   0.0795 & 0.880 & 1.87 & 2.28 &   4.67e-09 & 1.28e-09 & 7.79e-10 & 7.13e-10 & $ 0.282$ \\
n253$^{d}$ & 49 & 1.0-2.0 &   0.0606 & 0.278 & 1.60 & 2.28 &   1.05e-08 & 6.89e-09 & 1.78e-09 & 1.33e-09 & $ 0.445$ \\

NGC 3628 & 284 & 1.0-2.0 &   0.969 & 2.66 & 7.41 & 9.58 &   7.01e-10 & 4.14e-10 & 1.68e-10 & 1.41e-10 & $ 0.224$ \\
NGC 3628 & 104 & 1.0-2.0 &   0.930 & 2.67 & 4.73 & 5.72 &   5.42e-10 & 2.97e-10 & 2.09e-10 & 1.68e-10 & $ 0.181$ \\

NGC 3079 & 345 & 1.0-2.0 &   0.568 & 2.59 & 4.95 & 6.54 &   1.03e-08 & 3.69e-09 & 2.27e-09 & 1.72e-09 & $ 3.69$ \\
NGC 3079 & 165 & 1.0-2.0 &   0.509 & 2.81 & 4.89 & 6.33 &   8.12e-09 & 2.46e-09 & 1.69e-09 & 1.39e-09 & $ 1.49$ \\

NGC 4945 & 223 & 1.0-2.0 &   0.513 & 2.58 & 4.35 & 5.70 &   1.89e-09 & 5.90e-10 & 4.11e-10 & 3.21e-10 & $ 0.368$ \\
NGC 4945 & 43 & 1.0-2.0 &   0.414 & 1.28 & 2.57 & 3.02 &   1.51e-09 & 7.56e-10 & 3.47e-10 & 3.29e-10 & $ 0.209$ \\

NGC 4631 & 266 & 1.0-2.0 &   3.21 & 6.60 & 11.0 & 12.8 &   2.77e-10 & 2.07e-10 & 1.42e-10 & 1.40e-10 & \nodata \\
NGC 4631 & 86 & 1.0-2.0 &   1.53 & 2.45 & 5.12 & 6.55 &   4.34e-10 & 3.85e-10 & 2.03e-10 & 1.89e-10 & \nodata \\

NGC 891 & 203 & 1.0-2.0 &   0.458 & 1.12 & 1.72 & 1.94 &   3.98e-10 & 3.10e-10 & 2.50e-10 & 2.39e-10 & \nodata \\
NGC 891 & 23 & 1.0-2.0 &   1.06 & 1.88 & 2.31 & 2.43 &   3.54e-10 & 3.24e-10 & 3.16e-10 & 3.21e-10 & \nodata \\

\enddata
\tablecomments{Fixed enclosed flux and isophotal {\em radii} 
	measured along the {\em major} axis of each galaxy.
	Apart from being the radius from the 
nucleus, instead of height above the plane, the meaning of the table columns
are identical to those in Table~\ref{tab:enclosed_flux_fractions:minor}.}
\tablenotetext{a}{ACIS-I observations of M82, ObsIDs 361 and 1302.}
\tablenotetext{b}{ACIS-S observation of M82, ObsID 2933.}
\tablenotetext{c}{NGC 253 observation ObsID 969, nucleus and disk on chip S3, northern halo on S4.}
\tablenotetext{d}{NGC 253 observation ObsID 790, northern halo on chip S3, disk and nucleus on S2.}
\end{deluxetable}

\subsubsection{Sizes at fixed enclosed-light fractions}

In Tables~\ref{tab:enclosed_flux_fractions:minor} and 
\ref{tab:enclosed_flux_fractions:major} we present the spatial
extent of the diffuse emission measured along the minor and major axes
from the nucleus, within which a fixed fraction (50, 75, 90 and 95\%)
of the total number diffuse X-ray counts were found within the 1-dimensional
slices. The mean X-ray surface
brightness $\Sigma$ within these regions is also tabulated. 
As an example, the value of $\Sigma_{0.5}$ in the first row of
Table~\ref{tab:enclosed_flux_fractions:minor} is the surface brightness
in the 0.3 -- 1.0 keV energy band within a region 10 kpc in major axis
extent, and extending 0.99 kpc (the minor axis half-light height $z_{0.5}$) 
along the minor axis of M82 to the NNE
(along position angle 335).
The total number of counts is evaluated the extreme 
edge of each slice, \ie we assume 
all emission is contained within $20$ kpc of the nucleus. There
exists the possibility that X-ray emission may extended significantly
further than this at lower surface brightness than these observations
are sensitive to, although this would have relatively little impact
on the 50 and 75\%-enclosed light radii and heights we report.

The diffuse X-ray emission in NGC 6503 and NGC 4244
is so faint that direct measurements of the half light and isophotal
sizes can not be robustly measured from the 1-D minor axis, or major
axis, surface brightness profiles. In this case, we simply report
a measurement of the surface brightness within a circular apertures
of radius 1 and 4 kpc, values typical of the 50 and 90\% enclosed light
heights and radii of the other galaxies in the sample.
It is possible that at these very low surface brightness levels
we are looking at unresolved faint X-ray point source emission, and
not truly diffuse hot gas.

\subsubsection{Sizes at fixed surface brightness levels}

Isophotal sizes, measured from the 1-dimensional surface brightness
profiles at the point $\Sigma_{X}$ reaches $10^{-9}$ \photsurfb, 
are given in 
Tables~\ref{tab:enclosed_flux_fractions:minor} and 
\ref{tab:enclosed_flux_fractions:major}.

Given the spectrum the X-ray emitting halo gas 
(see \S~\ref{sec:spectra:fits}), a surface brightness of
$10^{-9}$ \photsurfb in the 0.3--1.0 keV band corresponds
to an intrinsic emitted energy flux (\ie correcting for Galactic 
and intrinsic absorption) of approximately $\sim 2\times 10^{-18}$ \ergsurfb
in either the 0.3--1.0 or 0.3--2.0 keV energy band. In the 1.0--2.0
keV energy band, $10^{-9}$ \photsurfb corresponds to 
$\approx 1.5\times10^{-17}$
\ergsurfb in the 0.3--2.0 keV energy band, and 
$\approx 2\times 10^{-18}$ \ergsurfb in the 1.0--2.0 keV energy band.
These conversion factors are the typical, approximately average,
value for all of the galaxies with halo X-ray emission, and are in fact
based on the NGC 3079 halo spectra.
Note that these conversions take the contamination of the ACIS optical
blocking filter into account.  

The range of X-ray spectral hardness, foreground absorption, and level of
blocking filter contamination in this sample introduces
a range of a factor two into this conversion. More specifically, the
the intrinsic energy flux per photon is a factor $\sim2$ higher for the
ACIS-S observation of M82 and
a factor $\sim 2$ lower for the NGC 253 observations.

Diffuse X-ray emission can be robustly detected at lower
X-ray surface brightness levels than $10^{-9}$ \photsurfb, 
in both the 0.3--1.0 and 1.0--2.0 keV
energy bands, but we chose this isophote as it is well-matched
to deep studies of optical extra-planar emission (for \halpha~emission
from gas at $T = 10^{4}$ K, a surface brightness of 
$2\times 10^{-18}$ \ergsurfb corresponds
to an emission measure [EM] of 1 pc cm$^{-6}$).

\begin{deluxetable}{lrcclrrrrrrrr}
\tabletypesize{\scriptsize}%
\tablecolumns{13} 
\tablewidth{0pc} 
\tablecaption{Exponential, Gaussian and power law
  model fits to minor axis diffuse emission
	\label{tab:exponential_models}} 
\tablehead{ 
\colhead{Galaxy} & \colhead{PA} & \colhead{Energy} 
	& \colhead{$z_{\rm range}$} & \multicolumn{5}{l}{Exponential model}
        	& \multicolumn{2}{l}{Gaussian model}
        	& \multicolumn{2}{l}{Power law} \\
\colhead{} & \colhead{($\degr$)} & \colhead{(keV)} 
	& \colhead{(kpc)} & \colhead{$H_{\rm eff}$} & \colhead{$\Sigma_{0}$}
	& \colhead{$\chi^{2}$} & \colhead{$\nu$} & \colhead{$\chi^{2}/\nu$}
	& \colhead{$FWHM$} & \colhead{$\chi^{2}/\nu$}
	& \colhead{$\alpha$} & \colhead{$\chi^{2}/\nu$} \\
\colhead{(1)} & \colhead{(2)} & \colhead{(3)} &  \colhead{(4)} 
	& \colhead{(5)} & \colhead{(6)} & \colhead{(7)} & \colhead{(8)} & \colhead{(9)} 
	& \colhead{(10)} & \colhead{(11)} & \colhead{(12)} & \colhead{(13)}  
}
\startdata

M82$^{a}$ & 335 & 0.3-1.0
	& 0.5--5 & $1.42_{-0.05}^{+0.06}$ & $29.16^{+1.53}_{-1.48} \times 10^{-9}$ 
	& 354.2 & 30 & \underline{11.8}
	& $4.13_{-0.13}^{+0.13}$ & 25.8 
	& $1.20_{-0.03}^{+0.03}$ & 14.1 \\
M82$^{a}$ & 335 & 0.3-1.0
	& 1--5 & $1.53_{-0.08}^{+0.09}$ & $25.51^{+2.25}_{-2.08} \times 10^{-9}$ 
	& 329.3 & 27 & 12.20 	
	& $4.83_{-0.17}^{+0.18}$ & 18.8 
	& $1.53_{-0.07}^{+0.07}$ & \underline{8.5} \\
M82$^{a}$ & 155 & 0.3-1.0
	& 0.2--10 & $0.73_{-0.02}^{+0.02}$ & $103.4^{+3.52}_{-3.42} \times 10^{-9}$ 
	& 1318.5 & 41 & \underline{32.16} 	
	& $2.42_{-0.05}^{+0.05}$ & 71.1 
	& $1.44_{-0.01}^{+0.01}$ & 86.6 \\
M82$^{a}$ & 155 & 0.3-1.0
	& 2--6 & $1.10_{-0.15}^{+0.17}$ & $32,21^{+1.62}_{-0.98} \times 10^{-9}$ 
	& 308.8 & 25 & 12.35 	
	& $4.87_{-0.48}^{+0.48}$ & 13.9 
	& $3.11_{-0.17}^{+0.18}$ & \underline{11.2} \\

M82$^{b}$ & 335 & 0.3-1.0
	& 0.35--16 & $1.54_{-0.05}^{+0.05}$ & $36.22^{+1.50}_{-1.45} \times 10^{-9}$ 
	& 498.2 & 56 & \underline{8.9} 	
	& $4.78_{-0.15}^{+0.15}$ & 14.0 
	& $1.39_{-0.02}^{+0.02}$ & 30.6 \\
M82$^{b}$  & 335 & 0.3-1.0
	& 1--5 & $0.64_{-0.02}^{+0.02}$ & $175.50^{+16.14}_{-14.77} \times 10^{-9}$ 
	& 453.5 & 27 & 16.8 
	& $4.66_{-0.15}^{+0.16}$ & 20.4 
	& $0.94_{-0.03}^{+0.03}$ & \underline{5.8} \\
M82$^{b}$ & 155 & 0.3-1.0
	& 0.2--10 & $0.78_{-0.02}^{+0.02}$ & $104.0^{+2.8}_{-2.8} \times 10^{-9}$ 
	& 674.3 & 36 & 18.7
	& $2.35_{-0.04}^{+0.04}$ & \underline{18.3} 
	& $1.27_{-0.01}^{+0.01}$ & 121.2 \\
M82$^{b}$ & 155 & 0.3-1.0
	& 2--6 & $1.04_{-0.32}^{+0.30}$ & $27.98^{+56.98}_{-12.71} \times 10^{-9}$ 
	& 145.4 & 21 & 6.92 
	& $4.92_{-0.77}^{+0.69}$ & 7.4 
	& $3.71_{-0.39}^{+0.47}$ & \underline{6.3} \\

NGC 1482 & 13 & 0.3-1.0
	& 1--17 & $1.19_{-0.33}^{+0.38}$ & $17.59^{+12.11}_{-6.98} \times 10^{-9}$ 
	& 21.8 & 13 & 1.67 
	& $4.62_{-0.70}^{+0.70}$ & \underline{1.52} 
	& $2.21_{-0.42}^{+0.42}$ & 2.59  \\
NGC 1482 & 13 & 0.3-1.0
	& 2--17 & $0.77_{-0.27}^{+0.34}$ & $108.5^{+532.0}_{-78.4} \times 10^{-9}$ 
	& 10.1 & 9 & 1.12 
	& $3.76_{-0.83}^{+0.87}$ & \underline{0.85} 
	& $4.32_{-0.82}^{+1.57}$ & 1.50 \\
NGC 1482 & 103 & 0.3-1.0
	& 0.2--19 & $0.81_{-0.10}^{+0.11}$ & $64.61^{+11.40}_{-10.11} \times 10^{-9}$ 
	& 28.3 & 24 & \underline{1.18} 
	& $2.07_{-0.23}^{+0.26}$ & 1.94 
	& $1.39_{-0.06}^{+0.06}$ & 5.46 \\
NGC 1482 & 103 & 0.3-1.0
	& 2--19 & $1.14_{-0.49}^{+0.75}$ & $29.23^{+123.6}_{-20.33} \times 10^{-9}$ 
	& 14.9 & 10 & 1.49 
	& $4.81_{-1.41}^{+1.80}$ & \underline{1.42} 
	& $3.17_{-0.72}^{+1.40}$ & 1.71 \\

NGC 253$^{c}$ & 319 & 0.3-1.0
	& 4--8 & $1.39_{-0.45}^{+0.77}$ & $45.11^{166.4}_{-32.5} \times 10^{-9}$ 
	& 43.8 & 8 & 1.68 
	& $6.31_{-1.21}^{+1.66}$ & \underline{1.67} 
	& $2.93_{-0.22}^{+0.27}$ & 2.93 \\
NGC 253$^{d}$ & 319 & 0.3-1.0
	& 4--10 & $2.91_{-0.45}^{+0.60}$ & $7.46^{2.96}_{-2.10} \times 10^{-9}$ 
	& 34.8 & 20 & 1.74 
	& $9.65_{-0.81}^{+0.99}$ & \underline{1.37} 
	& $1.92_{-0.07}^{+0.07}$ & 2.20 \\
NGC 253$^{c}$ & 139 & 0.3-1.0
	& 0.3--4 & $0.58_{-0.03}^{+0.03}$ & $44.20^{+3.39}_{-3.19} \times 10^{-9}$ 
	& 35.3 & 11 & 3.21 
	& $1.69_{-0.07}^{+0.08}$ & \underline{3.14} 
	& $1.38_{-0.03}^{+0.03}$ & 29.9 \\
NGC 253$^{d}$ & 139 & 0.3-1.0
	& 0.3--4 & $0.58_{-0.02}^{+0.02}$ & $34.37^{+1.91}_{-1.82} \times 10^{-9}$ 
	& 73.6 & 14 & \underline{5.26} 
	& $1.68_{-0.07}^{+0.06}$ & 9.76 
	& $1.86_{-0.02}^{+0.02}$ & 77.4 \\

NGC 3628 & 14 & 0.3-1.0
	& 1--17 & $2.00_{-0.27}^{+0.31}$ & $5.42^{+1.21}_{-1.00} \times 10^{-9}$ 
	& 41.9 & 42 & 1.00 
	& $6.33_{-0.65}^{+0.67}$ & \underline{0.95} 
	& $1.57_{-0.16}^{+0.17}$ & 2.01 \\
NGC 3628 & 14 & 0.3-1.0
	& 2--17 & $1.94_{-0.34}^{+0.41}$ & $5.81^{+2.50}_{-1.73} \times 10^{-9}$ 
	& 40.4 & 32 & 1.26 
	& $6.75_{-0.78}^{+0.86}$ & \underline{1.04} 
	& $2.08_{-0.22}^{+0.28}$ & 1.92 \\
NGC 3628 & 194 & 0.3-1.0
	& 2--19 & $3.72_{-0.86}^{+1.09}$ & $1.42^{+0.56}_{-0.44} \times 10^{-9}$ 
	& 83.2 & 32 & 2.60 
	& $11.39_{-1.69}^{+1.86}$ & \underline{2.05} 
	& $1.45_{-0.24}^{+0.35}$ & 3.46  \\
NGC 3628 & 194 & 0.3-1.0
	& 1--19 & $4.21_{-0.93}^{+1.15}$ & $1.07^{+0.38}_{-0.27} \times 10^{-9}$ 
	& 90.7 & 37 & 2.45 
	& $11.72_{-1.70}^{+1.84}$ & \underline{1.87} 
	& $1.05_{-0.20}^{+0.26}$ & 3.45 \\

NGC 3079 & 75 & 0.3-1.0
	& 1--18 & $1.82_{-0.33}^{+0.41}$ & $29.00^{+9.67}_{-7.12} \times 10^{-9}$ 
	& 88.6 & 42 & 2.11 
	& $5.04_{-0.76}^{+0.97}$ & 3.39 
	& $1.67_{-0.14}^{+0.15}$ & \underline{0.96} \\
NGC 3079 & 75 & 0.3-1.0
	& 2--18 & $3.23_{-0.70}^{+0.85}$ & $11.44^{+4.81}_{-3.19} \times 10^{-9}$ 
	& 28.3 & 30 & \underline{0.94} 
	& $10.67_{-2.00}^{+2.14}$ & 1.37 
	& $1.67_{-0.15}^{+0.19}$ & 1.01 \\
NGC 3079 & 255 & 0.3-1.0
	& 0--18 & $5.20_{-0.98}^{+1.29}$ & $4.85^{+0.94}_{-0.86} \times 10^{-9}$ 
	& 46.7 & 38 & \underline{1.23} 
	& $12.89_{-1.68}^{+1.94}$ & 1.33 
	& $0.55_{-0.08}^{+0.07}$ & 1.78 \\
NGC 3079 & 255 & 0.3-1.0
	& 2--18 & $4.96_{-1.12}^{+1.66}$ & $5.38^{+1.92}_{-1.51} \times 10^{-9}$ 
	& 23.1 & 26 & 0.89 
	& $13.33_{-1.93}^{+2.44}$ & \underline{0.77} 
	& $1.08_{-0.14}^{+0.17}$ & 1.26 \\

NGC 4945 & 313 & 0.3-1.0
	& 0.5--5 & $1.06_{-0.21}^{+0.30}$ & $5.01^{+1.17}_{-0.96} \times 10^{-9}$ 
	& 12.9 & 15 & \underline{0.86} 
	& $2.50_{-0.32}^{+0.41}$ & 1.24 
	& $0.88_{-0.07}^{+0.08}$ & 1.06 \\
NGC 4945 & 133 & 0.3-1.0
	& 0.5--5 & $0.59_{-0.13}^{+0.15}$ & $3.11^{+0.82}_{-0.66} \times 10^{-9}$ 
	& 11.7 & 12 & \underline{0.98} 
	& $1.61_{-0.35}^{+0.39}$ & 1.83 
	& $1.22_{-0.08}^{+0.07}$ & 2.74 \\

NGC 4631 & 356 & 0.3-1.0
	& 0.5--17 & $2.42_{-0.18}^{+0.19}$ & $5.70^{+0.46}_{-0.44} \times 10^{-9}$ 
	& 57.5 & 51 & \underline{1.12} 
	& $6.66_{-0.45}^{+0.46}$ & 1.91 
	& $0.92_{-0.05}^{+0.05}$ & 4.03 \\
NGC 4631 & 356 & 0.3-1.0
	& 2--17 & $2.37_{-0.27}^{+0.29}$ & $5.99^{+1.33}_{-1.05} \times 10^{-9}$ 
	& 49.9 & 39 & 1.28 
	& $7.85_{-0.63}^{+0.68}$ & \underline{1.08} 
	& $1.63_{-0.12}^{+0.14}$ & 1.83 \\
NGC 4631 & 176 & 0.3-1.0
	& 0--17 & $2.33_{-0.21}^{+0.23}$ & $4.33^{+0.34}_{-0.33} \times 10^{-9}$ 
	& 46.1 &  38 & \underline{1.21} 
	& $5.27_{-0.38}^{+0.41}$ & 1.28 
	& $0.44_{-0.03}^{+0.03}$ & 8.02 \\
NGC 4631 & 176 & 0.3-1.0
	& 2--17 & $1.89_{-0.38}^{+0.50}$ & $5.76^{+2.76}_{-1.75} \times 10^{-9}$ 
	& 21.8 & 23 & 0.95 
	& $5.97_{-0.61}^{+0.59}$ & \underline{0.94} 
	& $1.82_{-0.23}^{+0.27}$ & 1.27 \\

NGC 6503 &  32 & 0.3-1.0 
	& 0--10 & $0.33_{-0.17}^{+0.33}$ & $3.57^{+2.68}_{-1.72} \times 10^{-9}$ 
	& 7.3 & 13 & \underline{0.56} 
	& $0.72_{-0.29}^{+0.64}$ & 0.59 
	& $1.04_{-0.18}^{+0.12}$ & 0.81 \\
NGC 6503 &  212 & 0.3-1.0 
	& 0--5 & $0.18_{-0.18}^{+0.36}$ & $3.72^{+0.11}_{-2.56} \times 10^{-9}$ 
	& 1.1 & 5 & \underline{0.22} 
	& $0.50_{-0.45}^{+0.43}$ & \underline{0.22} 
	& $2.24_{-0.25}^{+0.14}$ & 0.56 \\

NGC 891 &  293 & 0.3-1.0 
	& 2--8 & $2.44_{-1.34}^{+3.94}$ & $2.21^{+6.77}_{-1.43} \times 10^{-9}$ 
	& 9.7 & 10 & 0.97 
	& $8.12_{-4.16}^{+6.49}$ & 1.18 
	& $1.57_{-0.46}^{+0.76}$ & \underline{0.83} \\
NGC 891 &  293 & 0.3-1.0 
	& 0.4--8 & $1.97_{-0.47}^{+0.65}$ & $3.33^{+1.01}_{-0.85} \times 10^{-9}$ 
	& 19.2 & 21 & \underline{0.91} 
	& $4.46_{-0.74}^{+1.03}$ & 0.99 
	& $0.88_{-0.17}^{+0.15}$ & 1.63 \\
NGC 891 &  113 & 0.3-1.0 
	& 0.4--5 & $1.35_{-0.35}^{+0.50}$ & $3.64^{+1.33}_{-1.05} \times 10^{-9}$ 
	& 15.4 & 14 & 1.10 
	& $3.57_{-0.66}^{+0.82}$ & \underline{0.67} 
	& $1.03_{-0.21}^{+0.17}$ & 2.31 \\

\enddata
\tablecomments{See \S~\ref{sec:surfb_fits} for more information. All
	errors are $1\sigma$ for 2 interesting parameters. For each row
	the lowest reduced $\chi^{2}$ value has been underlined, to aid the reader
	in assessing which form of surface model provides the best description of the data..
	The columns are
	(2): Position angle of the minor axis surface brightness slice used.
	(3): Energy band from which the data was extracted.
	(4): Range of heights above or below the plane of the galaxy used in fitting
	the surface brightness profiles.
	(5): Best-fit effective exponential scale height, in kpc.
	(6): Extrapolated central surface brightness (\ie at $z = 0$), 
	in units of \photsurfb.
	(7, 8 and 9): Total $\chi^{2}$ value for best-fit exponential model,
	number of degrees of freedom $\nu$ (this is the same for the Gaussian and
	power law models) and reduced $\chi^{2}$ value.
	(10): Best fit Gaussian model FHWM in kpc.
	(11): Best fit reduced $\chi^{2}$ for the Gaussian surface brightness model.
	(12): Best fit power law slope.
	(13): Best fit reduced $\chi^{2}$ for the power law surface brightness model.
	}
\tablenotetext{a}{ACIS-I observations of M82, ObsIDs 361 and 1302.}
\tablenotetext{b}{ACIS-S observation of M82, ObsID 2933.}
\tablenotetext{c}{NGC 253 observation ObsID 969, nucleus and disk on chip S3, northern halo on S4.}
\tablenotetext{d}{NGC 253 observation ObsID 790, northern halo on chip S3, disk and nucleus on S2.}
\end{deluxetable}

\subsubsection{Model fits}
\label{sec:surfb_fits}

We used the {\sc Ciao} fitting routine {\sc Sherpa} to fit exponential,
Gaussian and power law models to the 0.3--1.0 keV energy band
1-D X-ray minor axis surface brightness
profiles, after splitting the minor axis data into two separate profiles for
positive and negative $z$. 

We used the 0.3--1.0 keV energy band as it 
contains the greatest number of diffuse
X-ray counts, and traces the diffuse X-ray emission to greater distances from the
plane of the galaxy, than the 1.0--2.0 keV energy band. 
The obvious disadvantage
with using this energy band is that absorption by gas and dust in
the plane of the host galaxy can be significant (as can be seen in the higher
values of hardness ratio $Q_{A}$ near the plane in Fig.~\ref{fig:minor_axis_profiles}).

However, as the primary aim of this paper is to quantify the properties of
the {\em extra}-planar diffuse X-ray emission, it is advantageous to
exclude the data within 1 or 2 kpc of the plane from the fit in any case.
Results of the model fitting,
along with the range of $z$ values used, is given in Table~\ref{tab:exponential_models}
(to conserve space we have not tabulated the
best-fit model normalizations for the Gaussian and power law models).
We find that exponential, and/or Gaussian, models provide the best description of the
observed minor axis surface profiles. 

\subsubsubsection{Spectral variation and other caveats}

Note that these 1-dimensional fits are intended as a simple empirical measure
of the general way in which the surface brightness drops off with increasing
height. It is clear that there is real 2-or-3-dimensional 
structure in the X-ray-emitting gas, with edge-brightened limbs, 
filaments and clumps apparent in several 
of Figs.~\ref{fig:images:m82} -- \ref{fig:images:n4244}, 
in addition to the 1st-order
decrease in brightness with $z$. 

In several galaxies, notably M82, NGC 253 (northern halo), and NGC 3079,
the hardness ratios $Q_{A}$ and $Q_{B}$ show no significant spatial
variation within increasing distance from the mid-plane, beyond a
height of $|z|\sim 2$ kpc (Fig.~\ref{fig:minor_axis_profiles}).
This implies that no significant variations in plasma temperature 
or intervening absorption exist, and hence that changes 
in $\Sigma_{X}$ are directly proportional to changes in $\int n_{e}^{2} 
\delta l$. 
The typical best-fit surface-brightness 
exponential scale heights are $H_{\rm eff} \sim$ a few kpc, implying
density scale heights twice as large.

Some galaxies do show intriguing hints of spectral variation 
within their halos,
\eg NGC 4631, or to the southern side of NGC 3628 (negative z-axis 
in Fig.~\ref{fig:minor_axis_profiles}). A gentle decrease in hardness
ratio $Q_{\rm A}$ is apparent in both of these cases, consistent
with either a decrease in effective plasma temperature and/or intrinsic
or intervening absorption. Peculiarly, the northern side of NGC 3628
shows no similar spectral softening.

\begin{deluxetable}{llclrrrrrrr}
 \tabletypesize{\scriptsize}%
\tablecolumns{11} 
\tablewidth{0pc} 
\tablecaption{Nuclear, disk and halo count rates
	\label{tab:spectral:rates_and_areas}} 
\tablehead{ 
\colhead{Galaxy} 
	& \colhead{Region} & \colhead{Size}
	& \colhead{Diffuse} & \colhead{Area} 
	& \multicolumn{2}{l}{Soft 0.3-1.0 keV}
	& \multicolumn{2}{l}{Medium 1.0-2.0 keV}
	& \multicolumn{2}{l}{Hard 2.0-8.0 keV} \\
	& & \colhead{maj.$\times$min.} & \colhead{only?} & & \colhead{cts/s}
	& \colhead{tot cts} & \colhead{cts/s} 
	& \colhead{tot cts}
	& \colhead{cts/s} & \colhead{tot cts} \\
\colhead{(1)} & \colhead{(2)}
	& \colhead{(3)}
	& \colhead{(4)}
	& \colhead{(5)} & \colhead{(6)}
	& \colhead{(7)} & \colhead{(8)} 
	& \colhead{(9)} & \colhead{(10)} & \colhead{(11)} 
	}
\startdata
M82$^{a}$ & nucl & 1.90 & diff &   0.663 & $0.2424\pm{0.0023}$ &    11757 & $0.5666\pm{0.0034}$ &    27482 & $0.1445\pm{0.0018}$ &     7007 \\
M82$^{a}$ & nucl & 1.90 & wsrc &   0.711 & $0.2480\pm{0.0023}$ &    12028 & $0.6649\pm{0.0037}$ &    32250 & $0.3719\pm{0.0028}$ &    18038 \\
M82$^{a}$ & disk & $8.91\times3.81$ & diff &  30.448 & $0.5196\pm{0.0034}$ &    25206 & $0.3739\pm{0.0029}$ &    18137 & $0.0508\pm{0.0018}$ &     2463 \\
M82$^{a}$ & disk & $8.91\times3.81$ & wsrc &  31.181 & $0.4915\pm{0.0033}$ &    23845 & $0.3846\pm{0.0029}$ &    18658 & $0.0680\pm{0.0019}$ &     3300 \\
M82$^{a}$ & halo & $8.00\times15.50$ & diff &  90.024 & $0.1780\pm{0.0028}$ &     8635 & $0.0640\pm{0.0020}$ &     3105 & $0.0021\pm{0.0031}$ &      101 \\
M82$^{a}$ & halo & $8.00\times15.50$ & wsrc &  91.499 & $0.1878\pm{0.0028}$ &     9111 & $0.0816\pm{0.0021}$ &     3960 & $0.0140\pm{0.0032}$ &      677 \\

M82$^{b}$  & nucl & 1.90 & diff &   0.642 & $0.3955\pm{0.0048}$ &     7116 & $0.6057\pm{0.0059}$ &    10898 & $0.1848\pm{0.0033}$ &     3325 \\
M82$^{b}$  & nucl & 1.90 & wsrc &   0.712 & $0.4879\pm{0.0053}$ &     8779 & $0.8471\pm{0.0069}$ &    15242 & $0.5104\pm{0.0054}$ &     9184 \\
M82$^{b}$  & disk & $8.91\times3.81$ & diff &  31.183 & $1.1697\pm{0.0084}$ &    21048 & $0.5490\pm{0.0058}$ &     9879 & $0.0744\pm{0.0035}$ &     1339 \\
M82$^{b}$  & disk & $8.19\times3.81$ & wsrc &  31.435 & $1.1961\pm{0.0085}$ &    21524 & $0.5889\pm{0.0060}$ &    10597 & $0.1018\pm{0.0037}$ &     1832 \\
M82$^{b}$  & halo & $8.00\times15.50$ & diff &  35.144 & $0.2589\pm{0.0044}$ &     4658 & $0.0482\pm{0.0023}$ &      866 & $-0.0002\pm{0.0030}$ &       -4 \\
M82$^{b}$  & halo & $8.00\times15.50$ & wsrc &  35.426 & $0.2668\pm{0.0045}$ &     4800 & $0.0581\pm{0.0025}$ &     1045 & $0.0063\pm{0.0031}$ &      114 \\

NGC 1482 & nucl & 0.312 & diff &   0.016 & $0.0073\pm{0.0006}$ &      170 & $0.0077\pm{0.0006}$ &      180 & $0.0027\pm{0.0004}$ &       64 \\
NGC 1482 & nucl & 0.312 & wsrc &   0.020 & $0.0092\pm{0.0007}$ &      215 & $0.0120\pm{0.0008}$ &      280 & $0.0128\pm{0.0008}$ &      300 \\
NGC 1482 & disk & $2.19\times0.62$ & diff &   1.343 & $0.0359\pm{0.0014}$ &      843 & $0.0171\pm{0.0010}$ &      400 & $0.0039\pm{0.0008}$ &       91 \\
NGC 1482 & disk & $2.19\times0.62$ & wsrc &   1.343 & $0.0359\pm{0.0014}$ &      843 & $0.0171\pm{0.0010}$ &      400 & $0.0039\pm{0.0008}$ &       91 \\
NGC 1482 & halo & $1.00\times1.80$ & diff &   1.178 & $0.0099\pm{0.0008}$ &      231 & $0.0020\pm{0.0005}$ &       46 & $-0.0000\pm{0.0006}$ &       -1 \\
NGC 1482 & halo & $1.00\times1.80$ & wsrc &   1.178 & $0.0099\pm{0.0008}$ &      231 & $0.0020\pm{0.0005}$ &       46 & $-0.0000\pm{0.0006}$ &       -1 \\

NGC 253$^{c}$ & nucl & 2.65 & diff &   1.310 & $0.1816\pm{0.0042}$ &     1995 & $0.1532\pm{0.0039}$ &     1683 & $0.0406\pm{0.0022}$ &      446 \\
NGC 253$^{c}$ & nucl & 2.65 & wsrc &   1.374 & $0.2294\pm{0.0047}$ &     2520 & $0.2585\pm{0.0050}$ &     2839 & $0.1197\pm{0.0035}$ &     1315 \\
NGC 253$^{c}$ & disk & $20.42\times5.29$ & diff &  40.380 & $0.4296\pm{0.0069}$ &     4719 & $0.0865\pm{0.0037}$ &      951 & $0.0228\pm{0.0046}$ &      251 \\
NGC 253$^{c}$ & disk & $20.42\times5.29$ & wsrc &  40.993 & $0.5298\pm{0.0075}$ &     5820 & $0.2422\pm{0.0053}$ &     2661 & $0.1237\pm{0.0055}$ &     1359 \\
NGC 253$^{c}$ & halo & $12.00\times24.00$ & diff &  27.770 & $0.0895\pm{0.0037}$ &      983 & $0.0077\pm{0.0021}$ &       84 & $0.0029\pm{0.0037}$ &       32 \\
NGC 253$^{c}$ & halo & $12.00\times24.00$ & wsrc &  28.137 & $0.0950\pm{0.0038}$ &     1043 & $0.0111\pm{0.0022}$ &      122 & $0.0062\pm{0.0037}$ &       68 \\

NGC 253$^{d}$ & halo & $12.00\times24.00$ & diff & 69.1181 & $0.1755\pm{0.0030}$ &     6720 & $0.0194\pm{0.0019}$ &      744 & $0.0043\pm{0.0032}$ &      163 \\
NGC 253$^{d}$ & halo & $12.00\times24.00$ & wsrc & 69.9867 & $0.1962\pm{0.0031}$ &     7513 & $0.0328\pm{0.0020}$ &     1258 & $0.0146\pm{0.0032}$ &      559  \\

NGC 3628    & nucl & 0.69 & diff &   0.086 & $0.0018\pm{0.0002}$ &       98 & $0.0024\pm{0.0002}$ &      134 & $0.0012\pm{0.0002}$ &       66 \\
NGC 3628    & nucl & 0.69 & wsrc &   0.094 & $0.0020\pm{0.0002}$ &      107 & $0.0043\pm{0.0003}$ &      233 & $0.0037\pm{0.0003}$ &      201 \\
NGC 3628    & disk & $10.47\times1.37$ & diff &  11.267 & $0.0195\pm{0.0010}$ &     1068 & $0.0075\pm{0.0007}$ &      410 & $0.0056\pm{0.0010}$ &      305 \\
NGC 3628    & disk & $10.47\times1.37$ & wsrc &  11.485 & $0.0258\pm{0.0010}$ &     1409 & $0.0372\pm{0.0010}$ &     2034 & $0.0368\pm{0.0013}$ &     2013 \\
NGC 3628    & halo & $4.00\times7.00$ & diff &  17.842 & $0.0291\pm{0.0012}$ &     1592 & $0.0036\pm{0.0007}$ &      196 & $-0.0002\pm{0.0012}$ &      -12 \\
NGC 3628    & halo & $4.00\times7.00$ & wsrc &  17.901 & $0.0381\pm{0.0013}$ &     2085 & $0.0105\pm{0.0008}$ &      576 & $0.0044\pm{0.0013}$ &      241 \\

NGC 3079    & nucl & 0.40 & diff &   0.032 & $0.0050\pm{0.0005}$ &      131 & $0.0101\pm{0.0007}$ &      268 & $0.0071\pm{0.0006}$ &      187 \\
NGC 3079    & nucl & 0.40 & wsrc &   0.032 & $0.0050\pm{0.0005}$ &      131 & $0.0101\pm{0.0007}$ &      268 & $0.0071\pm{0.0006}$ &      187 \\
NGC 3079    & disk & $5.50\times0.80$ & diff &   4.266 & $0.0579\pm{0.0017}$ &     1549 & $0.0209\pm{0.0010}$ &      553 & $0.0032\pm{0.0010}$ &      86 \\
NGC 3079    & disk & $5.50\times0.80$ & wsrc &   4.390 & $0.0682\pm{0.0018}$ &     1809 & $0.0322\pm{0.0012}$ &      854 & $0.0105\pm{0.0011}$ &      279 \\
NGC 3079    & halo & $4.80\times6.60$ & diff &  23.274 & $0.0491\pm{0.0021}$ &     1301 & $0.0099\pm{0.0013}$ &      259 & $0.0026\pm{0.0020}$ &      69 \\
NGC 3079    & halo & $4.80\times6.60$ & wsrc &  23.401 & $0.0530\pm{0.0021}$ &     1406 & $0.0174\pm{0.0014}$ &      461 & $0.0096\pm{0.0021}$ &      256 \\

NGC 4945    & nucl & 1.86 & diff &   0.667 & $0.0124\pm{0.0008}$ &      356 & $0.0230\pm{0.0010}$ &      659 & $0.0432\pm{0.0013}$ &     1238 \\
NGC 4945    & nucl & 1.86 & wsrc &   0.681 & $0.0133\pm{0.0008}$ &      380 & $0.0240\pm{0.0010}$ &      688 & $0.0438\pm{0.0014}$ &     1254 \\
NGC 4945    & disk & $13.80\times3.72$ & diff &  31.143 & $0.0967\pm{0.0038}$ &     2771 & $0.0519\pm{0.0032 }$ &     1489 & $0.0662\pm{0.0055}$ &     1897 \\
NGC 4945    & disk & $13.80\times3.72$ & wsrc &  31.514 & $0.1347\pm{0.0040}$ &     3861 & $0.1750\pm{0.0038}$ &     5016 & $0.1712\pm{0.0058}$ &     4908 \\
NGC 4945    & halo & $8.20\times8.20$ & diff &  26.119 & $0.0057\pm{0.0029}$ &      165 & $0.0045\pm{0.0025}$ &      130 & $0.0003\pm{0.0046}$ &        8 \\
NGC 4945    & halo & $8.20\times8.20$ & wsrc &  26.249 & $0.0069\pm{0.0030}$ &      198 & $0.0058\pm{0.0025}$ &      167 & $0.0016\pm{0.0046}$ &       46 \\

NGC 4631   & nucl & 0.92 & diff &   0.122 & $0.0009\pm{0.0002}$ &       49 & $0.0006\pm{0.0001}$ &       31 & $0.0003\pm{0.0001}$ &       15 \\
NGC 4631   & nucl & 0.92 & wsrc &   0.165 & $0.0014\pm{0.0002}$ &       78 & $0.0026\pm{0.0003}$ &      145 & $0.0021\pm{0.0003}$ &      117 \\
NGC 4631   & disk & $10.47\times1.83$ & diff &  15.107 & $0.0808\pm{0.0016}$ &     4496 & $0.0175\pm{0.0009}$ &      977 & $0.0033\pm{0.0013}$ &      181 \\
NGC 4631   & disk & $10.47\times1.83$ & wsrc &  15.889 & $0.1120\pm{0.0017}$ &     6235 & $0.0600\pm{0.0013}$ &     3340 & $0.0431\pm{0.0016}$ &     2401 \\
NGC 4631   & halo & $8.20\times8.20$ & diff &  45.585 & $0.0571\pm{0.0019}$ &     3181 & $0.0030\pm{0.0012}$ &      164 & $0.0018\pm{0.0021}$ &       98 \\
NGC 4631   & halo & $8.20\times8.20$ & wsrc &  46.231 & $0.0673\pm{0.0020}$ &     3746 & $0.0135\pm{0.0013}$ &      754 & $0.0118\pm{0.0022}$ &      659 \\

NGC 6503    & nucl & 1.32 & diff &   0.323 & $0.0027\pm{0.0007}$ &       29 & $0.0008\pm{0.0005}$ &        8 & $0.0010\pm{0.0007}$ &       10 \\
NGC 6503    & nucl & 1.32 & wsrc &   0.343 & $0.0086\pm{0.0011}$ &       92 & $0.0082\pm{0.0010}$ &       88 & $0.0057\pm{0.0010}$ &       61 \\
NGC 6503    & disk & $5.58\times2.65$ & diff &  14.280 & $0.0177\pm{0.0030}$ &      188 & $0.0020\pm{0.0023}$ &       22 & $0.0076\pm{0.0042}$ &       81 \\
NGC 6503    & disk & $5.58\times2.65$ & wsrc &  14.323 & $0.0249\pm{0.0031}$ &      264 & $0.0173\pm{0.0026}$ &      184 & $0.0174\pm{0.0044}$ &      185 \\
NGC 6503    & halo & $5.00\times7.00$ & diff &  17.141 & $-0.0002\pm{0.0031}$ &      -2 & $0.0039\pm{0.0026}$ &       42 & $0.0006\pm{0.0047}$ &      6 \\
NGC 6503    & halo & $5.00\times7.00$ & diff &  17.141 & $<0.0068$           &    $<73$ & $<0.0095$           &   $<101$ & $<0.0112$           &     $<119$ \\
NGC 6503    & halo & $5.00\times7.00$ & wsrc &  18.294 & $0.0083\pm{0.0033}$ &       88 & $0.0154\pm{0.0029}$ &      164 & $0.0097\pm{0.0050}$ &      103 \\

NGC 891     & nucl & 0.72 & diff &   0.087 & $0.0003\pm{0.0002}$ &        8 & $0.0005\pm{0.0002}$ &       16 & $0.0001\pm{0.0002}$ &        4 \\
NGC 891     & nucl & 0.72 & wsrc &   0.101 & $0.0002\pm{0.0002}$ &        8 & $0.0006\pm{0.0002}$ &       18 & $0.0009\pm{0.0003}$ &       27 \\
NGC 891     & disk & $9.33\times1.43$ & diff &  12.727 & $0.0235\pm{0.0020}$ &      722 & $0.0041\pm{0.0016}$ &      125 & $-0.0074\pm{0.0027}$ &     -226 \\
NGC 891     & disk & $9.33\times1.43$ & wsrc &  13.190 & $0.0383\pm{0.0022}$ &     1177 & $0.0560\pm{0.0021}$ &     1720 & $0.0345\pm{0.0030}$ &     1059 \\
NGC 891     & halo & $4.40\times4.40$ & diff &  12.949 & $0.0117\pm{0.0020}$ &      360 & $0.0000\pm{0.0016}$ &        1 & $-0.0005\pm{0.0028}$ &      -17 \\
NGC 891     & halo & $4.40\times4.40$ & wsrc &  13.050 & $0.0204\pm{0.0020}$ &      627 & $0.0083\pm{0.0017}$ &      253 & $0.0033\pm{0.0028}$ &      101  \\

NGC 4244    & nucl & 1.90 & diff &   0.674 & $0.0003\pm{0.0002}$ &       16 & $0.0002\pm{0.0002}$ &        8 & $-0.0004\pm{0.0002}$ &      -18 \\
NGC 4244    & nucl & 1.90 & wsrc &   0.681 & $0.0006\pm{0.0002}$ &       27 & $0.0002\pm{0.0002}$ &        8 & $-0.0004\pm{0.0002}$ &      -18 \\
NGC 4244    & disk & $10.23\times3.81$ & diff &  30.685 & $0.0049\pm{0.0015}$ &      231 & $0.0023\pm{0.0012}$ &      110 & $0.0070\pm{0.0023}$ &      335 \\
NGC 4244    & disk & $10.23\times3.81$ & wsrc &  30.968 & $0.0123\pm{0.0016}$ &      583 & $0.0087\pm{0.0013}$ &      415 & $0.0112\pm{0.0023}$ &      530 \\
NGC 4244    & halo & $8.20\times8.20$ & diff &  33.428 & $0.0002\pm{0.0016}$ &       10 & $-0.0003\pm{0.0013}$ &      -14 & $-0.0004\pm{0.0024}$ &      -17 \\
NGC 4244    & halo & $8.20\times8.20$ & diff &  33.428 & $<0.0034$           &   $<159$ & $<0.0024$         &      $<112$ & $0.0045$           &     $<213$ \\
NGC 4244    & halo & $8.20\times8.20$ & wsrc &  33.658 & $0.0063\pm{0.0017}$ &      299 & $0.0057\pm{0.0014}$ &      269 & $0.0037\pm{0.0024}$ &      174 \\
\enddata
\tablecomments{Errors are $1\sigma$, calculated using the 
	\citet{gehrels86} approximation, and include the uncertainties 
	associated with the background subtraction. Upper limits are exact 
	Poissonian $3\sigma$ (99.73\% confidence) upper limits.
	Column 2: Nuclear (nucl), disk or halo region, as defined in
	\$~\ref{sec:spectra:regions}
	Column 3: Size of the spectral region in arcminutes. For nuclear
	regions the value quoted is the angular diameter 
        of the circular extraction region, corresponding to a 
	physical diameter of 2 kpc. For the disk and halo regions the
	values are the total $r$ and $z$-axis angular widths of the
	rectangular extraction regions. All regions are centered on the
	positions given in Table~\ref{tab:galaxies}. 
	The nuclear region
	is excluded from the disk region, and that both the nuclear and
	disk regions are excluded from the halo region. Note also that the
	final area used is the fraction of this geometrical region that
	falls within the boundaries of the field of view of the S3 chip. 
	Column 4: Count rates where the emission from point sources has been 
	excluded (\ie diffuse X-ray emission only) are noted as diff.
	Count rates including both diffuse and point source emission
	are denoted as wsrc.
	Column 5: Total resulting area, in units of arcmin$^{2}$.
	Columns 6 through 11: Background-subtracted soft energy band 
	count rate, count rate error, and total number of counts
	accumulated after background subtraction. Values are given for the
	soft, medium and hard energy bands.	
	}
\tablenotetext{a}{ACIS-I observations of M82, ObsIDs 361 and 1302. Data from ACIS-I chips.}
\tablenotetext{b}{ACIS-S observation of M82, ObsID 2933. Data from the S3 chip only.}
\tablenotetext{c}{NGC 253 observation ObsID 969, Data from S3 chip only.}
\tablenotetext{d}{NGC 253 observation ObsID 790, Data from S3 chip only.}
\end{deluxetable}

\section{Spectral properties of the diffuse emission}
\label{sec:spectra}

\subsection{Separation into nuclear, disk and halo components}
\label{sec:spectra:regions}

We created background, 
and point-source, subtracted 
ACIS spectra of the nucleus, the disk and the halo for each galaxy 
and observation thereof. In this paper we will concentrate primarily on
the faint extra-planar diffuse emission from the halo region,
although we do tabulate fluxes and luminosities of the diffuse X-ray
emission from all three regions for each galaxy.
These three regions are defined as:
\begin{itemize}
\item {\bf Nucleus:} a circular aperture 1 kpc in radius, centered
  on the nuclear position given in Table~\ref{tab:galaxies}.
\item {\bf Disk:} a rectangular aperture oriented along the major
  axis of the disk, of sufficient major-axis length
  to encompass all of the observed diffuse emission, and extending
  between $-2 \le z~(\kpc)\le 2$ along the minor-axis. The nuclear
  region defined above is {\em excluded} from the disk region.
\item {\bf Halo:} A rectangular aperture oriented along the minor
  axis of the galaxy, chosen to enclose all observed diffuse emission,
  but excluding the region with $|z| \le 2$ kpc of the plane of the
  galaxy (\ie excluding the disk and nuclear regions defined above).
\end{itemize}
For all ACIS-S observations (Table~\ref{tab:xray_obs_log}), 
only events from the back-illuminated
S3 chip that are also within the regions defined above were used to create
the X-ray spectra, as the spectral responses of the neighboring 
front-illuminated S2 and S4 chips are substantially different.
We used data from all four ACIS-I chips in creating spectra
from the ACIS-I observation of M82.

We have chosen a projected height of $z=2$ kpc as the 
dividing line between what we term
as the halo and disk regions, based on the minor-axis optical
extent of the galaxies, and that the X-ray spectral hardness profiles
we observe are relatively uniform for $|z| \ge 2$ kpc. 
This ensures negligible contamination of
diffuse X-ray emission seen within the halo region by hot gas 
actually within the disk of the galaxy. Note that this does {\em not}
automatically imply 
that X-ray emission in what we call the disk region is actually
within the plane of the galaxy. For the starburst galaxies, it is clear
that the majority of the diffuse X-ray emission in both the 
disk and nuclear regions is directly associated with outflowing gas.

The angular size of these regions, the area of sky covered, and the
observed background subtracted count rate for the nuclear,
disk and halo regions for all galaxies are given in 
Table~\ref{tab:spectral:rates_and_areas}, both with and without
point-source removal.

\begin{figure*}[!t]
\epsscale{2.0}
\plotone{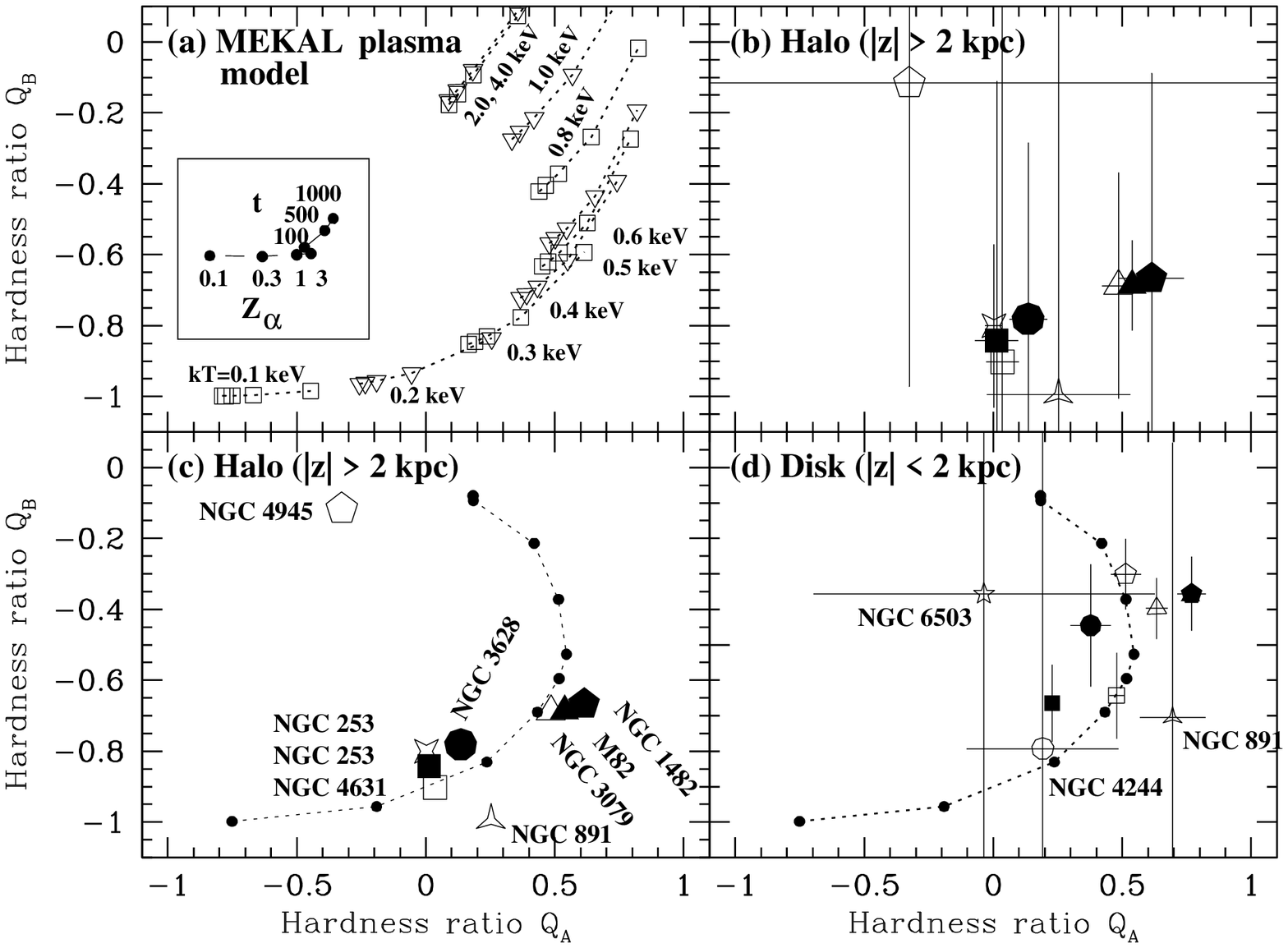}
\caption{Halo and disk diffuse X-ray emission hardness ratios, compared to
	hardness ratios from a MEKAL hot plasma code simulation. The hardness
	ratios $Q_A$ and $Q_B$ are defined in Equation~\ref{equ:qa_qb} 
	(see \S~\ref{sec:results:extent}). Only data from the back-illuminated
	S3 chip is used. 
	Panel (a) shows the hardness ratios expected from a single 
	temperature hot plasma in collisional ionization equilibrium,
	calculated using the MEKAL model. The plasma is assumed to have 
	Solar ratios of the alpha elements (\ie $Z_{\rm \sim O}=1 \Zsol$), 
	with the relative abundance of iron and other refractory elements
	being 1/3 Solar. The use of square and inverted triangular symbols
	is alternated to aid the reader in distinguishing between predictions
	for different temperature plasmas in regions where hardness
	ratios overlap.
	For each temperature, hardness ratios for
	hydrogen columns of $\nH = 0, 1, 3, 10$ and $30\times10^{20} \pcmsq$ 
	are shown. The lowest column always has the lowest value of 
	$Q_A$ and $Q_B$. The insert shows the magnitude of hardness ratio 
	changes due to different metal abundances ($Z_{\rm \sim Fe}$ is 
	always assumed to be 1/3 of the alpha-element abundance), 
	and due to the increasing contamination of the ACIS optical blocking
	filter ($t$ is days since launch).
	Panel (b) displays the observed S3-chip hardness ratios for the
	diffuse emission within the halo region 
	(defined in \S~\ref{sec:spectra:hardness}), with $1\sigma$ error 
	bars. A different symbol is used for each galaxy --- the same 
	symbol is used in panels c and d.
	Panel (c) displays the same data, sans error bars. Individual
	galaxies are identified by name. Predicted hardness ratios for a
	single temperature plasma with temperatures (from lower left) 
	$kT = 0.1, 0.2, 0.3, 0.4, 0.5, 0.6, 0.8, 1.0, 2.0$ and 4.0 keV, 
	$Z_{\rm \sim O}=1 \Zsol$, $Z_{\rm \sim Fe}=0.33 \Zsol$ and  
	$\nH = 3 \times 10^{20} \pcmsq$ are shown as the connected set
	of small circles.
	Panel (d) shows the observed S3-chip hardness ratios, with errors,
	for the diffuse emission found within $z \le 2$ kpc the disk. 
	The connected set of small circles represent the same
	predicted hardness ratios as used in panel c.}
\label{fig:hardness_ratios}
\end{figure*}

\subsection{Hardness ratio analysis}
\label{sec:spectra:hardness}

To provide a quantitative cross-comparison of the spectra, the diffuse
emission $Q_A$ and $Q_B$ hardness ratios for both halo and disk
regions are shown in Fig.~\ref{fig:hardness_ratios}\footnote{For 
the more distant
nuclear starburst galaxies in this sample (\eg NGC 3628, NGC 3079 
or NGC 1482), it is impossible to achieve point source removal in the nuclear
regions to same depth as that obtained in the nearer galaxies such as M82 or 
NGC 253, \eg see Table~\ref{tab:xray_bgprops}. For this reason
we have not included nuclear region hardness ratios in 
Fig.~\ref{fig:hardness_ratios}}.

The spectral hardness of the diffuse halo emission 
of all the galaxies, both the starburst galaxies
(with the exception of the poorly-constrained halo spectrum of NGC 4945) and
NGC 891, are generally very similar to each other. M82, NGC 1482 and
NGC 3079 are spectrally harder than the other galaxies, but from this
form of analysis it is impossible to distinguish whether this is
due to a higher mean plasma temperature, or to higher intrinsic absorption.

Disk diffuse emission is significantly spectrally harder than the
halo emission, as mentioned earlier. Explaining this as entirely
due to increased absorption requires hydrogen columns 
$\nH \sim 3 \times 10^{21} \pcmsq$ (as can be seen from
Fig.~\ref{fig:hardness_ratios}). This is reasonable
given we would expect some fraction of the diffuse X-ray emission to
experience absorption 
by gas and dust in the highly inclined disk, and consistent
with the X-ray measurement of the absorption column through 
an edge-on spiral galaxy disk by \citet{clarke03}.

\subsection{A single common spectral model for all halo emission}
\label{sec:spectra:fits}

Given the apparent spectral similarity of the halo diffuse emission
spectra, it is worth asking if they are, in fact, spectrally identical?
To answer this question we simultaneously fit all 
the halo diffuse emission spectra with a two temperature 
MEKAL \citep*{mekal} hot plasma model. This approach is  
motivated by the results from many {\it ROSAT} PSPC and {\it ASCA}
X-ray spectroscopy studies (to name but a few, 
\citealt*{dellaceca96,dellaceca97,ptak97,tsuru97};
\citealt{dwh98}; \citealt*{zezas98,moran99}), 
and our detailed {\it Chandra}
study of the diffuse halo emission from NGC 253 \citep{strickland02}.
These studies found that the X-ray spectra of starburst galaxies
are usually best-fit by low temperature ($kT \la 0.2$ keV)
and a medium temperature ($kT \sim 0.6$ keV) thermal plasma (in addition
to a spectrally hard power law component representing X-ray emission from
X-ray binaries and low luminosity AGN, in the {\it ROSAT} PSPC 
and/or {\it ASCA} observations
which lack the spatial resolution to remove point source emission).

Our aim with this approach is {\em not} to find a set of models
that perfectly fit each of the halo spectra, but rather to see if all of the 
halo spectra can be fit by the same spectral model, and if not,
in what ways does this approach fail? In reality, the halo plasma
is unlikely to be in the ionization equilibrium state assumed
by the MEKAL model \citep[see the discussion in][]{ss2000}. Similarly, 
although two temperature models fit
well, a more complex range of temperatures is likely to exist.
These, indeed all, best-fit X-ray spectral models of diffuse
emission in star-forming galaxies, should only be interpreted as empirical 
characterizations of the spectra, and not as representing the true
physical state of the emitting plasma.

\subsubsection{Spectral model description}

The basic model we use is a two-temperature, variable-abundance 
{\sc Mekal} hot plasma model \citep{mekal}, incorporating the effects
of foreground absorption (using the Solar abundance 
photoelectric absorption model of \citealt{morrison83}) and
the molecular contamination of the ACIS optical blocking filters
(using version 1.1 of the {\sc Acisabs} model by G.~Chartas \& 
K.~Getman, and the observation dates given in Table~\ref{tab:xray_obs_log}).
In terms of the syntax used in the {\sc Xspec} spectral fitting
program, the model specification is {\tt acisabs*wabs*(vmekal+vmekal)}.

Model parameters that are fit for, and which are forced to have the same value 
in each spectrum, are the two plasma temperatures
$kT_{\rm S}$ and $kT_{\rm M}$, and elemental abundances with respect to 
hydrogen. In other words, the best-fit values of these parameters
are those that provide maximum goodness-of-fit to the collective
ensemble of spectra. For each individual spectrum the
normalization of each of the two thermal components 
($K_{\rm S}$ and $K_{\rm M}$), and the equivalent hydrogen
column density of foreground absorption $\nH$ 
each spectrum experiences, are not tied to a 
universal, collectively-determined value, but instead are 
free to assume any best-fit value. As we do not expect large amounts of
intrinsic absorption within the halos of these galaxies, we constrain
the fitted column to lie within the range from 
half to three times the Galactic foreground
absorption value (given in Table~\ref{tab:xray_bgprops}).

\subsubsection{Metal abundance fitting}

In general, the strongest spectral features visible
in {\it Chandra} ACIS spectra of star-forming galaxies are
unresolved emission line complexes due from highly ionized
O and Fe, and occasionally-visible weaker spectral emission features from
Mg, Ne, Si and S. Visual inspection of the many of the halo-region
diffuse emission spectra immediately revealed
that the strength of the Fe L-shell complex at $E\sim 0.8$--$1.0$ keV 
with respect to \ion{O}{7}/\ion{O}{8} emission
$E \sim 0.53$--$0.65$ keV is less than would be expected from a
plasma with Solar heavy element abundance ratios\footnote{We
use as Solar the abundance ratios given by \citet{anders89},
as this abundance set is most commonly employed within the 
X-ray community, and have converted any other abundances quoted
to this abundance scale. Note that more recent studies give 
significantly lower values for the Solar abundance of
both O and Fe with respect to hydrogen,
by 0.20 and 0.16 dex respectively \citep{grevesse98,holweger01}.}.
We therefore allow for non-Solar ratios of O to Fe in the
spectral fit. Given the weakness of spectral features due to other
elements, we constrain to abundance of Mg, Si, Ne, S, Ar, C and N 
with respect to hydrogen to track the fitted O abundance. This value,
now with respect to the Solar value, 
we will refer to as the approximate oxygen abundance $Z_{\rm \sim 0}$.
We similar constrain the abundances of Ni, Ca, Al and Na to track the
Fe abundance, and refer to this value as the approximate iron abundance
$Z_{\rm \sim Fe}$.

\subsubsection{Other spectral fitting details}

Fitting is only performed to data within the energy range 0.3 -- 3.0 keV
(slightly different energy ranges were used for 
the ACIS-I spectrum of M82 and the ACIS-S spectra of
NGC 253, of 0.5--3.0 keV and 0.3--2.0 keV respectively).
This approach is conservative, in the sense that had we included 
the spectral data at higher energies in the fit, we would have 
obtained {\emph{lower} values of best-fit reduced chi-squared 
given the very low 2 -- 8 keV count rates in the halo regions 
(see Table~\ref{tab:spectral:rates_and_areas}).

The spectral model was initialized by choosing parameter values
deemed reasonable, based on the discussion given above. Temperatures
were initialized at $kT_{\rm S} = 0.2$ and $kT_{\rm M} = 0.7$ keV,
the foreground hydrogen column at the Galactic value 
appropriate for each galaxy (see Table~\ref{tab:xray_bgprops}), and
the metal abundances at $Z_{\rm \sim O} = 1.00 \Zsol$
and $Z_{\rm \sim Fe} = 0.33 \Zsol$. Model normalizations
$K_{\rm S}$ and $K_{\rm M}$ were chosen so as to provide the
best fit to the observed spectra with all other parameters fixed
at their initial values. Rather than immediately fitting for all of the
model parameters, we initially only fit for the model normalizations
and temperatures, leaving other values at their initial values. 
Once this fit had converged, we let the metal abundance values to also be
determined by the fit. Finally we allowed the hydrogen column density
to assume a best-fit value.

The minimization technique employed within the {\sc Xspec} fitting
package is the standard Levenberg-Marquardt method. To avoid 
the fit getting trapped in local minima of the multi-dimensional 
fit-space (a common
occurrence with this method),
we perform a full confidence region calculation for all
model parameters after each invocation of the
fit command. 
This has the practical effect of extracting the 
fit from shallow local minima, as it ensures
that the final best-fit values lie in a valley that
is at least $\Delta \chi^{2} \ge 42.6$ deep (90\% confidence for
32 interesting parameters) in fit-space
for all parameters.
We also performed the fitting using the maximum likelihood C-statistic,
but obtained statistically identical results.

\begin{figure*}[!ht]
\epsscale{2.0}
\plotone{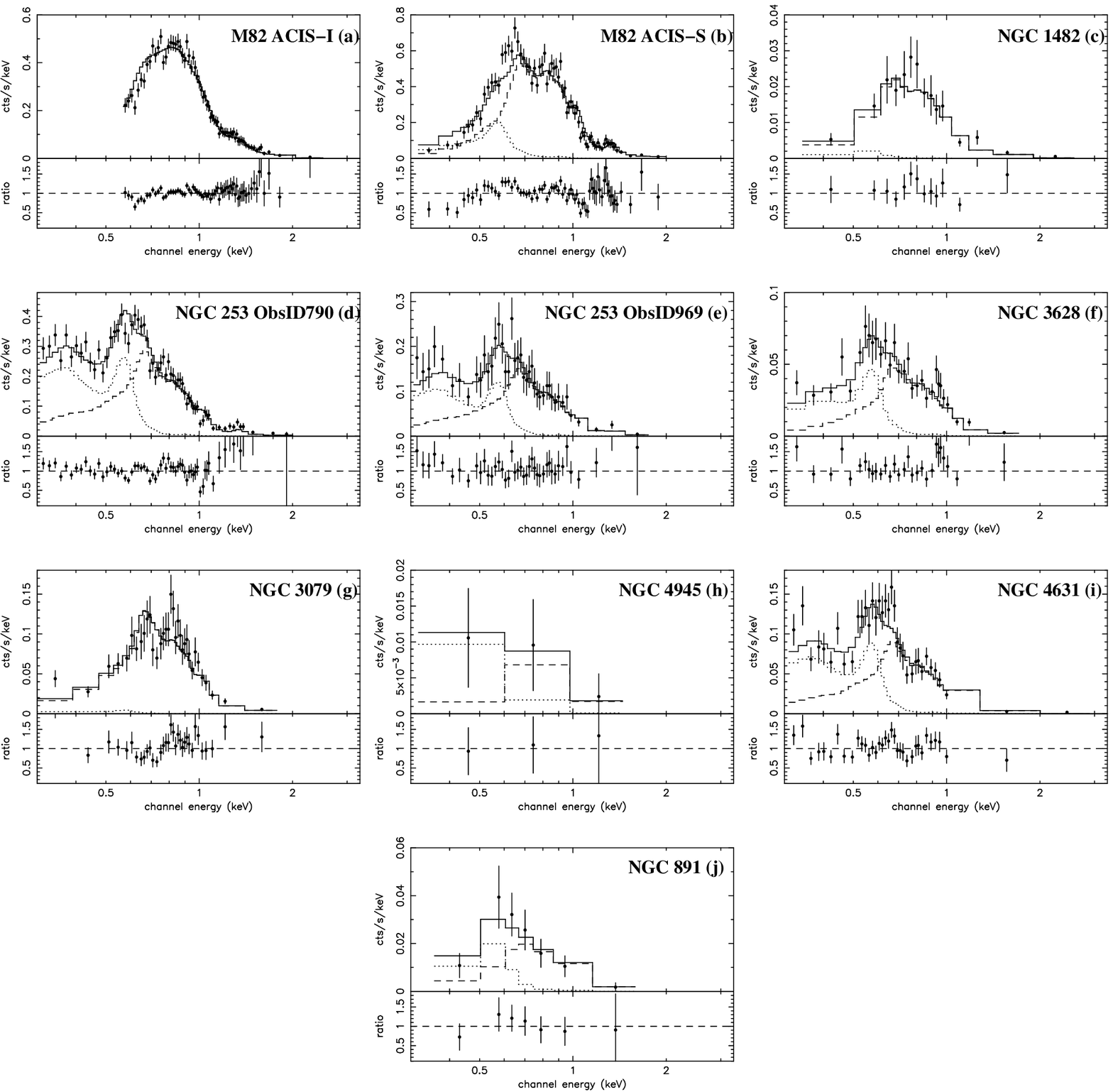}
\caption{{\it Chandra} ACIS X-ray spectra of the 
 diffuse X-ray emission from the halo
 regions ($|z| \ge 2$ kpc) of the sample galaxies. The top of each panel
 shows the observed spectrum (solid points with $1\sigma$ error bars),
 along with a solid line showing 
 the best-fit two thermal component model (simultaneously fit to
 all the spectra --- see \S~\ref{sec:spectra:fits}). 
 The individual $kT=0.37$ keV and $kT=0.11$ keV thermal components
 are represented by the dashed and dotted lines respectively.
 The lower part of each panel is the ratio of the data to the model.}
\label{fig:spectral_halo_diffuse_fits}
\end{figure*}

\begin{figure*}[!ht]
\epsscale{2.0}
\plotone{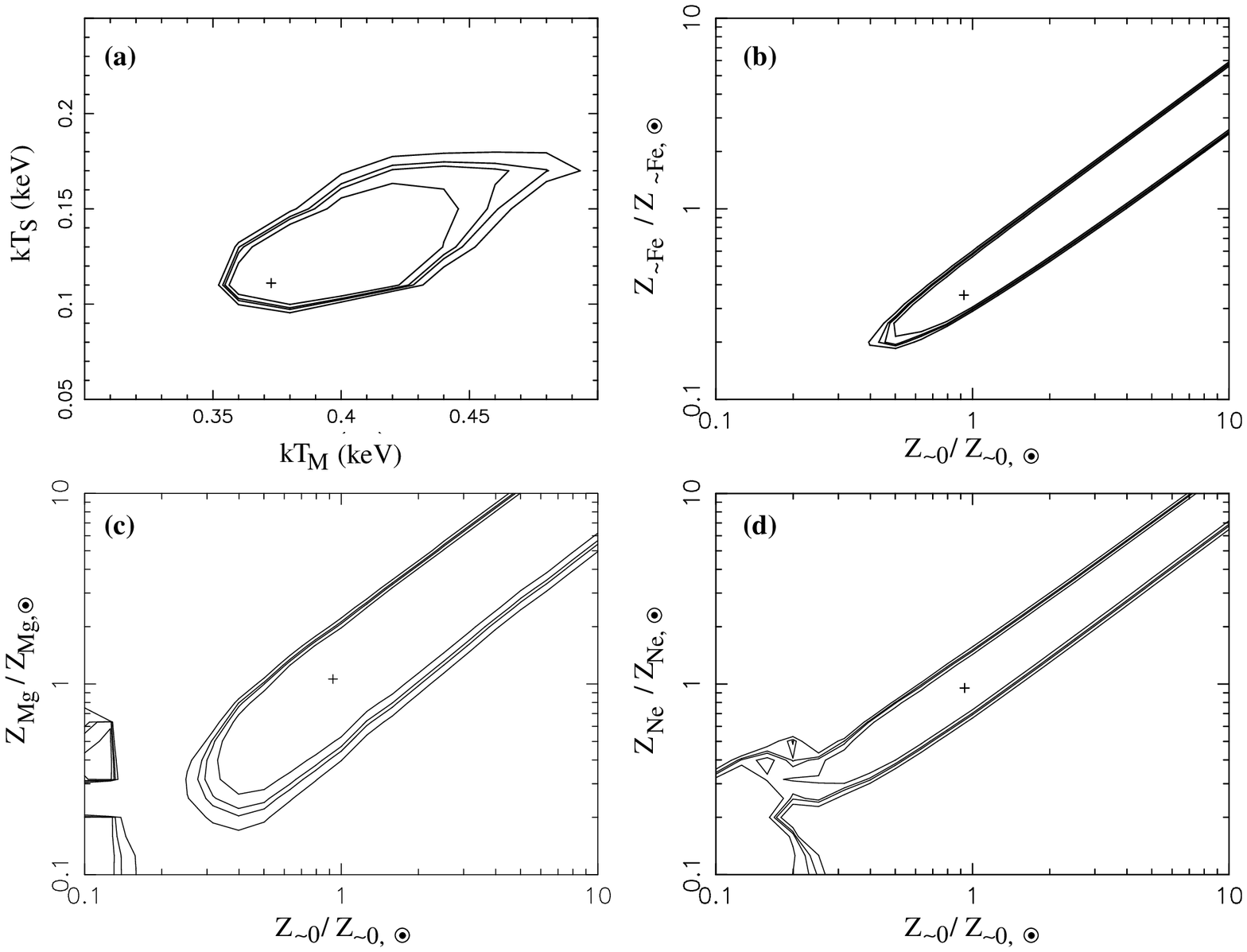}
\caption{Two-dimensional confidence regions from the combined
best-fit spectral model for the diffuse halo emission, 
for the temperatures of the two hot plasma components (panel a) 
and the abundance measures $Z_{\sim 0}$ and $Z_{\sim Fe}$ (panel b). 
The best-fit
value is marked as a cross. Solid contours are shown at
$\Delta \chi^{2}$ of 35.2, 42.6, 46.2 and 53.5 from the best-fit value 
(corresponding to the 68.3, 90.0, 95.0 and 99\%
confidence regions for 32 interesting parameters).
In panels c and d the confidence regions are of the magnesium abundance,
and the neon abundance, with respect to oxygen and the other 
$\alpha$-elements respectively. Again contours corresponding to
68.3, 90.0, 95.0 and 99\% confidence are plotted, now for 33 interesting
parameters.}
\label{fig:confidence_contours}
\end{figure*}

\begin{deluxetable}{lrrrrrrrr}
 \tabletypesize{\scriptsize}%
\tablecolumns{9} 
\tablewidth{0pc} 
\tablecaption{Halo diffuse X-ray emission spectral fit results and mean surface brightness
	\label{tab:spectral:fits}} 
\tablehead{ 
\colhead{Galaxy}
	& \colhead{$\nH$} & \multicolumn{2}{c}{Normalizations}
	& \colhead{$<kT>$}
	& \colhead{$\chi^{2}_{\nu}$} & \colhead{$\nu$} & \multicolumn{2}{c}{$\Sigma_{\rm HALO}$} \\
\colhead{} & \colhead{}
	& \colhead{$K_{S}$} & \colhead{$K_{M}$}
	& \colhead{} 
	& \colhead{} & \colhead{} 
	& \colhead{0.3--1.0 keV} & \colhead{1.0--2.0 keV} \\
\colhead{(1)} & \colhead{(2)} & \colhead{(3)} & \colhead{(4)} 
	& \colhead{(5)} & \colhead{(6)} & \colhead{(7)} & \colhead{(8)} 
	& \colhead{(9)}
	}
\startdata
M82\tablenotemark{a} & 12.0(L)
	& 0.00(L) (<19.26) & $17.05_{-13.79}^{+28.31}$ & 0.37
	& 1.40 & 77 & $2.88E{-09}$ &   $5.12E{-10}$ \\
M82\tablenotemark{b} & 12.0(L)
	& $17.75_{-10.97}^{+23.44}$ & $11.57_{-8.25}^{+12.16}$ & 0.27 
	& 2.35 & 76 & $4.36E{-09}$ &   $8.06E{-10}$ \\
NGC 1482 & 8.79
	& $0.16_{-0.16}^{+1.71}$ & $0.41_{-0.26}^{+1.08}$ & 0.34 
	& 1.26 & 15 & $4.57E{-09}$ & $9.01E{-09}$ \\
NGC 253\tablenotemark{c} & 0.70(L)
	& $2.51_{-1.76}^{+8.40}$ & $1.28_{-0.52}^{+2.69}$ & 0.25 
	& 0.84 & 47 & $2.82E{-09}$ &   $1.49E{-10}$ \\
NGC 253\tablenotemark{d} &  0.70(L)
	& $5.39_{-2.24}^{+16.01}$ & $2.39_{-1.47}^{+5.38}$ & 0.25 
	& 1.78 & 76 & $2.28E{-09}$ &   $1.13E{-10}$ \\
NGC 3628 &  5.05
	& $1.81_{-1.55}^{+3.07}$ & $0.61_{-0.27}^{+1.53}$ & 0.23
	& 1.19 & 57 & $1.19E{-09}$ &   $9.81E{-11}$ \\
NGC 3079 &  2.4(L)
	& $0.15_{-0.15}^{+1.69}$ & $1.46_{-0.48}^{+2.55}$ & 0.36 
	& 1.17 & 53 & $1.30E{-09}$ &   $1.86E{-10}$ \\
NGC 4945 &  13.9
	& $0.99_{-0.99}^{+122.47}$ & $0.14_{-0.14}^{+4.73}$ & 0.18 
	& 0.49 & 7 & $1.80E{-10}$ &   $1.38E{-11}$ \\
NGC 4631 &  0.65(L)
	& $2.03_{-1.02}^{+8.05}$ & $0.81_{-0.20}^{+1.72}$ & 0.24
	& 1.28 & 58 & $1.08E{-09}$ &   $5.70E{-10}$ \\
NGC 6503 & \nodata 
	& \nodata & \nodata & \nodata
	& \nodata & \nodata & <$3.57E{-10}$ &   <$2.22E{-10}$ \\
NGC 891 &  7.80
	& $0.99_{-0.99}^{+15.70}$ & $0.31_{-0.31}^{+1.11}$ & 0.23 
	& 0.54 & 21 & $7.18E{-10}$ &   $6.84E{-11}$ \\
NGC 4244 & \nodata
	& \nodata & \nodata & \nodata 
	& \nodata & \nodata & <$9.14E{-11}$ &   <$2.88E{-11}$ \\
\enddata
\tablecomments{Results from the two temperature {\sc mekal} hot plasma
	model fit made simultaneously to all the halo region spectra. See
	\S~\ref{sec:spectra:fits} for further details.
 	Other best-fit model parameters are 
	$kT_{S}=0.11^{+0.03}_{-0.02}$ keV,
	$kT_{M}=0.37^{+0.03}_{-0.03}$ keV, 
	$Z_{\rm \sim O}=0.94^{+19.1}_{-0.64}\Zsol$,
	$Z_{\rm \sim Fe}=0.35^{+10.8}_{-0.22}\Zsol$ (the ratio 
	$Z_{\rm \sim O}/Z_{\rm \sim Fe}=2.69^{+0.64}_{-1.02}$ is better constrained)
	with a final reduced chi-squared value of $\chi^{2}_{\nu}=1.32$
        ($\chi^{2}=706.6$, number of degrees of freedom $\nu=534$).
  Unless specified otherwise below, the quoted errors 
  represent a $\Delta \chi^{2} = 42.58$ (90\% 
  confidence for 32 parameters of interest). Model parameters marked as (L) 
	represent a lower or upper limit imposed upon the fit.
	Column 2: Best-fit column density (assuming Solar abundances) in
	units of $10^{20} \pcmsq$.
	Columns 3 and 4: Model normalizations for the soft and medium
	temperature thermal components. The normalization
	$K = 10^{-10} EI/4\pi D^{2}$, where the volume emission integral
	$EI = \eta_{\rm X} n_{\rm e} n_{\rm H} V$ and $D$ the source
	distance in cm.
	Column (5): The X-ray luminosity weighted mean temperature in keV,
	<$kT$>$=(kT_{S} f_{X, S} + kT_{M} f_{X, M}) / (f_{X, S} + f_{X, M})$,
	where $f_{X, S}$ and $ f_{X, M}$ are the absorption-corrected
	X-ray fluxes produces by the two thermal components in the
	0.3--2.0 keV energy band.
	Columns 6 and 7: Reduced $\chi^{2}_{\nu}$ value and number of
	degrees of freedom, if the best-fit model was applied to this
	alone.
	Columns 8 and 9: Halo region X-ray mean surface brightness 
	(\photsurfb) derived from the ACIS spectra,
	in the 0.3--1.0 and 1.0--2.0 keV energy bands,
	averaged over the area given in 
	Table~\ref{tab:spectral:rates_and_areas}. These surface
	brightnesses have the same physical meaning as the surface
	brightness values derived from the ACIS images discussed in
	\S~\ref{sec:xray_and_optical_images}, \ie an estimate of the 
	X-ray photon flux in that energy band
	immediately {\em prior} to passing through the telescope's optics.                     
	}
\tablenotetext{a}{ACIS-I observations of M82, ObsIDs 361 and 1302. 
	Data from ACIS-I chips, covering
	the majority of the extra-planar diffuse emission, 
	but not extending as far as the northern
	cloud.}
\tablenotetext{b}{ACIS-S observation of M82, ObsID 2933. Data from the S3 chip only, which does not
	cover the full extent of the diffuse emission.}
\tablenotetext{c}{NGC 253 observation ObsID 969. Only a small 
	fraction of the Northern halo lies on the S3 chip.}
\tablenotetext{d}{NGC 253 observation ObsID 790, Data from S3 chip only, 
        covering most of the
	bright diffuse emission in the Northern halo.}
\end{deluxetable}

\subsubsection{Results of the halo-region combined spectral fitting}

We find, somewhat to our surprise, that it is possible to obtain
a reasonably acceptable model (reduced chi-squared 
$\chi^{2}_{\nu}=1.32$, with 534 degrees of freedom), provided that
the relative normalization of the soft and medium temperature thermal
components is not the same in all of the galaxies. The ACIS spectra, 
and the simultaneously-fit two
temperature model, are shown in Fig.~\ref{fig:spectral_halo_diffuse_fits}.
The best-fit halo model parameters are given in
Table~\ref{tab:spectral:fits}. Considered on a galaxy-by-galaxy basis,
this model does surprisingly well, yielding statistically acceptable fits
in several cases, although this is partially due to poorer photon statistics
in some of the better fits.

We do not wish to place undue significance on the best fit model
parameters, and hence will not convert emission integrals into gas
densities and masses. We note that the temperatures of the 
two thermal components ($kT_{\rm S} \sim 0.1$ keV 
and $kT_{\rm M} \sim 0.4$ keV, 
see Table~\ref{tab:spectral:fits}) are lower than
$kT \sim 0.25$ and $\sim0.7$ keV components typically found in two
temperature {\it ROSAT} and {\it ASCA} fits to the total X-ray spectra of
starbursts by \citet{dwh98}. This is not surprising,
as we are fitting only the halo-region emission, and the
hardness ratio profiles demonstrate that the halo emission is
spectrally softer than the brighter disk region (which
would have dominated the 
combined {\it ROSAT} and {\it ASCA} spectral fitting).

The metal abundances with respect to hydrogen are 
poorly constrained from the ACIS X-ray spectra 
(Fig.~\ref{fig:confidence_contours}b), as they
lack the spectral resolution necessary to resolve lines from
the continuum. Our model $\alpha$-to-iron
element ratio is significantly better constrained,
$Z_{\rm \sim O}/Z_{\rm \sim Fe}=2.69^{+0.64}_{-1.02}$ (errors are 68\% confidence
for 32 parameters of interest), statistically identical to the
value \citet{martin02} find for the diffuse X-ray emission of the
low-metallicity dwarf starburst galaxy NGC 1569.
Spectra of X-ray emission with apparently enhanced $\alpha$-to-iron 
element ratios in starburst galaxies were first
reported by \citet{ptak97} and \citet{tsuru97} based on the
{\it ASCA} spectroscopy of the entire emission from M82 and NGC 253, 
but these newer {\it Chandra} observations finally allow us to be
confident that they also apply to the diffuse extra-planar emission.

Although the physical processes behind the X-ray
emission are of interest in themselves, it is the gross radiative
cooling rate that is of importance in assessing the efficiency with
which superwinds can transport energy and metals out of galaxies. 
The model-derived 
absorption-and-contamination-corrected X-ray fluxes and luminosities
for the halo region of the sample galaxies
are given in Table~\ref{tab:fluxes}.
While the raw halo flux is a reasonably
robust value irrespective of the specific spectral model (provided the 
model reproduces the observed count rate reasonably well), the
absorption-and-contamination-corrected fluxes are model dependent. 
We estimate this systematic uncertainty (\ie not knowing what 
the true emission process is, and over-simplified modeling of probably
patchy absorption for the disk regions) makes the 
intrinsic, absorption corrected, fluxes $f_{\rm ABSC}$ quoted 
uncertain by a factor $\sim2$.  
This is likely to dominate over statistical uncertainties,
which is why we have not attempted to calculate the formal 
statistical uncertainties associated with the fitted spectral models.

\begin{deluxetable}{lrrrrrrrrrrrrrrrr}
 \tabletypesize{\scriptsize}%
\rotate
\tablecolumns{17} 
\tablewidth{0pc} 
\tablecaption{Halo, disk, nuclear and total diffuse thermal 
	X-ray fluxes and luminosities
	\label{tab:fluxes}} 
\tablehead{
	\colhead{Galaxy} & 
	\multicolumn{5}{l}{Halo region X-ray flux and luminosity} & 
	\multicolumn{4}{l}{Disk region X-ray flux and luminosity} & 
	\multicolumn{4}{l}{Nuclear region X-ray flux and luminosity} &
	\multicolumn{3}{l}{Total X-ray flux and luminosity}  \\
	& \colhead{$f_{\rm Raw}$} & 
	\colhead{$f_{\rm OBFC}$} & \colhead{$f_{\rm ABSC}$} & 
	\colhead{$L_{\rm X,H}$} & \colhead{$f_{X,H}/f_{\rm FIR}$} & 
	\colhead{$f_{\rm OBFC}$} & \colhead{$f_{\rm ABSC}$} & 
	\colhead{$L_{\rm X,D}$} & 
	\colhead{$f_{X,D}/f_{\rm FIR}$} &
	\colhead{$f_{\rm OBFC}$} & \colhead{$f_{\rm ABSC}$} & 
	\colhead{$L_{\rm X,N}$} & 
	\colhead{$f_{X, N}/f_{\rm FIR}$} &
 	\colhead{$f_{\rm ABSC}$} & \colhead{$L_{\rm X,T}$} & 
	\colhead{$f_{X, T}/f_{\rm FIR}$} \\
	\colhead{(1)} & 
	\colhead{(2)} & \colhead{(3)} & \colhead{(4)} & \colhead{(5)} &
	\colhead{(6)} & \colhead{(7)} & \colhead{(8)} &
	\colhead{(9)} & \colhead{(10)} & 
	\colhead{(11)} & \colhead{(12)} &
	\colhead{(13)} & \colhead{(14)} &
	\colhead{(15)} & \colhead{(16)} & \colhead{(17)} 
}
\startdata
M82\tablenotemark{a} & 1.4E{-12}  & 1.4E{-12}  & 2.7E{-12}
	& 4.1E{39} & 4.43E{-5}  
	& 4.0E{-12} & 1.0E{-11} 
	& 1.6E{40} & 1.73E{-4} 
	& 3.4E-12  &  1.5E-11  
	&  2.3E40  &  2.47E-4  
	&  2.8E-11  &  4.3E40  &  4.62E-4  \\
M82\tablenotemark{b} & 8.0E{-13}  & 1.2E{-12}  & 2.9E{-12}
	& 4.5E{39} & 4.82E{-5}
	& 5.9E{-12} & 1.8E{-11} 
	& 2.8E{40} & 2.97E{-4}  
	& 3.1E-12  &  1.8E-11   
	&  2.8E40  &  3.00E-4  
	&  3.9E-11  &  6.0E40  &  6.45E-4 \\
NGC 1482 & 2.9E{-14}  & 4.3E{-14}  & 7.4E{-14}
	& 4.3E{39} & 4.28E{-5}  
	& 1.6E{-13} & 4.8E{-13} 
	& 2.8E{40} & 2.78E{-4}  
	&  4.7E-14  &  7.4E-14  
	&  4.3E39  &  4.27E-5  
	&  6.3E-13  &  3.7E40  &  3.64E-4 \\
NGC 253\tablenotemark{c} & 2.7E{-13}  & 3.3E{-13}  & 3.5E{-13}
	& 2.9E{38} & 6.51E{-6} 
	& 1.2E{-12} & 4.1E{-12} 
	& 3.3E{39} & 7.57E{-5}  
	&  6.9E-13  &  1.2E-12  
	&  9.4E38  &  2.15E-5 
	&  \nodata  &  \nodata  &  \nodata \\
NGC 253\tablenotemark{d} & 5.3E{-13}  & 6.5E{-13} & 7.0E{-13}
	& 5.7E{38} & 1.30E{-5} 
	& \nodata & \nodata 
	& \nodata & \nodata  
	&  \nodata  &  \nodata  
	&  \nodata  &  \nodata  
	&  \nodata  &  \nodata  &  \nodata \\
NGC 253\tablenotemark{e} & \nodata  & \nodata  & 1.5E{-12}
	& 1.2E{39} & 2.72E{-5} 
	& 1.2E{-12} & 4.1E{-12}
	& 3.3E{39} & 7.57E{-5}
	&  6.9E-13  &  1.2E-12  
	&  9.4E38  &  2.15E-5  
	&  6.8E-12  &  5.5E39  &  1.25E-4 \\
NGC 3628 & 8.4E{-14}  & 1.3E{-13}  & 2.1E{-13}
	& 2.5E{39} & 6.65E{-5} 
	& 6.4E{-14} & 2.2E{-13} 
	& 2.6E{39} & 7.14E{-5}  
	&  9.0E-15  &  2.0E-14  
	&  2.4E38  &  6.46E-6  
	&  4.4E-13  &  5.3E39  &  1.44E-4 \\
NGC 3079 & 1.5E{-13}  & 2.0E{-13}  & 2.4E{-13}
	& 8.3E{39} & 8.05E{-5} 
	& 2.4E{-13} & 6.7E{-13} 
	& 2.3E{40} & 2.26E{-4}  
	&  1.4E-14  &  2.9E-14  
	&  1.0E39  &  9.90E-6  
	&  9.2E-13  &  3.2E40  &  3.14E-4 \\
NGC 4945 & 1.8E{-14}  & 2.2E{-14}  & 8.2E{-14}
	& 1.3E{38} & 2.22E{-6} 
	& 2.9E{-13} & 1.8E{-12} 
	& 2.9E{39} & 4.80E{-5}  
	&  9.0E-14  &  2.2E-13  
	&  3.7E38  &  6.10E-6  
	&  2.1E-12  &  3.4E39  &  5.62E-5 \\
NGC 4631 & 1.7E{-13}  & 2.3E{-13}  & 2.5E{-13}
	& 1.7E{39} & 4.81E{-5} 
	& 2.9E{-13} & 9.0E{-13} 
	& 6.1E{39} & 1.73E{-4}  
	&  6.2E-15  &  9.2E-15  
	&  6.2E37  &  1.78E-6  
	&  1.2E-12  &  7.8E39  &  2.23E-4 \\
NGC 6503\tablenotemark{f} & \nodata  & \nodata  & $<$4.9E{-14}
	& $<$1.6E{38} & $<$7.03E{-5}
	& 4.8E{-14} & 1.9E{-13} 
	& 6.0E{38} & 2.68E{-4}  
	&  1.5E-14  &  3.4E-14  
	&  1.1E38  &  4.91E-5  
	&  2.2E-13  &  7.1E38  &  3.17E-4 \\
NGC 891 & 3.7E{-14}  & 5.3E{-14}  & 1.1E{-13}
	& 1.2E{39} & 2.41E{-5} 
	& 9.1E{-14} & 4.2E{-13} 
	& 4.7E39 & 9.44E-5
	&  4.6E-15  &  1.1E-14  
	&  1.2E38  &  2.42E-6  
	&  5.4E-13  &  6.0E39  &  1.21E-4 \\
NGC 4244\tablenotemark{f} & \nodata  & \nodata  & $<$1.6E{-14}
	& $<$2.5E{37} & $<$4.67E{-5}
	& 1.4E{-14} & 3.4E{-14} 
	& 5.3E{37} & 9.91E{-5}  
	&  3.7E-15  &  7.7E-15  
	&  1.2E37  &  2.25E-5  
	&  4.1E-14  &  6.5E37  &  1.22E-4 \\
\enddata
\tablecomments{All X-ray fluxes and luminosities quoted are in the
	0.3--2.0 keV energy band, and refer to the diffuse thermal
	emission components of the best-fit
	spectral models alone.
	Columns 2, 3 and 4: Model-derived halo-region diffuse 
	X-ray flux, in units of
	$\ergps \pcmsq$ in the 0.3--2.0 keV energy band. No corrections
	for the absorption or the contamination of the ACIS optical
	blocking filter have been made to $f_{\rm Raw}$. Estimated 
	fluxes corrected 
	for the optical blocking filter contamination are given in
	column 3 ($f_{\rm OBFC}$), while 
	absorption-and-contamination-corrected fluxes are given in column
	4 ($f_{\rm ABSC}$). 
	Column 5: Halo region 0.3--2.0 keV energy band diffuse X-ray 
	luminosity ($\ergps$), based on $f_{\rm ABSC}$ and 
	the distances given in Table~\ref{tab:galaxies}.
	Column 6: Ratio of the X-ray luminosity from thermal emission
	in the halo region of each galaxy to the total galactic far-infrared
	flux estimated from the IRAS 60\micron~and 100\micron~flux 
	densities, where 
	$f_{\rm FIR} = 1.26\times10^{-11} \times (2.58\times f_{60} 	
	+ f_{100})$,
	in units of $\ergps \pcmsq$, where $f_{60}$ and $f_{100}$ 
	are given in units of Jy.
	Columns 7 to 10: Disk region diffuse thermal X-ray fluxes, 
	luminosities and X-ray to FIR flux ratios.
	Columns 11 to 14: Nuclear region diffuse thermal X-ray fluxes, 
	luminosities and X-ray to FIR flux ratios.
	Columns 15 to 17: Total thermal X-ray absorption-corrected 
	X-ray flux, luminosity and X-ray to FIR flux ratio, \ie
	the sum of the thermal emission from the 
	nuclear, disk and halo regions. For 
	NGC 6503 and NGC 4244 these totals ignore any contribution
	from the halo region.
}
\tablenotetext{a}{ACIS-I observations of M82, ObsIDs 361 and 1302. 
	Data from ACIS-I chips, covering
	the majority of the extra-planar diffuse emission, 
	but not extending as far as the northern
	cloud.}
\tablenotetext{b}{ACIS-S observation of M82, ObsID 2933. Data from the S3 chip only, which does not
	cover the full extent of the diffuse emission.}
\tablenotetext{c}{NGC 253 observation ObsID 969. Only a small 
	fraction of the Northern halo lies on the S3 chip.}
\tablenotetext{d}{NGC 253 observation ObsID 790, Data from S3 chip only, 
        covering most of the
	bright diffuse emission in the Northern halo.}
\tablenotetext{e}{Fluxes and luminosities covering all of NGC 253 and 
	its wind. We estimate that the northern halo region covered
	by the ObsID 790 observation accounts for $\sim48$\% of 
	the total halo region flux (\ie both sides of the planes of the galaxy)
        from NGC 253, based comparison of the estimated halo diffuse count
        rate in the {\it ROSAT} PSPC observation. Scaling the
	fluxes obtained from the {\it Chandra} observation, the
	 total 0.3--2.0 keV energy band 
	diffuse X-ray flux 
        from both the north and south of
        the plane is $f_{\rm ABSC} \sim 1.47\times10^{-12} \ergps \pcmsq$
	($L_{\rm X} \sim1.2\times10^{39}\ergps$).}
\tablenotetext{f}{Upper limits of the X-ray flux from diffuse gas in
  the halos of NGC 6503 and NGC 4244 were calculated from the $3\sigma$
  upper limits on the 0.3--2.0 keV diffuse emission count rate (0.0122 and
  0.0040 cts s$^{-1}$ respectively), and an ACIS S3 chip count rate to
  absorption-corrected flux conversion factor of $4\times10^{-12} \ergps 
  \pcmsq$ (cts s$^{-1}$)$^{-1}$.}
\end{deluxetable}

\subsection{Diffuse nuclear and disk emission}

For completeness we present diffuse X-ray fluxes and luminosities for
the nuclear and disk regions, along with totals (summing nuclear,
disk and any halo region diffuse X-ray emission) in Table~\ref{tab:fluxes}.
The nuclear and disk region diffuse flux are also based on simultaneous
fits. 

Simultaneous fitting of the nuclear and disk region
spectra was more difficult than for the halo as the spectra are
more complex, and in terms of reduced $\chi^{2}_{\nu}$ values these
nuclear and disk region combined
fits are worse than the combined halo region best-fit, with 
$\chi^{2}_{\nu} = 2.15$ (952 degrees of freedom) 
and $2.02$ (1200 degrees of freedom) respectively.
There are several reasons why model fitting to the nuclear 
and disk region spectra is more complicated 
than for the halo. Firstly these spectra typically 
have higher signal to noise than the halo region spectra. 
Secondly, residual contamination by point
sources is immediately apparent in the disk and nuclear spectra as a
hard power-law continuum above energies of $E \sim 2$ keV, which adds
an additional model component to the fits. Thirdly, the diffuse component
of the spectra is not well modeled as a two temperature plasma with
uniform foreground absorption. For both nuclear and disk spectral fits we
used an additional model component that physically 
represents patchy absorption, where a significant fraction of the diffuse 
X-ray emission passes through absorption equivalent to a hydrogen column of
$\nH \sim 3 \times 10^{21} \pcmsq$. The fit to the nuclear region spectra
also required the
presence of the much hotter thermal component, with $kT=5 \keV$, in
the M82 spectra. The normalization of this third thermal component
was found to be negligible in the other galaxies, except for NGC 1482 and
NGC 4945.

We shall reexamine the X-ray
temperature and abundance properties of the disk and nuclear
regions in more detail in a future paper, where the application of
updated spectral calibration should lead to better fits. 
For convenience we reproduce only
the model-derived thermal-component fluxes and luminosities (\ie we quote
only the fluxes associated with thermal emission components, ignoring the
power law model component used to fit for residual point source X-ray emission)
in this paper, which are given in Table~\ref{tab:fluxes},
along with the total diffuse component X-ray flux for each galaxy. The
primary source of uncertainty in these numbers is again systematic,
so the fluxes are unlikely to become more accurately known 
even after improved spectral calibrations of ACIS become available.

Our method treats
spectrally distinct regions (disk vs. halo, and 
nucleus vs. disk) separately, and thus goes
much of the way toward resolving out the spatial variations in 
spectral properties that can complicate and distort the results of
standard X-ray spectral fitting 
\citep*[see the discussion in][]{weaver00,dahlem00}.
The total diffuse-component X-ray fluxes quoted in
Table~\ref{tab:fluxes} should therefore be more robust and accurate 
than previous flux estimates based on fitting models 
to the summed X-ray emission from each galaxy in its entirety, 
the only method available given the low spatial resolution 
and sensitivity of {\it ROSAT} and {\it ASCA}.

\subsection{Distinguishing between $\alpha$-element enhancement
or depletion of iron onto dust}

\citet{martin02} present a plausible case for the direct observation
of metal-enriched X-ray-emitting gas in the ACIS observation of the low
mass dwarf starburst galaxy NGC 1569 ($v_{\rm rot} \sim 30\kmps$).
Despite having many more photons in our halo region spectra,
we can not automatically make the same claim.

\citeauthor{martin02}'s case is convincing as their best-fit oxygen 
abundance $Z_{\rm O} \sim 1 \Zsol$ (greater than $0.25 \Zsol$ at 99\% 
confidence) 
is higher than the ambient ISM oxygen abundance of $\sim0.2 \Zsol$ of
this dwarf galaxy. Our best-fit combined \emph{halo} region oxygen abundance
is similarly $Z_{\rm O} \sim 1 \Zsol$, but in the much more massive 
galaxies of our sample
the ambient ISM of the disk will have such a metallicity 
\citep{tremonti}. The high $\alpha$-to-Fe ratio can be explained as
some residual depletion of the iron on dust grains, assuming
the X-ray emission comes from gas originally part of the disk,
there being considerable evidence for dust in superwind outflows 
\citep[\eg][]{radovich,heckman01_dust,sugai03,shapley03}.
Dust destruction
by sputtering has a timescale comparable to the $10^{7}$--$10^{8}$ year
timescale appropriate for winds and fountains \citep{popescu00},
so it is plausible that some fraction of the refractory elements
are retained on dust grains in even hot halo gas.

To test this
hypothesis we investigated the abundances of the refractory $\alpha$-elements
magnesium and silicon, and the volatile element neon, with
respect to oxygen. 
For each of these three elements, we allowed its abundance to fit
independently of the oxygen-dominated $\alpha$-element group it had
previously been assigned to, and then measured its abundance 
relative to oxygen (see panels c and d in Fig.~\ref{fig:confidence_contours}).
As the confidence space is poorly behaved for low absolute abundances, 
we measure this ratio at $Z_{\sim O} = 1 Z_{\sim O, \odot}$.

The measured values are given in Table~\ref{tab:abundances}.
For comparison, we also reproduce the observed abundance pattern of
warm ionized gas in the halo of our own galaxy  
(from \citealt{savage_sembach96}), and theoretical predictions
for nucleosynthesis in SNe taken from \citet*{gibson97}.

The observed halo-region abundance pattern is consistent
with massive star enrichment, lying close to the broad range of
possible Type II SN ejecta abundance patterns 
collected by \citet{gibson97}, although the theoretical yields
predict a slightly higher ratio of iron to oxygen than observed.
Gas as heavily depleted as the warm halo gas of the Milky Way
is ruled out. However, without either knowing 
the Mg/O or Si/O ratios to within approximately $\pm{0.1}$ dex,
or (preferably) the abundances with
respect to hydrogen, it is currently impossible to rule out some form
of a depletion scenario for the hot X-ray-emitting gas.

Finally, some caution is warranted as this
analysis relies upon the temperature structure of emitting
plasma being accurately modeled.

\begin{deluxetable}{lllll}
 \tabletypesize{\scriptsize}%
\tablecolumns{5} 
\tablewidth{0pc} 
\tablecaption{Relative element abundances
	\label{tab:abundances}}
\tablehead{
	\colhead{} & \colhead{[Fe/O]}  & \colhead{[Mg/O]} 
	 & \colhead{[Si/O]}  & \colhead{[Ne/O]} 
}
\startdata
Observed hot halo emission
	& $-0.43^{+0.14}_{-0.08}$ 
	& $\phm{-}0.06^{+0.21}_{-0.29}$ 
	& $<0.60$ 
	& $\phm{-}0.01^{+0.14}_{-0.15}$ \\
Warm Galactic halo\tablenotemark{a}
	& $-0.80$ & $-0.55$ & $-0.30$ & \nodata \\
Type II SN averaged over IMF,  A96\tablenotemark{b}
	& $-0.08$ & $\phm{-}0.13$ & \nodata & $-0.05$ \\
Type II SN averaged over IMF,  T95\tablenotemark{b}
	& $-0.32$ & $-0.01$ & $\phm{-}0.01$ & $-0.17$ \\
Type II SN averaged over IMF,  W95;A;$10^{-4}\Zsol$\tablenotemark{b}
	& $-0.20$ & $-0.18$ & $\phm{-}0.25$ & $-0.21$ \\
Type II SN averaged over IMF,  W95;B;$10^{-4}\Zsol$\tablenotemark{b}
	& $-0.39$ & $-0.17$ & $\phm{-}0.04$ & $-0.10$ \\
Type II SN averaged over IMF,  W95;A;$\Zsol$\tablenotemark{b}
	& $-0.19$ & $-0.10$ & $\phm{-}0.14$ & $-0.11$ \\
Type II SN averaged over IMF,  W95;B;$\Zsol$\tablenotemark{b}
	& $-0.23$ & $-0.08$ & $\phm{-}0.07$ & $-0.08$ \\
Type Ia SN,  TNH93\tablenotemark{b}
	& $\phm{-}1.55$ & $-0.05$ & $\phm{-}1.16$ & $-0.75$ \\
\enddata
\tablecomments{All errors represent 90\% confidence, upper 
limits are 99.73\% confidence ($3\sigma$). Abundance ratios
are relative to the scale presented in \citet{anders89}.}
\tablenotetext{a}{From \citet{savage_sembach96}.}
\tablenotetext{b}{Taken from compilation of IMF-averaged yields 
	presented by \citet{gibson97}. See \citet{gibson97} for details
	and nomenclature.}
\end{deluxetable}

\section{Discussion}
\label{sec:discussion}

\subsection{A physical interpretation of the quasi-isothermal, 
	semi-exponential surface brightness profiles in the
	starburst galaxies}
\label{sec:discussion:phys_interp}

The general features of the extra-planar emission (namely
spatially-correlated X-ray and \halpha~emission, often with
apparent limb-brightening, soft thermal
spectra with little or no significant spectral variation with increasing
height above the plane, and a semi-exponential X-ray surface brightness
profile) are similar in all of
the starburst galaxies with detected halo diffuse X-ray emission.
These are the same empirical properties considered in our discussion of the
physical origin of the halo emission in NGC 253 
\citep[see \S~7 of][]{strickland02}, to which we refer the reader.

Our favored models for explaining the extra-planar X-ray and 
\halpha~emission was that the outflowing superwind was interacting with a
tenuous and dusty thick-disk/halo medium, itself possibly driven into the lower
halo by fountain activity.
The limb-brightened X-ray emission might either arise in a shell
of shock-heated ambient halo gas (see Fig.~11c of \citealt{strickland02}),
or from metal-enriched wind material that has been 
compressed in reverse shocks near the walls of
the outflow (Fig.~11d of \citealt{strickland02}).

The filamentary structures visible in even heavily smoothed X-ray images,
and their similarity and close spatial association with extra-planar
\halpha~filaments on both small and large scales (scales of 10's of pc 
and kiloparsecs respectively), 
are compelling evidence that the X-ray emission
is not from a volume-filling component of the wind
\citep[for a more detailed discussion of this, see][]{strickland00,strickland02}.

The observed abundances patterns are consistent with
either of these interpretations (\S~\ref{sec:spectra:fits}),
although with higher quality data and a better understanding
of the spectra they could be used to distinguish between
these two scenarios.

Our finding from this paper, of semi-exponential or 
Gaussian vertical X-ray surface brightness
distributions, are consistent with first scenario, and
do not rule out the second scenario. 
Exponential or Gaussian
minor-axis density distributions are expected
in both galactic fountain models 
\citep*[see][]{bregman80,collins02}
and the cosmological
accretion models \citep{sommerlarsen02}.
The distribution of X-ray emission from
halo gas shocked by a superwind will mirror the original 
halo gas distribution.
Optical studies of edge-on star-forming galaxies that have extra-planar
warm ionized gas 
\citep*[see][and references therein]{wang97,hoopes99,collins00,miller03} 
find emission scale heights of order a few kpc,
similar to those we have found in the diffuse X-ray emission.
Provided that the temperature of the gas
does not vary significantly with $z$, as we observe, then the
X-ray surface brightness profile expected from these models will also
be semi-exponential or Gaussian, with a characteristic scale-height half
that of the underlying density distribution.
Assessing the expected observational properties of the second scenario
is non-trivial, and we defer further consideration of this
to a future paper.

In contrast, X-ray emission from a volume-filling component of the
wind will produce a power-law X-ray surface brightness profile 
\citep{chevclegg}, with $\Sigma \propto z^{-3}$. 
Note that this applies to almost any model with
emission from a volume-filling component, including mass-loaded 
models with observationally-motivated conical outflow geometries
\citet{suchkov96}. This is further evidence that the observed X-ray 
emission in superwinds in not dominated by
a heavily-mass-loaded volume-filling
component of the wind, despite some 
claims to the contrary \citep[\eg][]{schurch02,pietsch2001}.

In either of the two scenarios discussed above we require the
presence of a pervasive gaseous halo. This suggests
that the disk/halo interaction may be more fountain-like 
\citep[\eg][]{shapiro76,bregman80}, rather
than the more localized chimney model of \citet{norman89},
given the need to produce a widespread halo medium.
In this case, it may be possible to use superwinds as a probe of
the pre-existing halo properties \citep[\eg][]{sofue_vogler}.

\subsection{No current need to propose AGN-driving}
\label{sec:discussion:agn}

When presenting this work, we are often asked ``NGC 4945 and NGC 3079
have AGN. Is there any evidence that AGN play a
role in creating the large-scale
winds you see?''

The best answer currently available is that they do not.  
Take the simplest assumption, that
the more radiatively-luminous AGN can also supply more mechanical
energy into the ISM. Then, if AGN contributed in any significant
way to powering the winds, we would expect galaxies with the most
powerful AGN to also to have the notably luminous, spatially extended and/or
kinematically powerful winds. Instead, there is no obvious correlation
between the presence and luminosity of any AGN and any superwind property.
Of the three intrinsically-most X-ray luminous, highest surface-brightness
galaxies in this sample, M82, NGC 1482 and NGC 3079, only NGC 3079 
has an AGN (which does not obviously have any direct relationship to the
wind, as discussed in \S~\ref{sec:results:gal_notes:n3079})
NGC 4945, the most powerful AGN in this sample, is in no way spectacular.
The nuclear outflow cone is very similar in X-ray luminosity
($L_{\rm X} \sim 10^{38} \ergps$) and size ($\sim 500$ pc, 
see \citealt{schurch02})
to that in NGC 253 (which has a low
luminosity AGN, see \citealt{weaver02}), and has lower outflow velocities
than M82 or NGC 3079. X-ray emission from the halo of NGC 4945 is
weak, if present at all.

A more quantitative answer to this question is difficult to present.
 \citet{ham90} show that the shape of the
optically-derived pressure profiles in the central few kpc of many
superwind galaxies \emph{exclude} AGN-driving, as expected for
clouds embedded in a \citet{chevclegg} style wind model 
with a distributed source of energy and mass (\ie a starburst) --- indeed, 
this is one of the reasons the \citeauthor{chevclegg} model
is still used despite the lack of a direct observational 
detection of the merged SN and stellar wind ejecta that drives superwinds.
No pressure profile is available for NGC 4945, but the pressure profile
for NGC 3079 is possibly consistent with distributed energy and mass 
injection \citep{ham90}.

Supernova-driving was originally
suggested for the winds in NGC 3079 and NGC 4945 for a variety of reasons,
but one significant physical
reason is that the estimated rates of \emph{mechanical} 
energy return to the ISM by SNe were more than sufficient to power
the outflows. The same can not be said for AGN. It is far from clear
what the relationship is between the power an AGN radiates in
any waveband, and energy it may transfer to the ISM, and/or to
the powering of jets.

Without further methods to quantitatively test or
directly falsify the AGN-driving hypothesis, we feel that the
question of whether AGN-driving is important in galaxies 
such as NGC 4945 and NGC 3079 is not a meaningful issue to raise at this 
time. In any case, the burden of proof lies with the advocates of 
AGN-driving for NGC 4945 and NGC 3079..

Seyfert nuclei and LLAGN 
may well drive weakly-collimated galactic outflows in galaxies without
obvious starbursts, \eg NGC 2110 and NGC 2992 
\citep*{weaver95,colbert96a,colbert96b,n2992}. 
Our argument is
that in the ``classic'' starburst galaxies of this sample, and most probably
in Seyfert galaxies with starbursts (for the reasons presented
in \citealt*{levenson01a,levenson01b}), 
it is the starburst that is the
dominant cause and driver of the outflow.

\subsection{The similarity in extra-planar gas properties
between NGC 891 and the starburst galaxies}
\label{sec:discussion:normal_vs_starburst}

The similarity between the properties of the 
extra-planar emission around the normal spiral NGC 891
and that around the starburst galaxies is intriguing.
The $f_{\rm X, HALO}/f_{\rm FIR}$ and $f_{\rm X, HALO}/f_{\rm 1.4 GHz}$ flux
ratios of NGC 891 are consistent with the values found in the starbursts.
In addition, the extra-planar X-ray emitting gas in NGC 891 
is accompanied by extra-planar optical emission, 
again similar to the apparently
ubiquitous coupling of \halpha~and X-ray emitting gas in the starbursts.
The implication is that the physical conditions in
the extra-planar gas around NGC 891 are very similar to those in the 
optical and X-ray emitting gas in starburst-driven superwinds.

\citet{read01} suggest that is it possible, using the spectral
properties of the diffuse X-ray emission, to distinguish between
``coronal'' systems (galaxies with static hot halos and/or fountains)
and ``wind'' systems (galaxies with starburst-driven superwinds).
In physical terms this is plausible --- the temperature of
hot gas in static or fountain-type halos should be less than or
similar to the halo virial temperature, while in winds the gas temperature
can easily be higher than virial temperature (although it might also
be lower, and show no correlation with the host galaxy mass at all).

Our best-fit halo spectral model,  
essentially dominated by the spectra of the starburst
galaxies, provides a good fit to the halo X-ray spectrum of the ``normal''
spiral NGC 891. When considered in terms of a simple 
hardness-ratio analysis (\S~\ref{sec:spectra:hardness}), 
or based on the spectral modeling in \S~\ref{sec:spectra:fits}
(\eg the luminosity-weighted gas temperatures shown in 
Table~\ref{tab:spectral:fits}), we find no significant spectral difference
between the X-ray-emitting gas halos of starburst galaxies and the
one normal galaxy (other than the Milky Way) currently known
to have hot gas in its halo.  

As our sample includes all of the edge-on 
galaxies on which \citeauthor{read01}'s
suggestion is based, why do we come to different conclusions?
\citeauthor{read01} assume that NGC 891 and NGC 4631 represent examples of 
semi-static halos. Of the approximately edge-on systems in their sample, 
these two galaxies have the lowest diffuse emission temperatures, based on
single temperature spectral fits to point-source-subtracted
{\it ROSAT} PSPC spectra of the \emph{entire} galaxy. In terms of
luminosity-weighted mean temperature, or spectral hardness, the {\it Chandra}
observation -derived \emph{halo}
spectra of NGC 891 and NGC 4631 are no cooler or spectrally softer
than the halos of NGC 253 or NGC 3628 --- galaxies with clear 
galactic winds. 

The physical interpretation of what drives the observed characteristic
halo emission temperature, or temperatures, remains an open, 
and important question.
Nevertheless, we would argue that evidence for a spectral difference
between hot gas in the halos of normal galaxies and that in the
halos of starburst galaxies, or evidence to allow
a separation between ``coronae'' (or fountains)
and ``winds,'' is currently lacking.


\section{Summary}
\label{sec:conclusions}
We present a detailed study of the diffuse X-ray emission in
a sample of 10 approximately edge-on disk galaxies 
(7 starburst galaxies, 3 normal spirals) that span the full range of
star formation found in disk galaxies. With the arcsecond spatial
resolution of the {\it Chandra} X-ray telescope we are able to robustly
separate the X-ray emission from point sources from the truly diffuse
component. {\it Chandra} is the first, and only, X-ray telescope with the
spatial resolution sufficient to allow a meaningful comparison of the
spatial location of the X-ray emission to observations
of the host galaxies at other wavelengths, in particular
to ground or space-based optical imaging. In addition to a detailed
discussion of each galaxy, we have presented a mini-atlas of
soft and hard X-ray, \halpha~and R-band images of each of the 10 galaxies,
shown at a common spatial and surface brightness scale to facilitate
cross-comparison.

We consider a variety of quantitative measures of the spatial extent, 
spectral hardness, and shape of the minor axis surface brightness profiles,
several of which can be applied to future samples
of galaxies with lower S/N data. We find that the spectral
properties of the diffuse X-ray emission show only weak, or some cases no, 
variation with increasing height (outside an absorbed
region within $z\sim2$ kpc of the disk mid-plane). 
The vertical decrease in surface brightness
of the extra-planar emission ($|z| \ge 2$ kpc) appears to better
described by exponential (effective surface brightness scale heights
are typically between 2 -- 4 kpc),
or Gaussian models, than the power law expected
of a freely expanding fluid.

For the eight galaxies with detections of extra-planar diffuse emission,
we find that a common spectral model, comprising a two-temperature
{\sc Mekal} hot plasma model with an enhanced $\alpha$-to-Fe 
element ratio, can fit the ACIS X-ray spectra. 
The X-ray-derived metal abundances show super-Solar
ratios of $\alpha$-process elements (primarily oxygen) to
iron. This is consistent
with the origin of the X-ray emission being either
(metal-enriched) merged SN ejecta, 
or from shocked halo gas (with 
moderate remaining depletion of refractory elements onto dust).
We wish to stress that with the present quality of X-ray spectroscopic
data, high $\alpha$/Fe ratios
in X-ray emitting plasma are not an automatic indication of massive
star SN enrichment.

The spectral and spatial properties of the one normal galaxy in which
extra-planar diffuse X-ray emission is detected (NGC 891), are consistent
with those of the starburst galaxies, many of which show unambiguous
evidence for galactic scale outflows (superwinds). Does this mean that
NGC 891 is driving a global wind, or is the similarity of its halo 
to the superwinds simply due to similar emission processes occurring
in both fountains and superwinds? More observation effort needs to be expended
on deep X-ray observations of edge-on normal spiral galaxies, before we
will fully understand the structure and physics of the disk/halo interface
in star-forming galaxies.

Our favored model for the origin of the extra-planar soft X-ray emission 
is that SN feedback in the disks
of star-forming galaxies create, via blow out and venting of hot
gas from the disk, tenuous exponential atmospheres
of density scale height $H_{\rm g} \sim 4$ -- 8 kpc.
AGN-driven winds do not
appear to be significant in this sample, in that there is no
obvious correlation between the presence and luminosity of any AGN
and the properties of the diffuse X-ray emission. 
The soft thermal X-ray emission observed in the halos of the starburst
galaxies is either pre-existing halo medium
(which has been swept-up and shock-heated by the starburst-driven superwinds)
or from \emph{a small fraction} (by volume)
of the merged-SN-ejecta that has been compressed
near the walls of the outflow (\eg by a reverse shock propagating back 
into the outflowing wind).
In either case the observed
exponential X-ray surface brightness distributions are an inheritance
from galactic fountain
activity prior to the currently-observed starburst phase.
This model is based on the qualitative 
2-D morphology of the diffuse X-ray and optical
\halpha~emission (in particular the filamentary, occasionally limb-brightened
morphology of both the X-ray and \halpha~emission), 
as well as interpretation of the more-quantitative
minor axis surface brightness and spectral
hardness profiles. One important implication of this model
is that galactic-scale gaseous halos may be common around 
star-forming disk galaxies. Observing starburst galaxies, in which superwinds
``light-up'' these pre-existing halos, may present the best method
for studying the gaseous halos of star-forming galaxies.


\acknowledgments

Over the several years this project has been in the making, we have
have been fortunate to benefit from the insightful comments and
questions of many astronomers, too numerous to mention 
individually, to whom we extend our thanks. The anonymous
referee deserves thanks for her or his constructive criticism.
We would also like to thank
S. Hameed for providing us with R-band and \halpha~images of NGC 1482,
and for drawing the attention of the community to this previously
ignored but very interesting superwind galaxy.

DKS is supported by NASA through {\it Chandra} Postdoctoral Fellowship Award
Number PF0-10012, issued by the {\it Chandra} X-ray Observatory Center,
which is operated by the Smithsonian Astrophysical Observatory for and on
behalf of NASA under contract NAS8-39073.

This research is partially based on data from the ING Archive.
This publication makes use of data products from the Two 
Micron All Sky Survey, which is a joint project
of the University of Massachusetts and the Infrared Processing 
and Analysis Center/California Institute of
Technology, funded by the National Aeronautics and 
Space Administration and the National Science
Foundation. Furthermore, this research has made use of 
the extremely-useful NASA/IPAC Extragalactic Database (NED) which is operated 
by the Jet Propulsion Laboratory, California Institute of 
Technology, under contract with the National 
Aeronautics and Space Administration.




\begin{thebibliography}{}
\bibitem[Aguirre \etal (2001)]{aguirre01}
	Aguirre, A., Hernquist, L., Schaye, J., 
	Weinberg, D.H., Katz, N. \& Gardner, J., 2001, \apj, 560, 590
\bibitem[\protect\citeauthoryear{Allen, Baldwin \& Sancisi}{Allen \etal}{1978}]{allen78}
	Allen, R.J., Baldwin, J.E. \& Sancisi, R., 1978, \aap, 62, 397
\bibitem[Anders \& Grevesse (1989)]{anders89}
	Anders, E. \& Grevesse, N., 1989, \gca, 53, 197
\bibitem[Armus \etal (1995)]{armus95}
	Armus, L., Heckman, T.M., Weaver, K.A. \& Lehnert, M.D.,
	1995, \apj, 445, 666 
\bibitem[Arp \etal (2002)]{arp02}
	Arp, H., Burbidge, E.M., Chu, Y., Flesch, E.,
	Patat, F., Rupprecht, G., 2002, \aap, 391, 833 
\bibitem[de Avillez (2000)]{avillez00}
	de Avillez, M.A., 2000, \mnras, 315, 479
\bibitem[Murakami \& Babul (1999)]{babul}
	Murakami, I. \& Babul, A., \mnras, 309, 161	
\bibitem[Bell \& de Jong (2001)]{bell01}
	Bell, E.F. \& de Jong, R.S., 2001, \apj, 550, 212
\bibitem[Benson \etal (2000)]{benson00}
	Benson, A.J., Bower, R.G., Frenk, C.S. \& White, S.D.M.,
	2000, \mnras, 314, 557 
\bibitem[Bergman \etal (1992)]{bergman92}
	Bergman, P., Aalto, S., Black, J.H. \& Rydbeck, G.,
	1992, \aap, 265, 403
\bibitem[\protect\citeauthoryear{Bessel, Castelli \& Plez}{Bessel \etal}{1998}]{bessel98}
	Bessel, M.S., Castelli, F. \& Plez, B., 1998, \aap, 333, 231
\bibitem[Bland \& Tully (1988)]{bland88}
	Bland, J. \& Tully, R.B., \nat, 334, 43
\bibitem[Bottema \& Gerritsen (1997)]{bottema97}
	Bottema, R. \& Gerritsen, J.P.E., 1997, \mnras, 290, 585
\bibitem[Boulanger \& P\'{e}rault (1988)]{boulanger88}
	Boulanger, F. \& P\'{e}rault, M., 1998, \apj, 330, 964
\bibitem[Braine \etal (1997)]{braine97}
	Braine, J., Guelin, M., Dumke, M., Brouillet, N.,
 	Herpin, F. \& Wielebinski, R. 1997, \aap, 326, 963
\bibitem[Bregman (1980)]{bregman80}
	Bregman, J.N., 1980, \apj, 236, 577
\bibitem[Bregman \& Glassgold (1982)]{bregman82}
	Bregman, J.N. \& Glassgold, A.E., 1982, \apj, 263, 564
\bibitem[Bregman \& Houck (1997)]{bregman97}
	Bregman, J.N. \& Houck, J.C., 1997, \apj, 485, 159
\bibitem[Bregman \& Irwin (2002)]{bregman_and_irwin}
	Bregman, J.N. \& Irwin, J.A., 2002, \apjl, 56, L13
\bibitem[Bregman \& Pildis (1994)]{bregman_and_pildis}
	Bregman, J.N. \& Pildis, R.A., 1994, \apj, 420, 570
\bibitem[\protect\citeauthoryear{Bregman, Schulman \& Tomisaka}{Bregman \etal}{1995}]{bregman95}
	Bregman, J.N., Schulman, E. \& Tomisaka, K., 1995, \apj, 439, 155
\bibitem[Brock \etal (1988)]{brock88}
	Brock, D., Joy, M., Lester, D.F., Harvey, P.M. \& Benton Eliss., 
	H., Jr., 1988, \apj, 329, 208
\bibitem[Burstein \& Heiles (1982)]{burstein82}
	Burstein, D. \& Heiles, C., 1982, \aj, 87, 1165
\bibitem[Carpenter (2001)]{carpenter01}
	Carpenter, J.M., 2001, \aj, 121, 2851
\bibitem[\protect\citeauthoryear{Carral, Turner \& Ho}{Carral \etal}{1990}]{carral90}
	 Carral, P., Turner, J.L., Ho, P.T.P., 1990, \apj, 362, 434
\bibitem[Cecil \etal (2001)]{cecil01}
	Cecil, G., Bland-Hawthorn, J., Veilleux, S. \&
	Filippenko, A.V., 2001, \apj, 555, 338
\bibitem[\protect\citeauthoryear{Cecil, Bland-Hawthorn \& Veilleux}{Cecil \etal}{2002}]{cecil02}
	Cecil, G., Bland-Hawthorn, J. \& Veilleux, S., 2002, \apj, 576, 745
\bibitem[Chen \etal (1997)]{chen97}
	Chen, K.P., Collins, N., Angione, R., Talbert, F.,
	Hintzen, P., Smith, E.P., Stecher, T. \& The UIT Team,
	1997, ``UV/Visible Sky Gallery on CDROM''
\bibitem[Chevalier \& Clegg (1985)]{chevclegg}
	Chevalier, R. \& Clegg, A., 1985, \nat, 317, 44
\bibitem[\protect\citeauthoryear{Chiang, Ryu \& Vishniac}{Chiang \etal}{1988}]{chiang88}
	Chiang, W.-H., Ryu, D. \& Vishniac, E.T., 1988, \pasp, 100, 1386
\bibitem[Clarke \etal (2003)]{clarke03}
	Clarke, T.E., Uson, J.M., Sarazin, C.L. \& Blanton, E.L.,
	2003, ApJ, in press (astro-ph/0310508)
\bibitem[Colbert \etal (1996a)]{colbert96a}
	Colbert, E.J.M., Baum, S.A., Gallimore, J.F., O'Dea, C.P.,
	Lehnert, M.D., Tsvetanov, Z.I.,  Mulchaey, J.S. \& Caganoff, S.,
	1996a, \apjs, 105, 75
\bibitem[Colbert \etal (1996b)]{colbert96b}
 	Colbert, E.J.M., Baum, S.A., Gallimore, J.F., O'Dea, C.P. \&
	Christensen, J.A., 1996b, \apj, 467, 551
\bibitem[\protect\citeauthoryear{Cole, Mundell \& Pedlar}{Cole \etal}{1998}]{cole98}
	Cole, G.H.J., Mundell, C.G. \& Pedlar, A., 1998, \mnras, 300, 656
\bibitem[\protect\citeauthoryear{Collins, Benjamin \& Rand}{Collins \etal}{2002}]{collins02}
	Collins, J.A., Benjamin, R.A. \& Rand, R.J., 2002, \apj, 578, 98
\bibitem[Collins \etal (2000)]{collins00}
	Collins, J.A., Rand, R.J., Duric, N. \& Walterbos, R.A.M.,
	2000, \apj, 536, 645
\bibitem[Cox (1981)]{cox81}
	Cox, D.P., \apj, 245, 534
\bibitem[Crillon \& Monnet (1969)]{crillon69}
	Crillon, R. \& Monnet, G., 1969, \aap, 2, 1
\bibitem[Dahlem (1997)]{dahlem97}
	Dahlem, M., 1997, \pasp, 109, 1298
\bibitem[Dahlem \etal (1993)]{dahlem93}
	Dahlem, M., Golla, G., Whiteoak, J. B., Wielebinski, R.,
	 Huettemeister, S. \&  Henkel, C., 1993, \aap, 270, 29
\bibitem[Dahlem \etal (1996)]{dahlem96}
	Dahlem, M., Heckman, T.M., Fabbiano, G., Lehnert, M.D. \&
	Gilmore, D., 1996, \apj, 461, 724
\bibitem[\protect\citeauthoryear{Dahlem, Lisenfeld \& Golla}{Dahlem \etal}{1995}]{dahlem95}
	Dahlem, M., Lisenfeld, U. \& Golla, G., 1995, \apj, 444, 119
\bibitem[Dahlem \etal (2001)]{dahlem01}
	Dahlem, M., Lazendic, J.S., Haynes, R.F., Ehle, M. \& 
	Lisenfeld, U., 2001, \aap, 374, 42
\bibitem[Dahlem \etal (2000)]{dahlem00}
	Dahlem, M.,  Parmar, A., Oosterbroek, T., 
	Orr, A., Weaver, K.A. \& Heckman, T.M.,
	2000, \apj, 538, 555
\bibitem[\protect\citeauthoryear{Dahlem, Weaver \& Heckman}{Dahlem \etal}{1998}]{dwh98}
        Dahlem, M., Weaver, K.A. \& Heckman, T.M. 1998,
        \apjs, 118, 401
\bibitem[Dale \etal (2000)]{dale2000}
	Dale, D.A., \etal, 2000, \aj, 120, 583
\bibitem[Deharveng \etal (1986)]{deharveng86}
	Deharveng, J.-M., Bixler, J., Joubert, M., Bowyer, S.
	\& Malina, R., 1986, \aap, 154, 119
\bibitem[Dekel \& Silk (1986)]{dekel86}
	Dekel A., Silk J., 1986, \apj, 303, 39
\bibitem[Della Ceca \etal (1996)]{dellaceca96}
	Della Ceca, R., Griffiths, R.E., Heckman, T.M. \& MacKenty, J.W.,
	1996, \apj, 469, 662
\bibitem[\protect\citeauthoryear{Della Ceca, Griffiths \& Heckman}{Della Ceca \etal}{1997}]{dellaceca97}
	Della Ceca, R., Griffiths, R.E. \& Heckman, T.M.,
	1997, \apj, 485, 581
\bibitem[\protect\citeauthoryear{de Mello, Wiklind \& Maia}{de Mello \etal}{2002}]{demello02}
	de Mello, D.F., Wiklind, T. \& Maia, M.A.G., 2002, \aap, 381, 771
\bibitem[\protect\citeauthoryear{D\'{e}sert, Boulanger \& Puget}{D\'{e}sert \etal}{1990}]{dust_temp}
	D\'{e}sert, F.-X., Boulanger F. \& Puget, J.L., 1990, \aap, 237, 215
\bibitem[Dettmar (1990)]{dettmar90}
	Dettmar, R.-J., 1990, \aap, 232, L15
\bibitem[de Vaucouleurs \& Caulet (1982)]{devaucouleurs82}
	de Vaucouleurs, G. \& Caulet, A., 1982, \apjs, 49, 515
\bibitem[de Vaucouleurs \etal (1991)]{rc3}
	de Vaucouleurs G., de Vaucouleurs A., Corwin Jr. H.G., Buta R.J.,
        Paturel G. \& Fouque P., 1991, {\it Third Reference Catalogue 
	of Bright Galaxies (RC3)} (Springer-Verlag: New York)
\bibitem[Devine \& Bally (1999)]{devine99}
	Devine, D. \& Bally, J., 1999, \apj, 510, 197
\bibitem[De Young \& Heckman (1994)]{deyoung94}
	De Young, D.S. \& Heckman T.M., 1994, \apj, 431, 598
\bibitem[Dickey \& Lockman (1990)]{nhref}
	Dickey, J.M. \& Lockman, F.J., 1990, \araa. 28, 215. 
\bibitem[\protect\citeauthoryear{Done, Madejski \& Smith}{Done \etal}{1996}]{done96}
	Done, C., Madejski, G.M. \& Smith, D.A., 1996, \apjl, 463, L63
\bibitem[Douglas \etal (1996)]{douglas96}
	Douglas, J.N., Bash, F.N., Arakel Bozyan, F.,
        Torrence, G.W., Wolfe, C., 1996, \aj, 111, 1945
\bibitem[Ekers \& Sancisi (1977)]{ekers77}
	Ekers, R.D. \& Sancisi, R., 1977, \aap, 54, 973
\bibitem[Elfhag \etal (1996)]{elfhag96}
	Elfhag, T., Booth, R.S., Hoglund, B.,
	Johansson, L.E.B. \& Sandqvist, Aa., 1996, \aaps, 115, 439
\bibitem[Elmouttie \etal (1997)]{elmouttie97}
	Elmouttie, M., Haynes, R.F., Jones, K.L., Ehle, M.,
	Beck, R., Harnett, J.I. \& Wielebinski, R., 1997, \mnras, 284, 830
\bibitem[Engelbracht \etal (1998)]{engelbracht98}
        Engelbracht, C.W., Rieke, M. J., Rieke, G.H.,
        Kelly, D.M., Achtermann, J.M., 1998, \apj, 505, 639
\bibitem[\protect\citeauthoryear{Fabbiano, Heckman \& Keel}{Fabbiano \etal}{1990}]{fhk90}
	Fabbiano, G., Heckman, T.M., Keel, W.C., 1990, \apj, 355, 442
\bibitem[Fabbiano \& Juda (1997)]{fabbiano97}
	Fabbiano, G. \& Juda, J.Z., 1997, \apj, 476, 666
\bibitem[Fabbiano \& Trinchieri (1984)]{fabbiano84}
	Fabbiano, G. \& Trinchieri, G., 1984, \apj, 286, 491
\bibitem[\protect\citeauthoryear{Ferrara, Pettini \& Shchekinov}{Ferrara \etal}{2000}]{fps00}
	Ferrara, A., Pettini, M. \& Shchekinov, Y., 2000, \mnras, 319, 539
\bibitem[Ferrara \& Tolstoy (2000)]{ferrara_tolstoy00}
	Ferrara, A. \& Tolstoy, E., 2000, \mnras, 313, 291
\bibitem[Filippenko \& Sargent (1992)]{filippenko92}
	Filippenko, A.V. \& Sargent, W.L.W., 1992, \aj, 103, 28
\bibitem[Ford \etal (1986)]{ford86}
	Ford, H.C., Dahari, O., Jacoby, G.H., Crane, P.C. \&
	Ciadullo, R., 1986, \apjl, 311, L7
\bibitem[F\"{o}rster Schreiber \etal (2001)]{forster01}
	F\"{r}ster Schreiber, N.M., Genzel, R., Lutz, D., Kunze, D. 
	\& Sternberg, A., 2001, \apj, 552, 544
\bibitem[Freedman \etal (1994)]{freedman94}
	Freedman, W.L., \etal, 1994, \apj, 427, 628
\bibitem[Fry \etal (1999)]{fry99}
	Fry, A.M., Morrison, H.L., Harding, P. \& Boroson, T.A.,
	1999, \aj, 118, 1209
\bibitem[Garc\'{i}a-Burillo \& Gu\'{e}lin (1995)]{garciaburillo95}
	Garc\'{i}a-Burillo, S. \& Gu\'{e}lin, M., 1995, \aap, 299, 657
\bibitem[Gehrels (1986)]{gehrels86}
        Gehrels, N., 1986, \apj, 303, 336
\bibitem[Giacconni \etal (2001)]{giacconni2001}
        Giacconni, R., \etal, 2001, \apj, 551, 624
\bibitem[\protect\citeauthoryear{Gibson, Loewenstein \& Mushotsky}{Gibson \etal}{1997}]{gibson97}
	Gibson, B.K., Loewenstein, M. \& Mushotsky, R.F., 1997, 
	\mnras, 290, 623
\bibitem[Griffiths \etal (2000)]{griffiths2000}
	Griffiths, R.E., Ptak, A., Feigelson, E.D., Garmire, G.,  
	Townsley, L., Brandt, W.N.,
	Sambruna, R. \&  Bregman, J.N., Science, 290, 1325
\bibitem[Golla (1999)]{golla99}
	Golla, G., 1999, \aap, 345, 778
\bibitem[\protect\citeauthoryear{Golla, Allen \& Kronberg}{Golla \etal}{1996}]{golla96b}
	Golla, G., Allen, M.L. \& Kronberg, P.P., 1996, \apj, 473, 244
\bibitem[\protect\citeauthoryear{Golla, Dettmar \& Domg\"{o}rgen}{Golla \etal}{1996}]{golla96}
	Golla, G., Dettmar, R.-J. \& Domg\"{o}rgen, H., 1996, \aap, 313, 439
\bibitem[Golla \& Hummel (1994)]{golla94a}
	Golla, G. \& Hummel, E., 1994, \aap, 284, 777
\bibitem[Golla \& Wielebinski (1994)]{golla94b}
	Golla, G. \& Wielebinski, R., 1994, \aap, 286, 733
\bibitem[Grevesse \& Sauval (1998)]{grevesse98}
	Grevesse, N. \& Sauval, A.J., 1998, \ssr,85, 161 
\bibitem[Guianazzi \etal (2000)]{guianazzi00}
	Guianazzi, M., Matt, G., Brandt, W.N., Antonelli, L.A., Barr, P.
	\& Bassani, L., 2000, \aap, 356, 463
\bibitem[Hameed \& Devereux (1999)]{hameed99}
	Hameed, S. \& Devereux, N., 1999, \aj, 118, 730
\bibitem[Hardcastle (2000)]{hardcastle00}
	Hardcastle, M.J., 2000, \aap, 357, 884
\bibitem[Hawarden \etal (1995)]{hawarden95}
	Hawarden, T.G., Israel, F.P., Geballe, T.R. \& Wade, R.,
	1995, \mnras, 276, 1197
\bibitem[\protect\citeauthoryear{Haynes, Giovanelli \& Chincarini}{Haynes \etal}{1984}]{haynes84}
	Haynes, M.P.,  Giovanelli, R. \& Chincarini, G.L., 
	1984, \araa, 22, 445
\bibitem[Heckman (1980)]{heckman80}
	Heckman, T.M., 1980, \aap, 87, 152
\bibitem[Heckman (1999)]{heckman99}
	Heckman, T.M., 1999, \apss, 266, 3
\bibitem[\protect\citeauthoryear{Heckman, Armus \& Miley}{Heckman \etal}{1990}]{ham90}
        Heckman, T. M., Armus, L., \& Miley, G. K.
        1990, \apjs, 74, 833
\bibitem[Heckman (2001)]{heckman01_dust}
	Heckman, T.M., 2001, in Gas and Galaxy Evolution, 
	ed. J.E. Hibbard, M. Rupen, \& J. H. van Gorkom 
	(San Francisco: ASP), 345
\bibitem[Hollenbach \& Tielens (1997)]{hollenbach97}
	Hollenbach, D.J. \& Tielens, A.G.G.M., 1997, \araa, 35, 179
\bibitem[Holweger (2001)]{holweger01}
	Holweger, H., 2001, in AIP Conf. Ser. 598, Solar and Galactic 
	Composition, ed. R.F. Wimmer-Schweingruber (New Your: Springer), 23
\bibitem[\protect\citeauthoryear{Hoopes, Walterbos \& Greenawalt}{Hoopes \etal}{1996}]{hoopes96}
	Hoopes, C,G., Walterbos, R.A.M. \& Greenawalt, B.E., 1996,
	\aj, 112, 1429 
\bibitem[\protect\citeauthoryear{Hoopes, Walterbos \& Rand}{Hoopes \etal}{1999}]{hoopes99}
	Hoopes, C.G., Walterbos, R.A.M. \& Rand, R.J.,
	1999, \apj, 552, 669
\bibitem[Howk \& Savage (1999)]{howk97}
	Howk, J.C. \& Savage, B.D., 1997, \aj, 114, 2463
\bibitem[Howk \& Savage (2000)]{howk00}
	Howk, J.C. \& Savage, B.D., 1999, \aj, 119, 644
\bibitem[Ichikawa \etal (1995)]{ichikawa95}
	Ichikawa, T., Yanagisawa, K., Itoh, N., Tarusawa, K., 
	van Driel, W. \& Ueno, M., 1995, \aj, 109, 2038
\bibitem[Irwin \& Seaquist (1988)]{irwin88}
	Irwin, J.S. \& Seaquist, E.R., 1988, \apj, 335, 658
\bibitem[Irwin \& Seaquist (1991)]{irwin91}
	Irwin, J.S. \& Seaquist, E.R., 1991, \apj, 371, 111
\bibitem[Irwin \etal (1987)]{irwin87}
	Irwin, J.A., Seaquist, E.R., Taylor, A.R. \& Duric, N.,
	1987, \apjl, 313, L91
\bibitem[Irwin \& Sofue (1996)]{irwin96}
	Irwin, J.S. \& Sofue, Y., 1996, \apj, 464, 738
\bibitem[Israel \etal (1998)]{israel98}
	Israel, F.P., van der Werf, P.P., Hawarden, T.G. \&
	Aspin, C., 1998, \aap, 336, 433
\bibitem[Jarret \etal (2003)]{2mass_largegals}
	Jaaret, T.H., Chester, T., Cutri, R., Schneider, S. \& 
	Huchra, J.P., 2003, \aj, 125, 525
\bibitem[Kaaret \etal (2001)]{kaaret01}
	Kaaret, P., Prestwich, A.H., Zezas, A., Murray, S.S.,
	 Kim, D., Kilgard, R.E., Schlegel, E.M. \& Ward, M.J.,
	2001, \mnras, 321, L29
\bibitem[Karachentsev \& Sharina (1997)]{karachentsev97}
	Karachentsev I.D. \& Sharina M.E., 1997, \aap, 324, 457
\bibitem[Kennicutt (1998a)]{kennicutt98a}
	Kennicutt, R.C., 1998a, \apj, 498, 541
\bibitem[Kennicutt (1998b)]{kennicutt98b}
	Kennicutt, R.C., 1998b, \araa, 36, 189
\bibitem[\protect\citeauthoryear{Kennicutt, Tamblyn \& Congdon}{Kennicutt \etal}{1994}]{kennicutt94}
	Kennicutt, R.C, Tamblyn, P. \& Congdon, C.W., 1994, \apj, 435, 22
\bibitem[Kewley \etal (2001)]{kewley2001}
	Kewley, L.J., Heisler, C.A., Dopita, M.A. \& Lumsden, S.,
	2001, \apjs, 132, 37
\bibitem[Koorneef (1993)]{koorneef93}
	Koorneef, J., 1993, \apj, 403, 581
\bibitem[Kormendy \& Bahcall (1974)]{kormendy74}
	Kromendy, J. \& Bahcall, J.N., 1974, \aj, 79, 671
\bibitem[\protect\citeauthoryear{Krabbe, B\"{o}ker \& Maiolino}{Krabbe \etal}{2001}]{krabbe01}
	Krabbe, A., B\"{o}ker, T. \& Maiolino, R., 2001, \apj, 557, 626
\bibitem[van der Kruit (1984)]{kruit84}
	Kruit, P.C. van der, 1984, \aap, 140, 470
\bibitem[Larson (1974)]{larson74}
	Larson, R.B., 1974, \mnras, 169, 229
\bibitem[Lehnert \& Heckman (1995)]{lehnert95}
	Lehnert, M. \& Heckman, T.M., 1995, \apjs, 97, 89
\bibitem[Lehnert \& Heckman (1996)]{lehnert96}
	Lehnert, M. \& Heckman, T.M., 1996, \apj, 472, 546
\bibitem[\protect\citeauthoryear{Lehnert, Heckman \& Weaver}{Lehnert \etal}{1999}]{lhw99}
	Lehnert, M.D., Heckman, T.M. \& Weaver, K.A., 1999, \apj, 523, 575
\bibitem[Levenson \etal (2002)]{levenson02}
	Levenson, N.A., Krolik, J.H., \.{Z}ycki, P.T.,
	Heckman, T.M., Weaver, K.A., Awaki, H. \& Terashima, T.,
	2002, \apjl, 573, L81
\bibitem[\protect\citeauthoryear{Levenson, Weaver \& Heckman}{Levenson \etal}{2001a}]{levenson01a}
	Levenson, N.A., Weaver, K.A. \& Heckman, T.M., 2001a, \apjs, 133, 269
\bibitem[\protect\citeauthoryear{Levenson, Weaver \& Heckman}{Levenson \etal}{2001b}]{levenson01b}
	Levenson, N.A., Weaver, K.A. \& Heckman, T.M., 2001b, \apj, 550, 230
\bibitem[\protect\citeauthoryear{Lira, Johnson \& Lawrence}{Lira \etal}{2002}]{lira02}
	Lira, P., Johnson, R. \& Lawrence, A., 2002, \mnras 
	in press (astro-ph/0206123) 
\bibitem[Lynds \& Sandage (1963)]{lynds63}
	Lynds, C.R. \& Sandage, A.R., 1963, \apj, 137, 1005
\bibitem[McCarthy, Heckman \& van Bruegel (1987)]{mccarthy87}
        McCarthy, P.J., Heckman, T.M. \& van Breugel, W. 1987,
        \aj, 92, 264
\bibitem[McKee (1995)]{mckee95}
	McKee, C.F., 1995, in ASP Conf Series 80, The Physics of the
	Interstellar Medium, A. Ferrara, C.F. McKee, C. Heiles \&
	P.R. Shapiro (San Francisco: ASP), 292
\bibitem[McKee \& Ostriker (1977)]{mckee77}
	McKee, C.F. \& Ostriker, J.P., 1977, \apj, 218, 148 
\bibitem[McKeith \etal (1995)]{mckeith95}
	McKeith, C.D., Greve, A., Downes, D. \& Prada, F., 1995,
	\aap, 293, 703
\bibitem[Mac Low \& Ferrara (1999)]{maclow99}
	Mac Low, M.-M. \& Ferrara, A., 1999, \apj, 513, 142
\bibitem[Mac Low \& McCray (1988)]{maclow88}
	Mac Low, M.-M. \& McCray, R., 1988, \apj, 324, 776 
\bibitem[Maddox \etal (1990)]{apm}
        Maddox, S.J., Efstathiou, G., Sutherland, W.J., Loveday, J.,
        1990, \mnras, 243, 692
\bibitem[Madejski \etal (2000)]{madejski00}
	Madejski, G., \.{Z}ycki, O., Done, C., Valinia, A.,
	Blanco, P., Rothschild, R. \& Turek, B.,
	2000, \apjl, 535, L87 
\bibitem[Marconi \etal (2000)]{marconi00}
	Marconi, A., Oliva, E., van der Werf, P.P.,
	Maiolino, R., Schreier, E.J., Macchetto, F. \& Moorwood, A.F.M.,
	2000, \aap, 357, 24
\bibitem[Markevitch (2001)]{markevitch01}
	Markevitch, M., 2001, ``General discussion of the quiescent 
	and flare components of the ACIS background,'' (CXO: Cambridge),
	\url{http://asc.harvard.edu/cal/Links/Acis/acis/Cal\_prods/bkgrnd/current/background.html}
\bibitem[Martin \& Kern (2001)]{martin01}
	Martin, C. \& Kern, C., 2001, \apj, 555, 258
\bibitem[\protect\citeauthoryear{Martin, Heckman \& Kobulnicky}{Martin \etal}{2002}]{martin02}
	Martin, C.L., Heckman, T.M. \& Kobulnicky, H.A., 2002,
	\apj, 574, 663
\bibitem[Matsumoto \etal (2001)]{matsumoto01}
	Matsumoto, H., Tsuru, T.G., Koyama, K., Awaki, H.,
        Canizares, C.R., Kawai, N,. Matsushita, S.,
        Prestwich, A., Ward, M., Zezas, A.L. \& Kawabe, R.,
	2001, \apjl, 547, L25 
\bibitem[Matsushita \etal (2000)]{matsushita00}
	Matsushita, S., Kawabe., R., Matsumoto., H., Tsuru, T.G.,
	Kohno, K., Morite, K.-I., Okumura, S.K. \& Vila-Vilar\'{o}, B.,
	2000, \apjl, 545, L107
\bibitem[Mayya \& Rengarajan (1997)]{mayya97}
	Mayya, Y.D. \& Rengarajan, T.N., 1997, \aj, 114, 932
\bibitem[\protect\citeauthoryear{Mewe, Kaastra \& Liedahl}{Mewe \etal}{1995}]{mekal}
        Mewe R., Kaastra J. S., 
        Liedahl D. A., 1995, Legacy, 6, 16
\bibitem[Miller \& Veilleux (2003)]{miller03}
	Miller, S.T. \& Veilleux, S., 2003, \apjs, 148, 383
\bibitem[Moorwood \& Oliva (1994)]{moorwood94}
	Moorwood, A.F.M. \& Oliva, E., 1994, \apj, 429, 602
\bibitem[Moorwood \etal (1996)]{moorwood96}
	Moorwood, A.F.M., van der Werf, P.P., Kotilainen, J.K.,
	Marconi, A. \& Oliva, E., 1996, \aap, 308, L1
\bibitem[\protect\citeauthoryear{Moran, Lehnert \& Helfand}{Moran \etal}{1999}]{moran99}
	Moran, E.C., Lehnert, M.D., Helfand, D.J., 1999,
	\apj, 526, 649
\bibitem[Morrison \& McCammon (1983)]{morrison83}
        Morrison, R. \& McCammon, D., 1983, \apj, 270, 119
\bibitem[Muxlow \etal (1994)]{muxlow94}
	Muxlow, T.W.B., Pedlar, A., Wilkinson, P.N., Axon, D.J.,
	Sanders, E.M. \& de Bruyn, A.G., 1994, \mnras, 266, 455
\bibitem[Nakai (1989)]{nakai89}
	Nakai, N., 1989, \pasj, 41, 1107
\bibitem[Norman \& Ferrara (1996)]{norman96}
	Norman, C.A. \& Ferrara, A., 1996, \apj, 467, 280
\bibitem[Norman \& Ikeuchi (1989)]{norman89}
	Norman, C. A. \& Ikeuchi, S., 1989, \apj, 345, 372
\bibitem[Olling (1996)]{olling96}
	Olling, R.P., 1996, \aj, 112, 457
\bibitem[Ott \etal (2001)]{ott2001}
	Ott, M., Whiteoak, J.B., Henkel, K. \& Wielebinski, R.,
	2001, \aap, 372, 463
\bibitem[Pence (1981)]{pence81}
	Pence, W.D., 1981, \apj, 247, 473
\bibitem[\protect\citeauthoryear{Pietsch, Trinchieri \& Vogler}{Pietsch \etal}{1998}]{pietsch98}
	Pietsch, W., Trinchieri, G. \& Vogler, A., 1998,
	\aap, 340, 351
\bibitem[Pietsch \etal (2001)]{pietsch2001}
        Pietsch, W., \etal, 2001, \aap, 365, L174
\bibitem[Popescu \etal (2000)]{popescu00}
	Popescu, C.A., Tuffs, R.J., Fischera, J. \& V\"{o}lk, H.,
	2000, \aap, 354, 480
\bibitem[Ptak \etal (1997)]{ptak97}
	Ptak, A., Serlemitsos, P., Yaqoob, T., Mushotzky, R.
	\& Tsuru, T., 1997, \aj, 113, 1286
\bibitem[Puche \& Carignan (1988)]{puche88}
        Puche, D., Carignan, C., 1988, \aj, 95, 1025
\bibitem[Puche, Carignan \& van Gorkom (1991)]{puche91}
        Puche, D., Carignan, C., van Gorkom, J.H., 1991, \aj, 101, 456
\bibitem[Pudritz \& Fiege (2000)]{pudritz00}
	Pudritz, R.E. \& Fiege, J.D., 2000, in ASP Conf. Series 168,
	New Perspectives on the Interstellar Medium,
	A.R. Taylor, T.L. Landecker \& G. Joncas, (San Francisco: ASP), 235
\bibitem[Radovich, Kahanp\"a\"a \& Lemke (2001)]{radovich}
        Radovich, M., Kahanp\"a\"a, J., Lemke, D., 
        2001, \aap, 377, 73
\bibitem[Rand (1994)]{rand94}
	Rand, R., 1994, \aap, 285, 833
\bibitem[Rand (2000)]{rand00}
	Rand, R.J., 2000, \apj, 535, 663
\bibitem[\protect\citeauthoryear{Rand, Kulkarni \& Hester}{Rand \etal}{1990}]{rand90}
	Rand, R.J., Kulkarni, S.R. \& Hester, J.J., 1990, \apj, 352, 1
\bibitem[\protect\citeauthoryear{Rand, Kulkarni \& Hester}{Rand \etal}{1992}]{rand92}
	Rand, R.J., Kulkarni, S.R. \& Hester, J.J., 1992, \apj, 396, 97
\bibitem[Read \& Ponman (2001)]{read01}
	Read, A.M. \& Ponman, T.J., 2001, \mnras, 328, 127
\bibitem[\protect\citeauthoryear{Read, Ponman \& Strickland}{Read \etal}{1997}]{rps97}
	Read, A.M., Ponman, T.J., Strickland, D.K., 1997, \mnras, 286, 626
\bibitem[Rice \etal (1988)]{rice88}
	Rice, W., Lonsdale, C.J., Soifer, B.T., Neugebauer, G.,
	Kopan, E.L., Llyod, L.A., deJong, T. \& Habing, H.J.,
	1988, \apjs, 68, 91
\bibitem[Rots (1978)]{rots78}
	Rots, A. H., 1978, \aj, 83, 219
\bibitem[\protect\citeauthoryear{Roth, Mould \& Davies}{Roth \etal}{1991}]{roth91}
	Roth, J., Mould, J.R. \& Davies, R.D., 1991, \aj, 102, 1303
\bibitem[Rupen (1991)]{rupen91}
	Rupen, M.P., 1991, \aj, 102, 48
\bibitem[Rupen \etal (1987)]{rupen87}
	Rupen, M.P., van Gorkom, J.H., Knapp, G.R. \& Gunn, J.E.,
	1987, \aj, 94, 61
\bibitem[Sanders \& Mirabel (1996)]{sander96}
	Sanders, D.B. \& Mirabel, I.F., 1996, \araa, 34, 749
\bibitem[Savage \& Sembach (1996)]{savage_sembach96}
	Savage, B.D. \& Sembach, K.R., 1996, \araa, 34, 279 
\bibitem[Sawada-Satoh \etal (2000)]{sawada00}
	Sawado-Satoh, S., Inoue, M., Shibata, K.M., Kameno, S.,
	Migenes, V., Nakai, N. \& Diamond, P.J., 2000, \pasj, 52, 421
\bibitem[Schiano (1985)]{schiano85}
	Schiano, A.V.R., 1985, \apj, 299, 24
\bibitem[Schlickeiser, Werner \& Wielebinski (1984)]{schlickeiser84}
	Schlickeiser, R., Werner W., \& Wielebinski, R., 1984, \aap, 140, 277
\bibitem[\protect\citeauthoryear{Schmidt, Angel \& Cromwell}{Schmidt \etal}{1976}]{schmidt76}
	Schmidt, G.D., Angel, J.R.P. \& Cromwell, R.H., 1976, \apj, 206, 888
\bibitem[Scoville \etal (1993)]{scoville93}
	Scoville, N.Z., Thakkar, D., Carlstrom, J.E. \& Sargent, A.I.,
	1993, \apjl, 404, L59
\bibitem[\protect\citeauthoryear{Schurch, Roberts \& Warwick}{Schurch \etal}{2002}]{schurch02}
	Schurch, N.J., Roberts, T.P. \& Warwick, R.S., 2002, \mnras, 335, 241
\bibitem[Seaquist, Davis \& Bignell (1978)]{seaquist78}
	Seaquist, E.R., Davis, L. \& Bignell, R.C., 1978, \aap, 63, 199
\bibitem[Shapiro \& Field (1976)]{shapiro76}
	Shapiro, P.R. \& Field, G.B., 1976, \apj, 205, 762
\bibitem[\protect\citeauthoryear{Shapiro, Giroux \& Babul}{Shapiro \etal}{1994}]{shapiro94}
	Shapiro, P.R., Giroux, M.L. \& Babul, A., 1994,  \apj, 427, 25
\bibitem[Shapley \etal (2003)]{shapley03}
	Shapley, A.E., Steidel, C.C., Pettini, M. \& Adelburger, K.L.,
	2003, \apj, 588, 65
\bibitem[Shopbell \& Bland-Hawthorn (1998)]{shopbell98}
	Shopbell, P.L., Bland-Hawthorn, J., 1998, \apj, 493, 129 
\bibitem[Silich \& Tenorio-Tagle (1998)]{silich98}
	Silich, S.A. \& Tenorio-Tagle, G., 1998, \mnras, 299, 249
\bibitem[Silich \& Tenorio-Tagle (2001)]{silich01}
	Silich, S.A. \& Tenorio-Tagle, G., 2001, \apj, 552, 91
\bibitem[Slavin \& Cox (1993)]{slavin93}
	Slavin, J.D. \& Cox, D.P., 1993, \apj, 417, 187
\bibitem[Smith \etal (2001)]{smith01}
	Smith, A.M., \etal, 2001, \apj, 546, 829
\bibitem[Snowden \etal (1995)]{snowden95_rass}
	Snowden, S.L., \etal, 1995, \apj, 454, 643
\bibitem[Snowden \& Pietsch (1995)]{snowden95}
	Snowden, S.L. \& Pietsch, W., 1995, \apj, 452, 627
\bibitem[Sofue (1997)]{sofue97}
	Sofue, Y., 1997, \pasj, 49, 17
\bibitem[Sofue \& Vogler (2001)]{sofue_vogler}
	Sofure, Y. \& Vogler, A., 2000, \aap, 370, 53
\bibitem[\protect\citeauthoryear{Sofue, Wakamatsu \& Malin}{Sofue \etal}{1994}]{sofue}
	Sofue, Y., Wakamatsu, K. \& Malin, D.F., 1994, \aj, 108, 2102
\bibitem[Soifer \etal (1989)]{soifer89}
	Soifer, B.T., \etal, 1989, \aj, 98, 1766
\bibitem[Soifer \etal (1987)]{soifer87}
	Soifer, B.T., Sanders, D.B., Madore, B.F., Neugebauer, G.,
	Danielson, G.E., Elias, J.H., Lonsdale, C.J. \&
	Rice, W.L., 1987, \apj, 320, 238
\bibitem[Sommer-Larsen \etal (2002)]{sommerlarsen02}
	Sommer-Larsen, J., Toft, S., Rasmussen, J., Pedersen, K.,
	Gotz, M. \& Portinari, L., 2003, \apss, 284, 693
\bibitem[Sorai \etal (2000)]{sorai2000}
	Sorai, K., Nakai, N., Kuno, N., Nishiyama, K. \&
	Hasegawa, T., 2000, \pasj, 52, 785
\bibitem[Spoon \etal (2000)]{spoon00}
	Spoon, H.W.W., Koornneef, J., Moorwood, A.F.M., Lutz, D.
	\& Tielens, A.G.G.M., 2000, \aap, 357, 898
\bibitem[Strickland (2002)]{strickland_vulcano}
	Strickland, D.K., 2002, in ASP Conf. Series 253 ``Chemical 
	Enrichment 
        of the ICM and IGM,'' Eds. R. Fusco-Femiano \& F. Matteucci, 
	(Sheridan Books: Michigan), 387
\bibitem[Strickland \etal (2000)]{strickland00}
	Strickland, D.K., Heckman, T.M., Weaver, K.A. \& Dahlem, M.,
	2000, \aj, 120, 2965
\bibitem[Strickland \etal (2003)]{strickland_iau}
	Strickland, D.K., Heckman, T.M., Colbert, E.J.M., Hoopes, C.G.
	\& Weaver, K.A., 2003, in 
	``A Massive Star Odyssey, from Main Sequence to Supernova,''
	Eds. K.A. van der Hucht, A. Herrero \& C. Esteban 
	(ASP: San Francisco), 612
\bibitem[Strickland \etal (2004)]{strickland03}
	Strickland, D.K., Heckman, T.M., Colbert, E.J.M., Hoopes, C.G.
	\& Weaver, K.A., 2004, submitted to ApJ (Paper II)
\bibitem[Strickland \etal (2002a)]{strickland02}
	Strickland, D.K., Heckman, T.M., Weaver, K.A., 
	Hoopes, C.G. \& Dahlem, M., 2002a, \apj, 568, 689
\bibitem[\protect\citeauthoryear{Strickland, Ponman \& Stevens}{Strickland \etal}{1997}]{sps97}
	Strickland, D.K., Ponman, T.J. \& Stevens, I.R., 1997, \aap, 320, 378
\bibitem[Strickland \& Stevens (1999)]{ss99}
	Strickland, D.K. \& Stevens, I.R., 1999, \mnras, 306, 43
\bibitem[Strickland \& Stevens (2000)]{ss2000}
	Strickland, D.K. \& Stevens, I.R., 2000, \mnras, 314, 511
\bibitem[Suchkov \etal (1994)]{suchkov94}
        Suchkov, A.A., Balsara, D,S.,
        Heckman, T.M., Leitherner, C., 1994, \apj, 430, 511
\bibitem[Suchkov \etal (1996)]{suchkov96}
	Suchkov A. A., Berman V. G., Heckman T. M., Balsara D. S., 
        1996, \apj, 463, 528
\bibitem[\protect\citeauthoryear{Sugai, Davies \& Ward}{Sugai \etal}{2003}]{sugai03}
	Sugai, H., Davies, R.I. \& Ward, M.J., 2003, \apjl, 584, 9
\bibitem[\protect\citeauthoryear{Swaters, Sancisi \& van der Hulst}{Swaters \etal}{1997}]{swaters97}
	Swaters, R.A., Sancisi, R. \& Hulst, J.M., van der, 1997, 
	\apj, 491, 140
\bibitem[Toft \etal (2002)]{toft02}
	Toft, S., Rasmussen, J., Sommer-Larsen, J. \& Pedersen, K.,
	2002, \mnras, 335, 799
\bibitem[Townsley \etal (2000)]{cti_correct}
	Townsley, L.K., Broos, P.S., Garmire, G.P. \&
	Nousek, J.A., 2000, \apjl, 534, L139
\bibitem[Tremonti \etal (in preparation)]{tremonti}
	Tremonti, C.A., \etal, in preparation
\bibitem[Trotter \etal (1998)]{trotter98}
	Trotter, A.S., Greenhill, L.J., Moran, J.M., Reid, M.J.,
	Irwin, J.A. \& Lo, K.-Y., 1998, \apj, 495, 740
\bibitem[Tsuru \etal (1997)]{tsuru97}
	Tsuru, T.G., Awaki, H., Koyama, K. \& Ptak, A., 1997, \pasj, 49, 619
\bibitem[Ulvestad (2000)]{ulvestad00}
	Ulvestad, J.S., 2000, \aj, 120, 670
\bibitem[Ulvestad \& Antonucci (1997)]{ua97}
        Ulvestad, J.S. \& Antonucci, R.R.J 1997, \apj, 488, 621
\bibitem[Veilleux \etal (2002)]{veilleux02b}
	Veilleux, S., Cecil, G., Bland-Hawthorn, J. \&
	Shopbell, P.L., 2002, Revista 
        Mexicana de Astronomia y Astrofisica, 13, 222
\bibitem[Veilleux \etal (1994)]{veilleux94}
	Veilleux, S., Cecil, G., Bland-Hawthorn, J.,
	Tully, R.B., Filippenko, A.V. \& Sargent, W.L.W.,
	1994, \apj, 433, 48
\bibitem[Veilleux \& Rupke (2002)]{veilleux02}
	Veilleux, S. \& Rupke, D.S., 2002, \apjl, 565, L63
\bibitem[\protect\citeauthoryear{Veilleux, Shopbell \& Miller}{Veilleux \etal}{2001}]{n2992}
	Veilleux, S., Shopbell, P.L. \& Miller, S.T., 2001. \aj, 121, 198
\bibitem[Vogler \& Pietsch (1996)]{vogler96}
	Vogler, A. \& Pietsch, W., 1996, \aap, 311, 35
\bibitem[\protect\citeauthoryear{Vogler, Pietsch \& Kahabka}{Vogler \etal}{1995}]{vogler95}
	Vogler, A., Pietsch, W. \& Kahabka, P., 1995, \aap, 305, 71
\bibitem[Voit (1996)]{voit96}
	Voit, G.M., 1996, \apj, 465, 548
\bibitem[\protect\citeauthoryear{Wainscoat, de Jong \& Wesselius}{Wainscoat \etal}{1987}]{wainscoat87}
	Wainscoat, R.J., de Jong, T. \& Wesselius, P.R., 1987, \aap, 181, 225
\bibitem[\protect\citeauthoryear{Wang, Heckman \& Lehnert}{Wang \etal}{1997}]{wang97}
	Wang, J., Heckman, T.M. \& Lehnert, M.D., 1997, \apj, 491, 114
\bibitem[Wang \etal (2001)]{wang2001}
	Wang, Q.D., Immler, S., Walterbos, R., Lauroesch, J.T. \&
	Breitschwerdt, D., 2001, \apjl, 555, L99
\bibitem[Wang \etal (1995)]{wang95}
	Wang, Q.D., Walterbos, R.A.M., Steakley, M.F.,
	Norman, C.A. \& Braun, R., 1995, \apj, 439, 176
\bibitem[\protect\citeauthoryear{Weaver, Heckman \& Dahlem}{Weaver \etal}{2000}]{weaver00}
	Weaver, K.A., Heckman, T.M. \& Dahlem, M., 2000,
	\apj, 534, 684
\bibitem[Weaver \etal (2002)]{weaver02}
	Weaver, K.A., Heckman, T.M., Strickland, D.K. \& Dahlem, M.,
	2002, \apjl, 576, L19
\bibitem[Weaver \etal (1995)]{weaver95}
	Weaver, K.A., Mushotzky, R.F., Serlemitsos, P.J.,
	Wilson, A.S., Elvis, M. \& Briel, U., 1995, \apj, 442, 597
\bibitem[Weiss \etal (1999)]{weiss99}
	Weiss, A., Walter, F., Neininger, N. \& Klein, U.,
	1999, \aap, 345, L23
\bibitem[Weliachew, Fomalont \& Greisen (1984)]{weliachew84}
	Weliachew, I., Fomalont, E.B. \& Greisen, E.W.,
	1984, \aap, 137, 335
\bibitem[Whiteoak (1986)]{whiteoak86}
	Whiteoak, J.B., 1986, Proc. Astronom. Soc. Australia., 6, 467
\bibitem[\protect\citeauthoryear{Williams, Baker \& Perry}{Williams \etal}{1999}]{williams99}
	Williams, R.J.R., Baker, A.C. \& Perry, J.J., 1999, \mnras, 310, 913
\bibitem[\protect\citeauthoryear{Wills, Pedlar \& Muxlow}{Wills \etal}{2002}]{wills02}
	Wills, K.A., Pedlar, A. \& Muxlow, T.W.B., 2002, \mnras, 331, 313
\bibitem[Wiseman \& Ho (1998)]{wiseman98}
	Wiseman, J.J. \& Ho, P.T.P., 1998, \apj, 502, 676
\bibitem[\protect\citeauthoryear{Yun, Ho \& Lo}{Yun \etal}{1993}]{yun93}
	Yun, M.S., Ho, P.T.P. \& Lo, K.Y., 1993, \apjl, 411, L17
\bibitem[\protect\citeauthoryear{Zezas, Georgantopoulos \& Ward}{Zezas \etal}{1998}]{zezas98}
	Zezas, A.L., Georgantopoulos, I. \& Ward, M.J., 1998, \mnras, 301, 915
\bibitem[Zhao \etal (1997)]{zhao97}
	Zhao, J.-H., Anantharamaiah, K.R., Goss, W.M. \& Viallefond, F.,
	1997, \apj, 482, 186
\end{thebibliography}
\end{document}